\documentclass[aps, preprint, groupedaddress, floatfix, onecolumn]{revtex4-2}
\usepackage{siunitx}
\usepackage{booktabs}
\usepackage{chemformula}
\usepackage{multirow}
\usepackage{graphicx}
\usepackage{array}
\usepackage{threeparttable}
\usepackage{longtable}
\usepackage{makecell}
\let\ce\ch
\usepackage{orcidlink}
\usepackage[normalem]{ulem}
\usepackage{lineno}
\usepackage{xcolor}
\newcommand{\be}{\begin{equation}}
\newcommand{\ee}{\end{equation}}

\usepackage{hyperref}

\newcommand{\etal}{{\it{et al.}}}

\newcommand{\eg}{{\it{e.g.}}}

\newcolumntype{Y}{>{\centering\arraybackslash}X}
\newcolumntype{Z}{>{\hsize=1.1\hsize\centering\arraybackslash}X}

\begin{document}

\title{
Nonlinear Hall Effect in Two-dimensional Materials
}
\author{Shuo Wang\orcidlink{0009-0006-1703-150X}$^{1}$}
\author{Wei Niu\orcidlink{0000-0002-4331-0928}$^{1}$}
\email{weiniu@njupt.edu.cn}
\author{Yue-Wen Fang\orcidlink{0000-0003-3674-7352}$^{2,3}$}
\email{yuewen.fang@ehu.eus}
\affiliation{$^1$ New Energy Technology Engineering Laboratory of Jiangsu Provence \& School of Science, Nanjing University of Posts and Telecommunications, Nanjing 210023, China \\
$^2$ Centro de F{\'i}sica de Materiales (CFM-MPC), CSIC-UPV/EHU, Manuel de Lardizabal Pasealekua 5, 20018 Donostia/San Sebasti{\'a}n, Spain
}

\begin{abstract}
Symmetry is a cornerstone of condensed matter physics, fundamentally shaping the behavior of electronic systems and inducing the emergence of novel phenomena. The Hall effect, a key concept in this field, demonstrates how symmetry breaking, particularly of time-reversal symmetry, influences electronic transport properties. Recently, the nonlinear Hall effect has extended this understanding by generating a transverse voltage that modulates at twice the frequency of the driving alternating current without breaking time-reversal symmetry. This effect is closely tied to the symmetry and quantum geometric properties of materials, offering a new approach to probing the Berry curvature and quantum metric. Here, we provide a review of the theoretical insights and experimental advancements in the nonlinear Hall effect, particularly focusing on its realization in two-dimensional materials. We discuss the challenges still ahead, look at potential applications for devices, and explore how these ideas might apply to other nonlinear transport phenomena. By elucidating these aspects, this review aims to advance the understanding of nonlinear transport effects and their broader implications for future technologies.
\end{abstract}
\maketitle

\section{1. Introduction}
 Symmetry is foundational in condensed matter physics, guiding the understanding and prediction of material behaviors. It plays a crucial role in determining the properties of structural, electronic, optical, and magnetic systems\cite{ruter2010observation,liu2023symmetry,dembowski2003observation,zhang2024light,structural-symmetry2024JMI-Zhang}. By considering the symmetry, one can classify phases of matter, comprehend phase transitions, and predict the emergence of new quantum phenomena\cite{ji2020categorical,bender2005introduction,li2024observation,Zhao-PRB-symmetry-Rashba2020}. For examples, in optics, symmetry dictates the selection rules for transitions and governs the behavior of light in various media\cite{brunschwig2002optical,chen2024defect}. In electronics, symmetry strongly affects the band structures of materials by modulating the band degeneracy, effective mass and topology, thus playing a significant role in determining electron transport properties~\cite{Niu-jpclett2022,Niu-acs.nanolett2017-GAOSTO,Hu2021-AdvEnergyMater,Hu2021-NatCommun-Pb7Bi4Se13,DXZheng-acsnano2021-halleffect}.
The Hall effects, characterized by the emergence of a transverse voltage in response to a longitudinal current, exemplify the profound influence of symmetry, manifesting in diverse and distinctive forms. The conventional Hall effect arises from the breaking of time-reversal (\textit{T}) symmetry under the influence of an external magnetic field, whereas the quantum Hall effect is marked by quantized conductance in the presence of strong magnetic field\cite{stormer1999nobel,popovic1989hall}. The anomalous Hall effect originates from intrinsic \textit{T} symmetry breaking driven by magnetic order\cite{nagaosa2010anomalous}, while the spin Hall effect leverages spin-orbit coupling, notably without violating \textit{T} symmetry\cite{sinova2015spin}.
 
\begin{figure}
    \centering
    \includegraphics[width=0.99\linewidth]{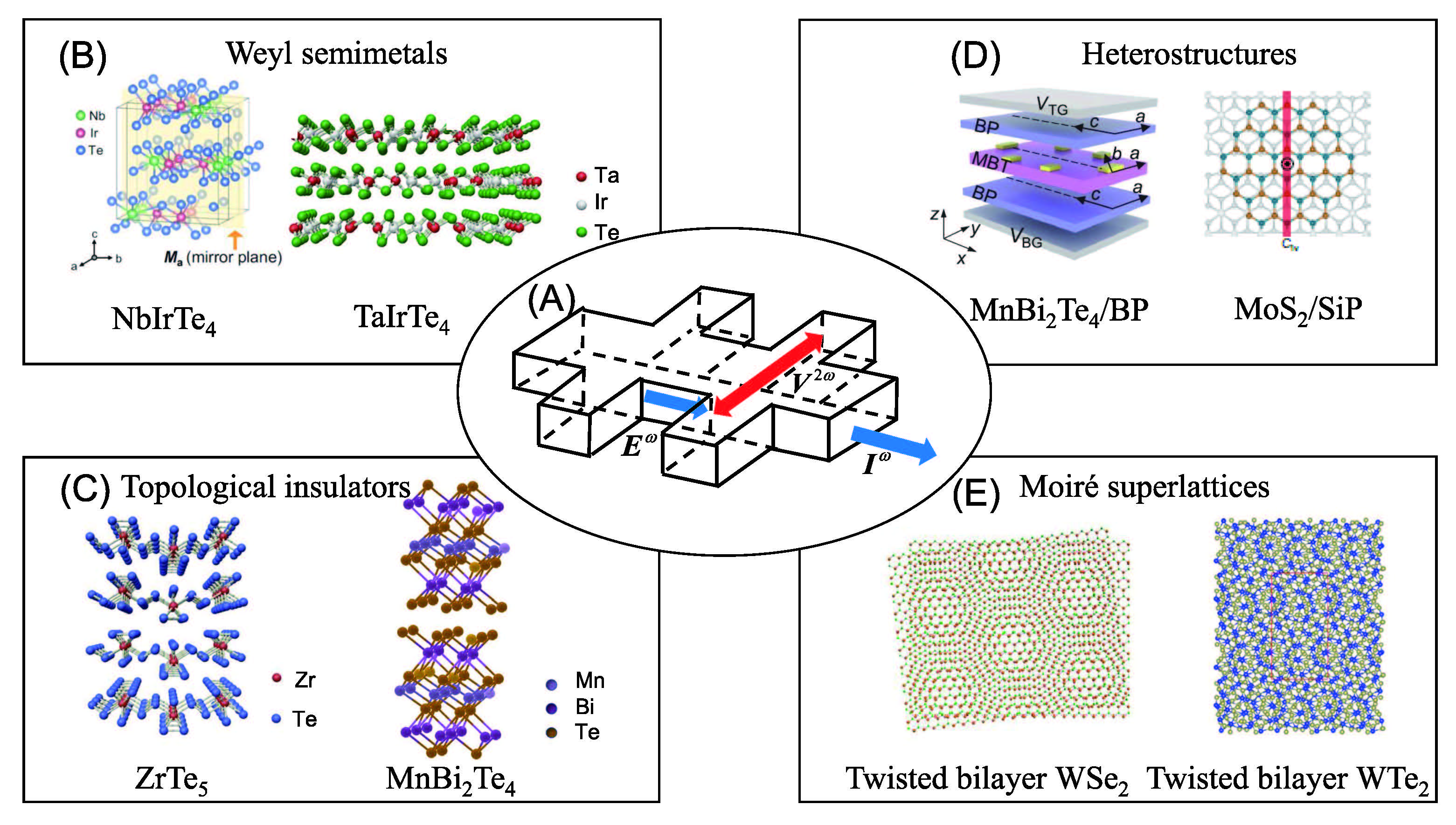}
    \caption{Advancements in the exploration of the nonlinear Hall response in various 2D systems. (A) Schematic of the measurement geometry of the second-order nonlinear Hall effect. (B, C) The nonlinear Hall response in 2D Weyl semimetals and 2D topological insulators with inherent symmetry breaking. (D, E) The nonlinear Hall response in heterostructures and moiré superlattices through structural engineering. Panel B reprinted with permission\cite{lee2024spin}, 2024, CC BY license. Panel D reprinted with permission\cite{gao2023quantum,duan2023berry}. Copyright 2023, The American Association for the Advancement of Science and 2023, CC BY license. Panel E reprinted with permission\cite{he2021giant,hu2022nonlinear}, 2021, CC BY license and 2022, CC BY license.}
    \label{fig:Experimental observation}
\end{figure}

 The nonlinear Hall effect, as depicted in Figure \ref{fig:Experimental observation}A, is a recently uncovered phenomenon characterized by the emergence of a Hall voltage that scales quadratically with the applied current. In non-magnetic systems, this effect arises exclusively from the breaking of parity inversion symmetry (\textit{P}), without requiring the violation of time-reversal symmetry, thereby providing profound insights into the topological and symmetry characteristics of materials\cite{sodemann2015quantum}. Furthermore, in magnetic systems, the realization of the nonlinear Hall effect, driven by the quantum metric, requires the simultaneous breaking of both \textit{P} and \textit{T} symmetries.
 Therefore, the nonlinear Hall effect serves as a novel probe for exploring quantum phases and is prevalent in a wide range of materials, particularly in two-dimensional (2D) systems, such as 2D Weyl semimetals and 2D elemental materials\cite{kang2019nonlinear,ma2019observation,tiwari2021giant,kumar2021room,wang2023quantum,gao2023quantum,rostami2020probing}. 
Although nonlinear Hall effect is usually observed in materials with metallic features, He and Law's recent theoretical suggested that this effect can also occur in insulators when the driving frequency approaches the band gap~\cite{He2024_nonlinearhalleffect_insulators}.
This ability to generate a transverse voltage, regardless of whether the material is metallic or insulating and without requiring an external magnetic field, opens new possibilities for applications in electronic devices and sensors.

 In this review, we initially introduce the fundamental symmetries underlying the nonlinear Hall effect, providing a theoretical framework to comprehend this phenomenon. Then the mechanisms and various factors resulting in this effect are discussed. Subsequently, experimental advancements in detecting the nonlinear Hall effect in 2D materials are explored, as illustrated in Figure \ref{fig:Experimental observation}B-E. This includes cases of inherent symmetry breaking in specific materials and instances resulting from structural engineering. Finally, we outline potential devices utilizing this effect and other nonlinear transport phenomena, highlighting promising avenues for future research and technological innovation.


\section{2. FUNDAMENTAL SYMMETRIES}
In electromagnetic phenomena, two fundamental symmetries play a crucial role: parity inversion and time-reversal symmetry. Parity inversion symmetry involves spatial inversion. It refers to the invariance of physical laws under inversion of spatial coordinates. 
Mathematically, this is represented as ($x \rightarrow -x, y \rightarrow -y, z \rightarrow -z$). Time-reversal symmetry, on the other hand, involves reversing the direction of time, defined by the operation ($t \rightarrow -t$). Therefore, the combination of parity and time-reversal (\textit{PT}) symmetry can impose both parity inversion and time-reversal transformations simultaneously ($x \rightarrow -x, y \rightarrow -y, z \rightarrow -z$, and $t \rightarrow -t$).

\begin{figure}[htb]
    \centering
    \includegraphics[width=0.63\linewidth]{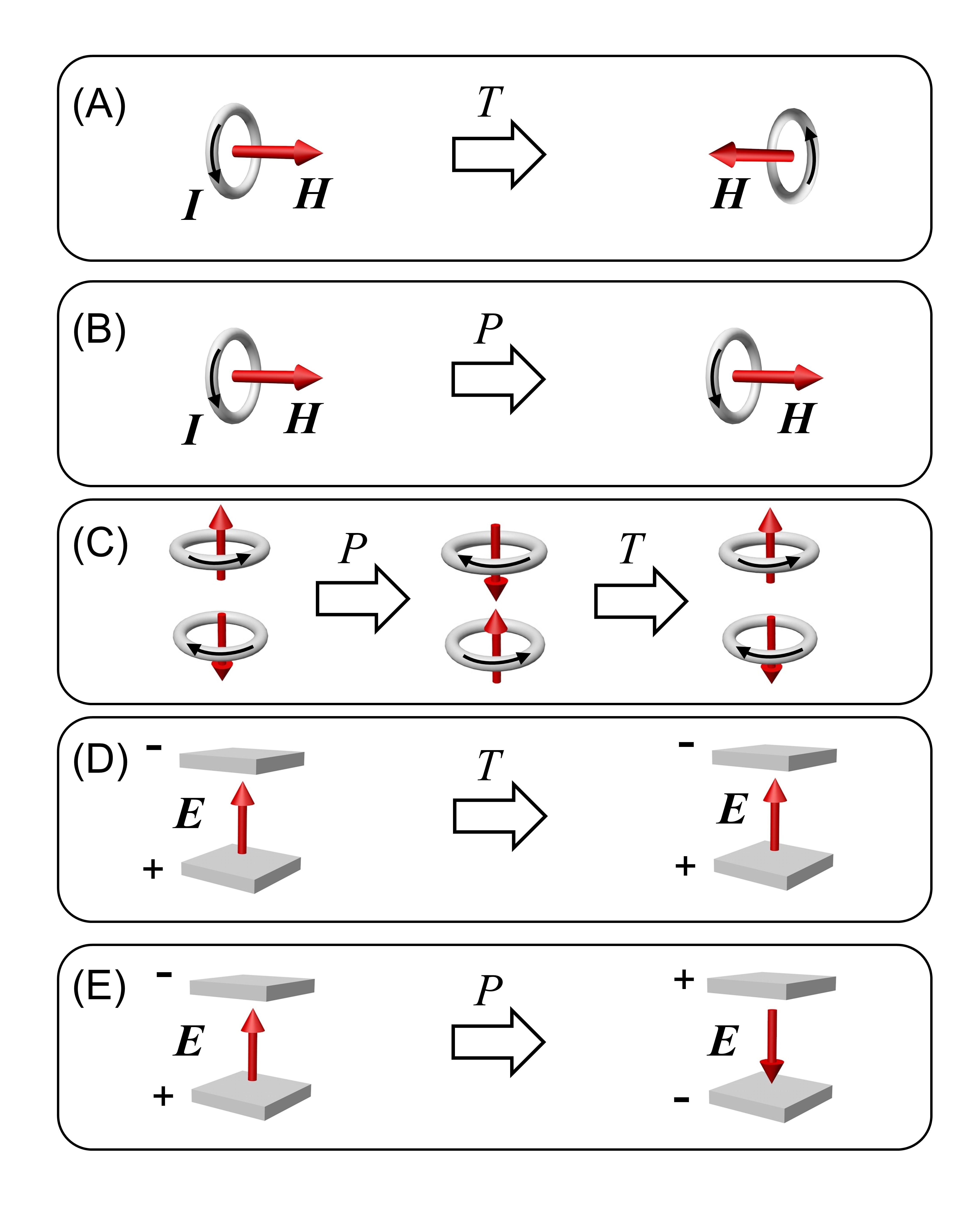}
    \caption{Fundamental symmetries for \textbf{\textit{E}} and \textbf{\textit{H}}. (A, B) \textbf{\textit{H}} is an axial vector. It is odd under time-reversal and even under parity inversion. (C) The symmetry of two antiparallel magnetic fields, both \textit{P} and \textit{T} symmetries are broken but \textit{PT} symmetry is preserved. (D, E) \textbf{\textit{E}} is a polar vector. It is even under time-reversal and odd under parity inversion.}
    \label{fig:fundamental-symmetry}
\end{figure}

According to the Maxwell equations, a magnetic field (\textbf{\textit{H}}) can be generated by a circular current loop, while an electric field (\textbf{\textit{E}}) arises from two oppositely charged plates. As illustrated in Figure \ref{fig:fundamental-symmetry}A, under time-reversal transformation, the current flow within the circular loop reverses, thus inverting the direction of magnetic field and confirming odd nature of \textbf{\textit{H}} under time-reversal. Consequently, the application of \textbf{\textit{H}} explicitly breaks the \textit{T} symmetry\cite{liu2023symmetry}. In contrast, as Figure \ref{fig:fundamental-symmetry}B demonstrates, \textbf{\textit{H}} remains even under parity inversion, as the orientation of current remains unchanged when spatial coordinates are inverted. As shown in Figure \ref{fig:fundamental-symmetry}C, in the case of two antiparallel magnetic fields, the \textit{P} and \textit{T} symmetry are individually broken, as the magnetic field direction reverses under both parity inversion and time-reversal. However, \textit{PT} symmetry remains intact, as the magnetic field direction is preserved when both transformations are applied sequentially\cite{gao2021layer}. The electric field, however, exhibits distinct symmetry behavior. Under time-reversal, the charge distributions on the plates remain static, leaving \textbf{\textit{E}} invariant, as depicted in Figure \ref{fig:fundamental-symmetry}D. In contrast, under parity inversion, \textbf{\textit{E}} reverses its direction due to the exchange in spatial position of the two oppositely charged plates, as illustrated in Figure \ref{fig:fundamental-symmetry}E\cite{liu2023symmetry,du2021nonlinear}.

Based on the symmetry considerations outlined above, the \textit{T} symmetry in the Hall effect is broken by applying a magnetic field\cite{klitzing1980new,yasuda2016geometric}. Additionally, this symmetry framework elucidates the conditions required for the nonlinear Hall effect, in which the induced current \textbf{\textit{J}} scales quadratically with the electric field \textbf{\textit{E}} rather than linearly, represented as:
\begin{equation}
     J_{a}=\chi_{abc}E_{b}E_{c},
     \label{con:1}
\end{equation}
where $\chi_{abc}$ denotes the tensor of nonlinear response \cite{sodemann2015quantum}. Here, both the current \textbf{\textit{J}} and electric field \textbf{\textit{E}} change sign under parity transformation. Consequently, $\chi_{abc}$ must also inverts its sign to preserve consistency across both sides of the equation. As a result, the second-order nonlinear response necessitates the breaking of \textit{P} symmetry\cite{du2021nonlinear}.

\section{3. MECHANISMS UNDERLYING THE NONLINEAR HALL EFFECT}
The nonlinear Hall effect arises from a rich interplay of fundamental symmetries that govern the behavior of charge carriers in materials. Through symmetry analysis, researchers have identified various mechanisms that can give rise to this effect. Among them, the quantum metric dipole and Berry curvature dipole are crucial, as they illustrate how the geometric properties of the band structure influence transport phenomena\cite{wang2023quantum,sodemann2015quantum,gao2014field,wang2021intrinsic}. Additionally, disorder could also play an important role in the nonlinear Hall response by various mechanisms such as side-jump and skew scattering \cite{du2019disorder,du2021quantum}. As a result, these factors create a diverse landscape for understanding and harnessing the nonlinear Hall effect across different materials.

\subsection{3.1 Quantum metric dipole}

In condensed matter physics, quantum geometry: $Q_{n}^{\mu v}=-\frac{i}{2}\Omega_{n}^{\mu v}+g_{n}^{\mu v}$ plays a significant role in understanding various phenomena, especially those related to the electronic properties of materials\cite{tokura2018nonreciprocal,provost1980riemannian}. The quantum geometry can typically be divided into two main components: the imaginary part is Berry curvature, which describes the phase difference between two quantum states.
The real component is the quantum metric, 
delineating the amplitude different between them. Both of these geometric quantities are crucial to determining the behavior of electrons in various nonlinear transport phenomena\cite{sodemann2015quantum,wang2021intrinsic}.

For the nonlinear Hall effect, the quantum metric is one of the indispensable points contributing to the second-order current response, whose conductivity tensor characterizes the second-order current response of the current \textbf{\textit{J}} to \textbf{\textit{E}}: 
\begin{equation}
    J^\alpha=\sum_{\beta,\gamma}\sigma^{\alpha\beta\gamma} E^\beta E^\gamma ,
     \label{con:2}
\end{equation}
($\alpha$, $\beta$ and $\gamma$ are Cartesian indices)\cite{sodemann2015quantum,wang2021intrinsic}. $\sigma$ can be divided into two components: the Ohmic component and the Hall component\cite{tsirkin2022separation}. The Ohmic component comprises a Drude conductivity of second-order, which depends on the relaxation time $\tau$\cite{holder2020consequences,watanabe2020nonlinear}. While, the Hall component contains a $\tau$ independent term\cite{gao2014field}. Based on the $\sigma-\tau$ dependencies, the contribution of the second-order nonlinear Hall effect is able to be separated into intrinsic ($\tau$-independent) and extrinsic ($\tau$-dependent) components\cite{gao2014field,du2019disorder}. It is noteworthy that, although the Berry curvature dipole actually depends on the relaxation time, it is generally regarded as the intrinsic contribution in research\cite{du2019disorder}. On the other hand, disorder-related side-jump and skew scattering are normally attributed to the extrinsic mechanisms (Table \ref{tab:summary mechanisms}).

The nonlinear Hall effect induced by the quantum metric dipole arises in the presence of an electric field. The electron velocity under electric field $\textbf{\textit{E}}$ consists of two contributions, the group velocity, and the anomalous velocity due to the Berry curvature:
\be
{\dot{\textbf{\textit{r}}}=\frac{1}{\hbar}\nabla_{k}\varepsilon+\frac{e}{\hbar}\textbf{\textit{E}}\times\Omega}.
\label{electron velocity}
\ee
Recently, Gao \etal~demonstrated that the electric field can introduce a gauge-invariant correction to the Berry connection: $A^{\textbf{\textit{E}}}(\textbf{\textit{k}})=G(\textbf{\textit{k}})\textbf{\textit{E}}$, where $G(\textbf{\textit{k}})$ is the Berry connection polarizability tensor\cite{gao2014field}, defined as
\be
G_{n}^{jk}(\textbf{\textit{k}})=2{\rm Re}\sum_{m\neq n}\frac{A_{nm}^{j}(\textbf{\textit{k}})A_{mn}^{k}(\textbf{\textit{k}})}{\varepsilon_{n}(\textbf{\textit{k}})-\varepsilon_{m}(\textbf{\textit{k}})}.
\ee
As a result, the Berry curvature can be rewritten as ${\widetilde{\Omega}}=\Omega+\nabla_{k}\times A^{\textbf{\textit{E}}}(\textbf{\textit{k}})$. Thus the electric velocity becomes 
\be
{\dot{\textbf{\textit{r}}}=\frac{1}{\hbar}\nabla_{k}\varepsilon+\frac{e}{\hbar}\textbf{\textit{E}}\times\Omega}+\frac{e}{\hbar}\times \textbf{\textit{E}}\times\nabla_{k}\times G(\textbf{\textit{k}})\textbf{\textit{E}}.
\ee
The third term represents an anomalous velocity that is proportional to the square of the electric field and is independent of the $\tau$. Therefore, it contributes to the intrinsic nonlinear Hall effect. Based on the Eq.(\ref{con:1}), the intrinsic nonlinear Hall tensor is\cite{liu2021intrinsic}:
\be
 \chi_{\rm INH} =e^{3}\sum_{n}\int\frac{d^{3}\textbf{\textit{k}}}{(2\pi)^{3}}v_{n}^{i}G_{n}^{jk}\frac{\partial f_{0}(\varepsilon_{n})}{\partial \varepsilon_{n}}-(i\leftrightarrow j).
\ee
$G_{n}^{jk}$ respresents the Berry connection polarizability in the \textit{n}-band situation. In the simplified two-band scenario, the Berry connection polarizability $G_{n}^{jk}$ simplifies as:
\be
G_{n}^{jk}=2\frac{{\rm Re}A_{nm}^{j}A_{mn}^{k}}{\varepsilon_{n}-\varepsilon_{m}}=2\frac{g_{n}^{jk}}{\varepsilon_{n}-\varepsilon_{m}},
\ee
where $g_{n}^{jk}$ is the quantum metric, and it is related to the Berry connection for non-diagonal terms. The quantum metric dipole, which characterizes the influence of the quantum geometry on transport phenomena, is given by the product of the group velocity and the quantum metric: $v_{n}^{i}g_{n}^{jk}$, reflecting the spatial variation of the quantum states in momentum space.

When considering symmetries, the emergence of the quantum metric requires the breaking of both \textit{P} and \textit{T} symmetries\cite{wang2023quantum}. In addition, \textit{PT} symmetry eliminates the contributions from the Berry curvature dipole, as the Berry curvature is zero\cite{sodemann2015quantum}. Consequently, a structure that violates both \textit{P} and \textit{T} symmetries while preserving \textit{PT} symmetry provides an ideal setting for investigating the quantum metric dipole induced nonlinear Hall effect\cite{wang2023quantum,gao2023quantum,han2024room}.

\begin{table}[!h]
\centering
\begin{threeparttable}[b]
    \tabcolsep=25pt
     \renewcommand\arraystretch{2.2}
      \caption{Summary of the scattering time $\tau$ dependence and the symmetry analysis for different mechanisms.}
    \begin{tabular}{ccccc}
        \hline
        \multirow{2}*{Mechanism} & \multirow{2}*{$\tau$-dependence} &\multicolumn{3}{c}{Symmetry}\\ 
        & & \textit{P}& \textit{T} & \textit{PT} \\ \hline
        Quantum metric dipole\cite{wang2023quantum} & $\tau^{0}$ & $\times$ & $\times$ & $\surd$\\ 
       Berry curvature dipole\cite{sodemann2015quantum} & $\tau^{1}$ & $\times$ & $\surd$ &$\times$ \\ 
       Side-jump\cite{du2019disorder} & $\tau^{2}$ or $\tau^{1}$ & $\times$ & $\surd$ &$\times$ \\ 
        Skew sacttering\cite{du2019disorder,isobe2020high} &  $\tau^{3}$ or $\tau^{2}$ & $\times$ & $\surd$ & $\times$ \\ \hline
    \end{tabular}
    \label{tab:summary mechanisms}
    \begin{tablenotes}
     \item[] \hfill($\times$ means forbidden, $\surd$ means allowed)
   \end{tablenotes}
    \end{threeparttable}
\end{table}

\subsection{3.2 Berry curvature dipole}
Besides the quantum metric dipole, the Berry curvature dipole can also give rise to the nonlinear transport\cite{kumar2021room,sodemann2015quantum}. The Berry curvature $\Omega$ can be viewed as an effective magnetic field in the parameter space (\eg, momentum space), which has been recognized as the origin of various novel electronic phenomena\cite{berry1984quantal,xiao2010berry,bohm2013geometric,nagaosa2010anomalous,karplus1954hall}. In the linear Hall effect, the Hall conductivity is defined as

\begin{equation}
    \sigma_{ab}^{in}=-\frac{e^{2}}{\hbar}\varepsilon_{abc}\int\frac{d^{n}\textbf{\textit{k}}}{(2\pi)^{n}}\Omega_{c}f_{0},
    \label{con:6}
\end{equation}
where $f_{0}$ is the Fermi distribution, \textit{n} is the dimension, $\varepsilon_{abc}$ is the Levi-Civita symbol\cite{du2021nonlinear}. The Berry curvature enables the calculation of the intrinsic anomalous Hall effect\cite{nagaosa2010anomalous,karplus1954hall}, which is invariant under \textit{P} symmetry but change its sign under \textit{T} symmetry. Therefore, in systems with \textit{T} symmetry, the positive and negative momentum contributions cancel out during integration, resulting in zero Hall conductivity. To achieve non-zero Hall conductivity, the \textit{T} symmetry must be broken by an external magnetic field or internal magnetization of the material.

The Berry curvature dipole induced nonlinear Hall effect can be derived from the Boltzmann equation under the relaxation time approximation, $-e\tau E_{a}\partial_{a}f+\tau \partial_{t}f=f_{0}-f$, where $E_{a}$ is the electric field and $f$ is the distribution function. The electric field is considered to take the form $E_{c}(t)={\rm Re}\{\varepsilon_{c}e^{ei\omega t}\}$\cite{sodemann2015quantum}. The distribution function can be expanded up to second order in terms of the electric field: $f=f_{0}+f_{1}+f_{2} $.
Substituting this expansion into the Boltzmann equation leads to:
\be
-e\tau E_{a}\partial_{a}f_{0}-e\tau E_{a}\partial_{a}f_{1}-e\tau E_{a}\partial_{a}f_{2}=-\tau\partial_{t}f_{1}-f_{1}-\tau \partial_{t}f_{2}-f_{2},
\ee
where it is assumed that $ \partial_{t}f_{0}=0$. This simplifies to: 
\be
    e\tau E_{a}\partial_{a}f_{0}=\tau \partial_{t}f_{1}+f_{1},
    \label{10}
\ee
\be
     e\tau E_{a}\partial_{a}f_{1}=\tau \partial_{t}f_{2}+f_{2}. 
\ee

The term $f_{1}$ could be separated into two parts as follows: $f_{1}=f_{1}^{\omega}e^{i\omega t}+f_{1}^{-\omega}e^{-i\omega t}$. Taking $f_{1}$ into Eq.(\ref{10}), it is found that:
\be 
f_{1}={\rm Re}\{f_{1}^{\omega}e^{i\omega t}\},~~~ f_{1}^{\omega}=\frac{e\tau\mathcal{E}_{a}\partial_{a}f_{0}}{1+i\omega\tau}.
\ee
The term $f_{2}$ can be rewritten as: $f_{2}=f_{2}^{2\omega+}e^{i2\omega t}+f_{2}^{0+}+f_{2}^{0-}+f_{2}^{2\omega-}e^{-i2\omega t}$, Thus, it can be expressed as:
\be
f_{2}={\rm Re}\{f_{2}^{0}+f_{2}^{2\omega}e^{i2\omega t}\},~~~
f_{2}^{0}=\frac{e^{2}\tau^{2}\mathcal{E}_{a}^{\ast}\partial_{ab}f_{0}}{2(1+\omega\tau)},
~~f_{2}^{2\omega}=\frac{e^{2}\tau^{2}\mathcal{E}_{a}\mathcal{E}_{a}\partial_{ab}f_{0}}{2(1+i\omega\tau)(1+2i\omega\tau)}.
\ee
Based on the current density formula $j_{a}=-e\int_{k}f(k)\{\partial_{a}\epsilon(k)-\varepsilon_{abc}\Omega_{b}eE_{c}(t) \}$, it is noted that $\partial_{a}\epsilon(k)\partial_{bc}f_{0}(k)$ is odd under time reversal within the approximation of a constant $\tau$, leading to its vanishing upon integration. The second order current corresponding currents are given by:
\be
j_{a}^{0}=\frac{e^{2}}{2}\int_{k}\varepsilon_{abc}\Omega_{b}\mathcal{E}_{c}^{\ast}f_{1}^{\omega},
\ee
which represents the rectified current, and
\be
j_{a}^{2\omega}=\frac{e^{2}}{2}\int_{k}\varepsilon_{abc}\Omega_{b}\mathcal{E}_{c}f_{1}^{\omega},
\ee
which corresponds to the second harmonic current. Based on the Eq.(\ref{con:1}),
 the tensor of nonlinear Hall response is written as:
\begin{equation}
    \chi_{abc}=\varepsilon_{abc}\frac{e^{3}\tau}{2(1+i\omega\tau)}\int f_{0}(\partial_{b}\Omega_{d}),
\end{equation}

It is clear the $\chi_{abc}$ is proportional to the dipole moment of the Berry curvature integrated over the occupied states, commonly referred to as the Berry curvature dipole\cite{sodemann2015quantum,low2015topological}:

\begin{equation}
    D_{ab}=\int_{k}f_{0}(\partial_{a}\Omega_{b}).
\end{equation}

In this vein, we can understand the requirement of the nonlinear Hall response induced by the Berry curvature dipole more clearly based on the analyses of the symmetry. Firstly, under \textit{T} symmetry, both the group velocity and Berry curvature reverse their signs, but their product remains unchanged, indicating the nonlinear Hall response can be observed with \textit{T} symmetry. Secondly, under \textit{P} symmetry, the group velocity reverses its sign, but the Berry curvature does not, resulting in a zero Berry curvature dipole moment. Therefore, to observe the nonlinear Hall effect, the \textit{P} symmetry must be broken.

Beside the the symmetries mentioned above, other symmetries significantly influence the nonlinear Hall effect. The chiral symmetry plays a pivotal role in the nonlinear Hall effect\cite{joseph2024chirality}. It refers to the absence of improper symmetries, such as inversion and mirror symmetries. The absence of these symmetries results in an inherent asymmetry, distinguishing between left- and right-handed configurations. The broken inversion symmetry, induced by chirality, ensures a finite valley-contrasting Berry curvature, which, in turn, gives rise to a Hall-like net transverse conductivity, contributing to the nonlinear Hall response\cite{zhu2024creating}. Moreover, when chiral symmetry is broken, the nonlinear Hall current can attain quantization\cite{peshcherenko2024quantized}. Beyond chiral symmetry, systems that break \textit{PT} symmetry also manifest a third-order nonlinear Hall effect, driven by the Berry curvature quadrupole. \textit{PT} symmetry enforces the vanishing of Berry curvature quadrupoles, so materials that break \textit{PT} symmetry, such as bulk \ce{MnBi2Te4}\cite{li2024quantum} and \ce{FeSn}\cite{sankar2024experimental} under an external magnetic field, exhibit a third-order nonlinear Hall effect. The mirror symmetry and cyclic rotations symmetry also affect the nonlinear hall effect. Under $C_{n}^{z}$ and $C_{3,4,6}^{z}T$ symmetry, both the quantum metric and berry curvature dipole are forbidden. For the combined $C_{2}^{z}T$ symmetry, the quantum metric dipole is allowed, but the Berry curvature dipole is forbidden. Conversely, $\sigma_{z}T$ symmetry permits only the Berry curvature dipole\cite{liu2021intrinsic}.
\subsection{3.3 Side-jump and skew scattering}
Besides the intrinsic mechanisms, disorder also plays an integral role in a variety of Hall effects, such as the extrinsic contributions of the anomalous, valley, and spin Hall effects\cite{nagaosa2010anomalous,sinova2015spin,mak2014valley,xu2006stability}. In the case of the nonlinear Hall effect, the intrinsic mechanism plays a crucial role, but disorder-related still cannot be overlooked. This is because this effect inherently requires the Fermi energy to intersect with the energy bands, making disorder scattering inevitable at the Fermi surface. In fact, disorder plays a role even at leading order, contributing critically to the overall response\cite{du2019disorder,du2021quantum}.

Utilizing the Boltzmann equation:
\begin{equation}
\frac{\partial f_{l}}{\partial_{t}}+{\dot{\textbf{\textit{k}}}}\cdot \frac{\partial f_{l}}{\partial \textbf{\textit{k}}}=-\sum_{l'}(W_{l'l}f_{l}-W_{ll'}f_{l'}),
\end{equation}
where $W_{l'l}$ is the average scattering rate from $l'$ to $l$,
in conjunction with the 2D tilted massive Dirac model:
\be
\widehat{H}=tk_{x}+v(k_{x}\sigma_{x}+k_{y}\sigma_{y})+m\sigma_{z},
\ee
Du \etal~systematically classified the contributions to the nonlinear Hall response under the influence of disorder. Their work identifies three key components: the intrinsic contribution, the extrinsic side-jump scattering, and the skew scattering mechanisms\cite{du2019disorder}.

The side-jump scattering occurs during the scattering of electrons off impurities or phonons in a material. When an electron scatters, its trajectory is deflected, not just by the usual scattering angle but also by an additional transverse displacement. The transverse displacement $\delta \textbf{r}$ is proportional to the spin-orbit coupling strength and the gradient of the impurity potential\cite{berger1970side,smit1955spontaneous,smit1958spontaneous}, so the accumulation of $\delta \textbf{r}$ over multiple scattering events generates an effective transverse velocity contributing to the nonlinear Hall conductivity. The side-jump velocity induced nonlinaer Hall response can be expressed as:
\begin{equation}
    \chi_{abc}^{sj,1}=-\sum_{l}\tau_{l}v_{a}^{sj}\partial_{c}g_{l}^{b},
\end{equation}
where $g_{l}^{a}=\tau_{l}\partial_{a}f_{l}^{(0)}$. The side-jump also modifies the distribution function, which consequently gives rise to a second-order response:
\begin{equation}
    \chi_{abc}^{sj,2}=-\hbar\sum_{l}\tau_{l}\{[\partial_{a}(\tau_{l}v_{c}^{sj})+{\widetilde{M}_{l}^{ac}}]v_{l}^{b}+\partial_{c}(\tau_{l}v_{l}^{a})v_{b}^{sj}\}\frac{\partial f_{l}^{(0)}}{\partial\varepsilon_{l}}
    .
\end{equation}

On the other hand, skew scattering occurs when electrons are scattered asymmetrically due to spin-orbit interaction. The spin-orbit interaction causes the scattering to be asymmetric, meaning the probability of scattering to one side is different from the probability of scattering to the other side\cite{smit1955spontaneous,smit1958spontaneous}. Over many scattering events, the asymmetry leads to a net transverse current, which causes the nonlinear Hall effect\cite{cheng2024giant,lu2024nonlinear}.
The skew scattering induced nonlinear Hall effect can also be separated into two parts:
\begin{equation}
    \chi_{abc}^{sk,1}=\sum_{ll'}w_{ll'}^{g}(\widetilde{u}_{ll'}^{ca}-\tau_{l}u_{ll'}^{a}\partial_{c})g_{l}^{b},
\end{equation}
and
\begin{equation}
   \chi_{abc}^{sk,2}=\sum_{ll'}w_{ll'}^{ng}(\widetilde{u}_{ll'}^{ca}-\tau_{l}u_{ll'}^{a}\partial_{c})g_{l}^{b},
\end{equation}
where $w_{ll'}^{g}$ and $w_{ll'}^{ng}$ represent the Gaussian and non-Gaussian antisymmetric scattering rates, respectively.

In conclusion, the mechanisms governing the nonlinear Hall effect encompass the quantum metric dipole, the Berry curvature dipole, and disorder-induced phenomena such as side-jump and skew scattering. Both the quantum metric dipole and the Berry curvature dipole give rise to nonlinear Hall responses intrinsically linked to the quantum geometry of the system. Meanwhile, disorder-induced nonlinear Hall effects offer a valuable avenue for investigating disordered systems. These mechanisms manifest distinct experimental behaviors. Notably, the Berry curvature dipole exhibits a response that closely mirrors that of disorder-induced mechanisms, particularly due to its dependence on relaxation time. Importantly, the Berry curvature dipole is prohibited in systems exhibiting threefold rotational symmetry\cite{he2021quantum}. Under such conditions, the nonlinear Hall effect is primarily driven by disorder and the quantum metric dipole, with the two contributions distinguishable by their contrasting relaxation-time dependencies.
\section{4. SYMMETRY BREAKING INDUCED NONLINEAR HALL EFFECT IN 2D MATERIALS}
2D materials exhibit a range of intriguing physical properties arising from their unique structural and electronic characteristics\cite{gao2021layer,novoselov2004electric,yang2022two,li2017graphene,chia2018characteristics}. Many of these materials possess intrinsic symmetry breaking\cite{tokura2018nonreciprocal,zhang2022controlled,yasuda2020large}, which gives rise to their emergent transport behaviors, including the nonlinear Hall effect. Beyond the intrinsic symmetry breaking, researchers can intentionally break the symmetry through structural engineering techniques, such as the creation of heterostructures and 2D moiré superlattices\cite{dean2013hofstadter,novoselov20162d,tong2017topological,finney2019tunable}. These engineered systems provide new avenues to explore and enhance the nonlinear Hall effect, leading to exciting potential applications in electronic and optoelectronic devices.

\subsection{4.1 Intrinsic symmetry broken in 2D materials}
The nonlinear Hall response is generally measured through a Hall-bar device (as illustrated in Figure \ref{fig:Experimental observation}A). To detect this effect, an alternating current $\textbf{\textit{I}}^{\omega}$ with a low frequency $\omega$ (typically between 10 and 1000 Hz) is applied to the sample. The resulting transverse Hall voltage, which appears at twice of the applied frequency $2\omega$, is then measured using a lock-in technique\cite{ma2019observation,kang2019nonlinear}. The low-frequency measurement in the nonlinear Hall effect provides a distinct advantage in separating the signal from other electronic noise, making it more feasible to study the nonlinear response in solid-state systems\cite{du2021nonlinear}.

\begin{figure}[htb]
    \centering
    \includegraphics[width=0.99\linewidth]{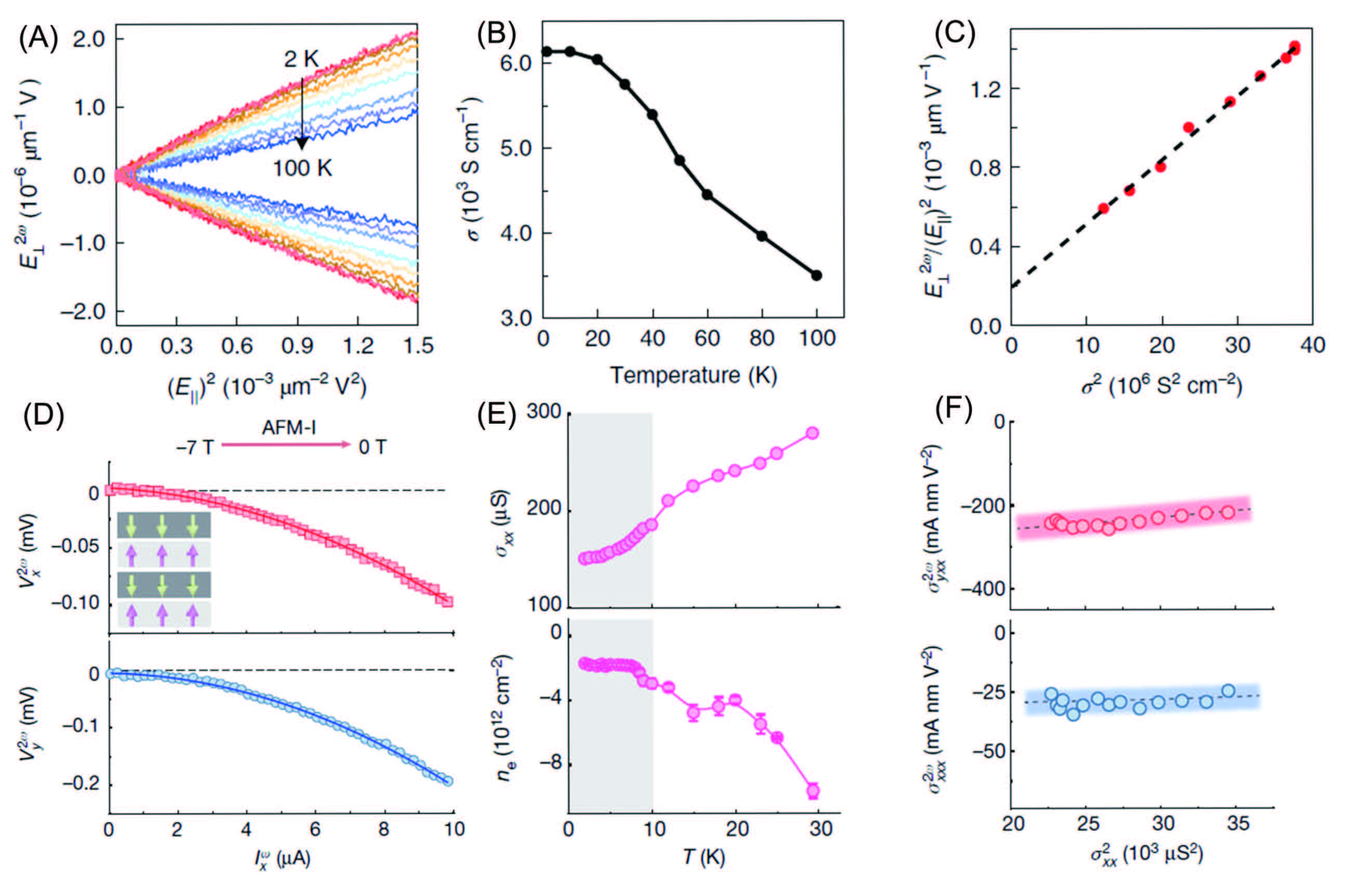}
    \caption{Experimental investigations of the nonlinear Hall response in \ce{WTe2} and \ce{MnBi2Te4}.  (A) the $E_{\bot}^{2\omega}$ depends linearly on the $(E_{\parallel})^{2}$ at different temperatures. (B) The temperature dependence of $\sigma$. (C) The nonlinear Hall signal as a function of the $\sigma^{2}$. (D) The nonlinear longitudinal $V_{x}^{2\omega}$ and transverse $V_{y}^{2\omega}$ responses as a function of the input current. (E) The conductivity and carrier density of 4 septuple
layers \ce{MnBi2Te4} at temperatures ranging from 2 K to 30 K. (F) The nonlinear conductivity $\sigma_{yxx}^{2\omega}$ ($\sigma_{xxx}^{2\omega}$) as a function of the $\sigma_{xx}^{2}$. Panels A-C reprinted with permission\cite{kang2019nonlinear}, 2019, CC BY license. Panels D-F reprinted with permission\cite{wang2023quantum}, 2023, CC BY license.}
    \label{fig:Experimental studies in WTe2}
\end{figure}

The nonlinear Hall effect was initially observed in $\textit{T}_{d}$-\ce{WTe2}, a 2D semimetal characterized by broken \textit{P} symmetry\cite{kang2019nonlinear,ma2019observation}. Through the typical measurement on a Hall-bar device, Kang \etal~measured the double-frequency voltage of the few-layer \ce{WTe2} at temperatures ranging from 2 K to 100 K\cite{kang2019nonlinear}. They found the transverse electric field with $2\omega~(E_{\bot}^{2\omega})$ scales linearly on the square of the longitudinal electric field ($E_{\parallel}$), as shown in Figure \ref{fig:Experimental studies in WTe2}A. The slope decreases consistently with increasing temperature. The temperature dependence of the longitudinal conductivity $\sigma$ was also measured (Figure \ref{fig:Experimental studies in WTe2}B). After a careful analysis, $\frac{E_{\bot}^{2\omega}}{(E_{\parallel})^{2}}$ scales linearly with $\sigma^{2}$ (Figure \ref{fig:Experimental studies in WTe2}C):

\begin{equation}
    \frac{E_{\bot}^{2\omega}}{(E_{\parallel})^{2}}=\xi\sigma^{2}+\eta,
    \label{WTe2}
\end{equation}

where $\xi$ and $\eta$ are constants. The anomalous Hall effect can result from Berry curvature, skew scattering, and side-jump mechanisms\cite{DXZheng-acsnano2021-halleffect,Berry-anomalous-Hall-infomat2024,nagaosa2010anomalous,tian2009proper,liu2018giant}. 
Accordingly, the nonlinear Hall effect in few-layer \ce{WTe2} is ascribed to contributions from both the Berry curvature dipole and scattering effects.

With further research into the nonlinear Hall effect, the mechanism of quantum metric dipole was proposed to induce the nonlinear Hall effect\cite{wang2021intrinsic}. This mechanism is referred to as the intrinsic nonlinear Hall effect since the $\tau$ independence\cite{gao2014field}. The first observation of the intrinsic nonlinear Hall effect was in topological antiferromagnet \ce{MnBi2Te4}\cite{wang2023quantum,gao2023quantum}, whose spins couple ferromagnetically within each Te-Bi-Te-Mn-Te-Bi-Te septuple layer (SL) with an out-of-plane easy axis, while adjacent septuple layers couple anti-parallel to each other\cite{deng2020quantum}. In the even layer \ce{MnBi2Te4}, the \textit{P} and \textit{T} symmetries are violated but the \textit{PT} symmetry remains intact. As shown in Figure \ref{fig:Experimental studies in WTe2}D, the nonlinear Hall effect of the 4 SLs \ce{MnBi2Te4} is measured along both the transverse and longitudinal directions. Significant responses are observed in the two directions, aligning with predictions that the quantum metric dipole can drive both nonlinear Hall response and non-reciprocal transport phenomena\cite{gao2024antiferromagnetic}. This is in stark contrast to the nonlinear Hall effect induced by the Berry curvature dipole, which only manifests the Hall response\cite{he2022graphene}. To explore the underlying cause of the nonlinear response, the temperature dependence of the longitudinal conductivity  $\sigma_{xx}^{\omega}$ and charge carrier density $\textit{n}_{\textit{e}}$ were measured. As shown in Figure \ref{fig:Experimental studies in WTe2}E, the charge carrier density stays almost unchanged between 1.6 K to 10 K, indicating the $\tau$ is the predominant component to determine the conductivity in this range. Then the $\sigma_{yxx}^{2\omega}$ and $\sigma_{xxx}^{2\omega}$ as a function of the $(\sigma_{xx}^{\omega})^{2}$ were plotted and fitted with the formula:

\begin{equation}
    \sigma^{2\omega}=\eta_{2}(\sigma_{xx}^{\omega})^{2}+\eta_{0},
    \label{MBT}
\end{equation}

as shown in Figure \ref{fig:Experimental studies in WTe2}F. They found the predominant contribution to both the $\sigma_{xxx}^{2\omega}$ and the $\sigma_{yxx}^{2\omega}$ is the quantum metric dipole ($\eta_{0}$, $\tau$-independence). This study introduces a technique for probing the quantum metric  and offers a framework for designing magnetic nonlinear devices.

Given the multitude of mechanisms that can induce the nonlinear Hall effect, an effective approach to classify them is essential. In experiments, one can apply the scaling law proposed by Du~\etal\cite{du2019disorder}:

\be
\frac{E_{yxx}^{2\omega}}{(E_{xxx}^{\omega})^2}-C_{1}\sigma_{x0}^{-1}\sigma_{a}^{2}=(C_{2}+C_{4}-C_{3})\sigma_{x0}^{-2}\sigma_{a}^{2}+(C_{3}-2C_{4})\sigma_{x0}^{-1}\sigma_{a}^{2}+C_{4}.
\ee
Here, the coefficients $C_{1,2,3,4}$ include the contributions from the Berry curvature dipole $(C^{in})$, side-jump $(C_{i}^{sj})$, intrinsic skew scattering $(C_{i}^{sk,1})$, and extrinsic skew scattering $(C_{i}^{sk,2})$ in the following forms:
\be
C_{1}=C^{sk,2}, C_{2}=C^{in}+C_{0}^{sj}+C_{00}^{sk,1}, C_{3}=2C^{in}+C_{0}^{sj}+C_{1}^{sj}+C_{01}^{sk,1}, C_{4}=C^{in}+C_{1}^{sj}+C_{11}^{sk,1}.
\ee
In the limit $(T\rightarrow 0)$, the scaling law simplifies to:
\be
\frac{E_{yxx}^{2\omega}}{(E_{xxx}^{\omega})^2}=C_{1}\sigma_{x0}+C_{2}.
\ee
From the calculated values of $C_{1}$ and $C_{2}$, we can identify the dominant mechanism driving the nonlinear Hall effect. In the case of the Weyl semimetal \ce{TaIrTe4}\cite{kumar2021room}, it was found that $C_{1}=-1.6\times10^{-15}$ m$^{2}$V$^{-1}$ and $C_{2}=2.6\times10^{-8}$ m$^{2}$V$^{-1}$, indicating that the contribution from extrinsic skew scattering to the nonlinear Hall effect is smaller than that of the Berry curvature dipole and static disorder scattering.

In addition, based on Eq.(\ref{WTe2}), since $J=\sigma E$, the second-order current can expressed as $J_{y}^{2\omega}=\sigma E_{y}^{2\omega}$, and thus: $\frac{E_{yxx}^{2\omega}}{(E_{xxx}^{\omega})^2}=\frac{\sigma_{yxx}^{(2)}}{\sigma}=\xi\sigma^{2}+\eta$. Given that $\sigma$ is linearly dependent on the scattering time $\tau$, the second-order conductivity $\sigma_{yxx}^{(2)}$ scales as $\tau^{3}$ and $\tau$ with the coefficients $\xi$ and $\eta$, respectively. The $\tau^{3}$ dependence contribution to the nonlinear Hall response originates from skew scattering, while the $\tau$ dependence arises from the Berry curvature dipole and side-jump mechanisms.
Through the scaling law, Lu~\etal~determined that the nonlinear Hall effect in \ce{BiTeBr} is primarily attributed to skew scattering and side-jump mechanisms, as the Berry curvature dipole is prohibited by the material’s threefold rotational symmetry\cite{lu2024nonlinear}. The coefficient $\xi$, which indicates the contribution from skew scattering, was calculated to be 6.9$\times$10$^{-15}$ m$^{3}$V$^{-1}$S$^{-2}$ for the 4-nm thick sample. Furthermore, the sign of $\eta$,  which represents the contribution from side-jump, was found to change with increasing thickness. After a comparison between the skew scattering and side-jump contributions, it was concluded that the nonlinear Hall effect in \ce{BiTeBr} is primarily dominated by skew scattering.

For the nonlinear Hall effect induced by the quantum metric dipole, a similar scaling analysis can be conducted. Based on the Eq.(\ref{MBT}), the contributions to the nonlinear Hall effect can be categorized into two distinct mechanisms: one that is independent of the scattering time $\tau$, and another that is dependent on $\tau$. Since the quantum metric dipole is independent of $\tau$, this distinction enables us to clearly differentiate the quantum metric dipole from other mechanisms.

\subsection{4.2 Symmetry design through structure engineering}
Due to the stringent symmetry requirements for the nonlinear Hall effect, the range of materials that can realize this effect is quite limited. Some of these materials require complex synthesis methods and may not even remain stable in air, which significantly restricts their practical applications\cite{kang2019nonlinear,tiwari2021giant}. Additionally, the electronic properties of these materials are strongly affected by their crystal structure and composition, making it difficult to modify them as desired\cite{li2017graphene,zhang2024broken}, which reduces the ability to optimize the magnitude of the nonlinear Hall effect signals. Recently, some researchers have attempted to break the symmetry of materials through doping\cite{wang2024nonlinear}. However, this approach has inherent limitations. The doping concentration is constrained, and achieving a controllable distribution of the dopants is often challenging. Significant emphasis has been placed on the study of oxide interfaces, where spontaneous symmetry breaking inherently fulfills the conditions required for the nonlinear Hall effect\cite{lesne2023designing,trama2022gate,groenendijk2020berry}. Likewise, there is growing emphasis on exploring heterostructures composed of 2D materials, where intentional symmetry breaking can be effectively achieved by stacking different materials\cite{duan2023berry,novoselov20162d,yankowitz2019van,li2023anisotropic}. Recent years have witnessed significant progress in the design and fabrication of heterostructures, opening new avenues for this research. Utilizing techniques such as mechanical exfoliation and the dry-transfer method\cite{wang2013one,kinoshita2019dry}, researchers can meticulously create 2D heterostructures tailored to specific experimental requirements. This innovative approach not only facilitates the investigation of fundamental physical phenomena but also paves the way for potential applications in next-generation devices, where engineered symmetry breaking may lead to novel functionalities.

\begin{figure}[htbp]
    \centering
    \includegraphics[width=0.99\linewidth]{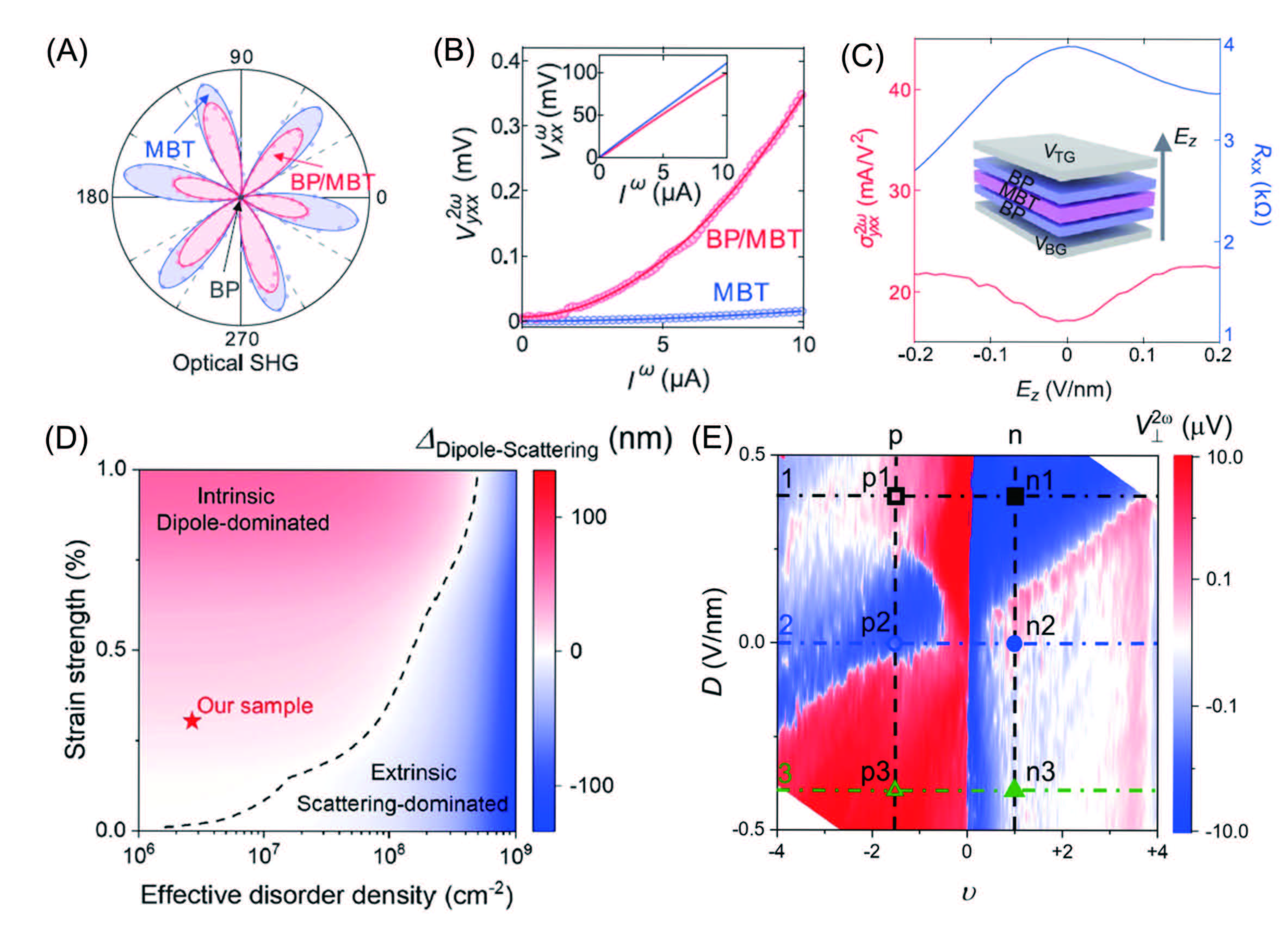}
    \caption{Experimental studies of the nonlinear Hall response in topological antiferromagnetic heterostructure (\ce{BP/MnBi2Te4}) and twisted bilayer graphene. (A) Optical second harmonic generation measurements of a 6 SLs \ce{MnBi2Te4} before and after being coupled with BP. (B) The $V_{yxx}^{2\omega}$ and $V_{xx}^{\omega}$ before and after being coupled with BP. (C) The $E_{Z}$ dependence of the $\sigma_{yxx}^{2\omega}$ and $R_{xx}$. (D) Phase diagram of the nonlinear Hall effect. the dominated mechanism changes with strain strength and effective disorder density, with red (blue) regions indicating dipole (scattering) dominance. (E) The $V_{\bot}^{2\omega}$ varies with $\upsilon$ and $\textit{D}$. Panels A-C reprinted with permission\cite{gao2023quantum}, Copyright 2023, The American Association for the Advancement of Science. Panels D-E reprinted with permission\cite{huang2023intrinsic}, Copyright 2023, The American Physical Society.}
    \label{fig:Experimental studies in BP/MnBi2Te4}
\end{figure}

Through this structure engineering, Gao \etal~investigated the nonlinear Hall effect in even layered \ce{MnBi2Te4} interfaced with black phosphorus (BP)\cite{gao2023quantum}. The lattice of \ce{MnBi2Te4} has $C_{3Z}$ rotational symmetry. To break this symmetry, a BP layer is stacked onto the \ce{MnBi2Te4}. As shown in Figure \ref{fig:Experimental studies in BP/MnBi2Te4}A, the polarization angle-dependent second harmonic generation (SHG) intensity of \ce{MnBi2Te4} exhibits six-fold rotational symmetry. In contrast, the SHG pattern of the BP/\ce{MnBi2Te4} heterostructure is asymmetric, indicating the BP breaks the $C_{3Z}$ symmetry of \ce{MnBi2Te4}. Figure \ref{fig:Experimental studies in BP/MnBi2Te4}B displays the nonlinear Hall signal of \ce{MnBi2Te4} before and after interfacing with BP. A prominent nonlinear Hall signal appears only after the introduction of BP, while the linear longitudinal voltage $V_{xx}^{\omega}$ exhibits only a slight decrease (inset of Figure \ref{fig:Experimental studies in BP/MnBi2Te4}B), further confirming the nonlinear Hall signal is induced by symmetry breaking. Since the quantum metric dipole induced nonlinear Hall response does not require explicit \textit{PT} symmetry breaking, applying a vertical $E_{Z}$ field via dual gating to break \textit{PT} symmetry has no impact on the nonlinear Hall signal. As shown in Figure \ref{fig:Experimental studies in BP/MnBi2Te4}C, the nonlinear Hall signal still remains at $E_{Z}$ = 0, confirming this is a quantum metric dipole induced nonlinear Hall effect.

With the advancement of 2D van der Waals stacking technology, the fabrication of 2D moiré superlattices has become feasible through techniques such as tear-and-stack\cite{kim2016van} and cut-and-stack\cite{saito2020independent} methods. Researchers have successfully synthesized 2D moiré superlattices in materials including twisted trilayer graphene\cite{saito2020independent,tian2023evidence,park2021tunable}, twisted bilayer BN, and twisted \ce{WSe2–MoSe2} heterobilayers\cite{mcgilly2020visualization}. These structures exhibit intriguing physical properties, including superconductivity\cite{park2021tunable,qiu2021recent,balents2020superconductivity}, topological phases\cite{tong2017topological,mak2022semiconductor} and correlated insulating states\cite{xu2020correlated,huang2021correlated,andrei2021marvels}. Furthermore, the spontaneous atomic reconstruction associated with these moiré patterns leads to symmetry breaking within the material, thereby facilitating the emergence of the nonlinear Hall effect.

Recently, Huang \etal~detected a nonlinear Hall signal in high-mobility monolayer twisted graphene samples\cite{huang2023intrinsic}. Two mechanisms for generating the nonlinear Hall effect in graphene superlattices have been proposed. As shown in Figure \ref{fig:Experimental studies in BP/MnBi2Te4}D, the Berry curvature dipole dominates in samples with high strain and low impurity concentration, whereas disorder dominates in samples with low strain and high impurity concentration. Additionally, the Berry curvature dipole can be effectively tuned by gate voltage, even allowing a change in the direction of the dipole (Figure \ref{fig:Experimental studies in BP/MnBi2Te4}E). This study presents an efficient method to control and manipulate the amplitude and direction of the Berry curvature dipole, thereby enabling control over the nonlinear Hall response.

In addition to the 2D materials previously mentioned, the nonlinear Hall effect has been identified in a variety of other materials in recent years (refer to Table \ref{tab:nonlinear Hall effect in various materials}). The Weyl semimetals, such as \ce{MoTe2}\cite{ma2022growth} and \ce{TaIrTe4}\cite{kumar2021room}, with a $T_{d}$ structure, exhibit mirror plane breaking on their surfaces, giving rise to the nonlinear Hall effect. Elemental materials like Te\cite{cheng2024giant,suarez2024odd} and Bi\cite{makushko2024tunable}, along with topological insulators such as 
\ce{Bi2Se3}\cite{han2024room}, distinguished by their unique surface states, can also significantly contribute to the generation of nonlinear signals. Similarly, Rashba materials such as BiTeBr\cite{lu2024nonlinear}, which exhibit significant Rashba-type band splitting, can generate a nonlinear Hall response. Furthermore, the spin-valley-locked Dirac material \ce{BaMnSb2}\cite{min2023strong}, with its non-centrosymmetric orthorhombic structure, serves as an exemplary platform for facilitating the nonlinear Hall effect. Additionally, advanced structural engineering approaches, including heterostructures\cite{han2024room,gao2023quantum} and 2D moiré superlattices\cite{huang2023giant,duan2022giant,he2022graphene,sinha2022berry} designed to break inversion symmetry, offer a compelling route for the realization of this intriguing phenomenon.

\setlength{\tabcolsep}{3pt}
\renewcommand\arraystretch{2}
\begin{longtable}{ccccccc}
    \caption{Nonlinear Hall effect in various materials.} \\
    \hline
    & Materials & Mechanisms & \makecell[c]{Temperature\\(K)} & \makecell[c]{Input\\current\\frequency\\(Hz)} & \makecell[c]{Input\\current\\maximum\\($\mu$A)} & \makecell[c]{Output\\voltage\\maximum\\($\mu$V)} \\ 
    \hline

   \multirow{5}*{\makecell[c]{Weyl semimetals}} &Bilayer \ce{WTe2}\cite{ma2019observation} & \makecell[c]{Berry curvature\\dipole} & 10-100 & 10-1000 & 1 & 200 \\
    &Few-layer \ce{WTe2}\cite{kang2019nonlinear} & \makecell[c]{Berry curvature\\dipole \& skew \\scattering} & 1.8-100 & 17-137 & 600 & 25 \\
   
    &Bilayer~\ce{MoTe2}\cite{ma2022growth} & \makecell[c]{Berry curvature\\dipole \& skew \\scattering} & 10-100 & \makecell[c]{13.373-\\133.33} & 200 & 45 \\
    
       &\ce{TaIrTe4}\cite{kumar2021room} & \makecell[c]{Berry curvature\\dipole} & 2-300 & 13.7-213.7 & 600 & 100 \\
      &\ce{NbIrTe4}\cite{lee2024spin} & \makecell[c]{Berry curvature\\dipole} & 2-300 & \makecell[c]{17.77-\\117.77} & 200 & 30 \\ 
\hline
      
      \multirow{1}*{\makecell[c]{Rashba materials}}&\ce{BiTeBr}\cite{lu2024nonlinear} & Skew scattering & 300-350 & \makecell[c]{7.777-\\277.77} & 5 & 100 \\ 
      \hline
      \multirow{1}*{\makecell[c]{Dirac semimetals}}&\ce{BaMnSb2}\cite{min2023strong} & \makecell[c]{Berry curvature\\dipole} & 200-400 & \makecell[c]{17.777-\\117.77} & 100 & 250 \\ 
      \hline
      \multirow{3}*{\makecell[c]{Elemental\\materials}} &Te \cite{cheng2024giant} & \makecell[c]{Side-jump \& \\skew scattering} & 200-300 & 53.7-313.7 & 50 & 200 \\ 
      &Te \cite{suarez2024odd} & \makecell[c]{Side-jump} & 10-100 & 31 & 1 & 220 \\ 
      &Bi \cite{makushko2024tunable} & \makecell[c]{Side-jump \& \\skew scattering} & 283-333 & 787 & 60 & 0.6\\
       \hline
\multirow{4}*{\makecell[c]{Topological \\insulators}}&\ce{MnBi2Te4}\cite{wang2023quantum} & \makecell[c]{Quantum metric\\dipole} & 1.6-30 & \makecell[c]{17.777-\\117.77} & 10 & 200 \\
       &\ce{Bi2Se3}\cite{he2021quantum} & \makecell[c]{Skew scattering} & 20-200 & 9-263 & 1500 & 15 \\
       &\ce{ZrTe5}\cite{wang2024non} & \makecell[c]{Berry curvature\\dipole} & 2-100 & 17.777 & 200 & 10 \\
       \hline
       &\makecell[c]{Twisted bilayer\\ \ce{WSe2}\cite{huang2023giant}} & \makecell[c]{Berry curvature\\dipole} & 1.5-40 & 4.579 & 0.06 & 20000 \\
       \multirow{5}*{\makecell[c]{Structure \\engineering}}&\makecell[c]{Twisted Bilayer\\Graphene\cite{duan2022giant}} & Skew scattering & 1.7-80 & \makecell[c]{13.777-\\33.777} & 5 & 1000 \\
       & \makecell[c]{hBN/graphene\\/hBN\cite{he2022graphene}} & Skew scattering & 1.65-210 & 31 & 5 & 125 \\
      & \makecell[c]{Twisted double\\bilayer graphene\cite{sinha2022berry}} & \makecell[c]{Berry curvature\\dipole} & 1.5-25 & \makecell[c]{18.03-\\177.81} & 0.2 & 60 \\
       & \makecell[c]{Strain tunable\\Monolayer \ce{WSe2}\cite{qin2021strain}} & \makecell[c]{Berry curvature\\dipole} & 50-140 & 17.777 & 4.5 & 10 \\
       &\ce{Mn3Sn/Pt}\cite{han2024room} & \makecell[c]{Quantum metric\\dipole} & 5-400 & 10-80 & 5 & 40 \\ 
       &\ce{BP/MnBi2Te4}\cite{gao2023quantum} & \makecell[c]{Quantum metric\\dipole} & 1.8-30 & \makecell[c]{17.77-\\1717.77} & 10 & 350 \\
      \hline
     \label{tab:nonlinear Hall effect in various materials}
\end{longtable}

\section{5. PROSPECTS}

\subsection{5.1 Other nonlinear transport effects}
The nonlinear Hall effect typically involves a quadratic response of the longitudinal Hall voltage to the applied transverse current. However, in some materials, there could be higher-order responses in Hall effect. The second-order derivatives of the Berry connection polarizability could also induce the third-order nonlinear Hall effect. Analogous to the second-order nonlinear Hall effect, the third
order nonlinear conductance can be expressed as\cite{li2024quantum}:
\begin{equation}
    \sigma_{abcd}^{3\omega}=\frac{e^{4}\tau}{\hbar}\left[\int_{k}(-\partial_{a}\partial_{b}G_{cd}+\partial_{a}\partial_{d}G_{bc}-\partial_{b}\partial_{d}G_{ac})f_{0}+\frac{1}{2}\int_{k}v_{a}v_{b}G_{cd}f_{0}^{''}\right]
.
\end{equation}
\begin{figure}[htb]
    \centering
    \includegraphics[width=0.95\linewidth]{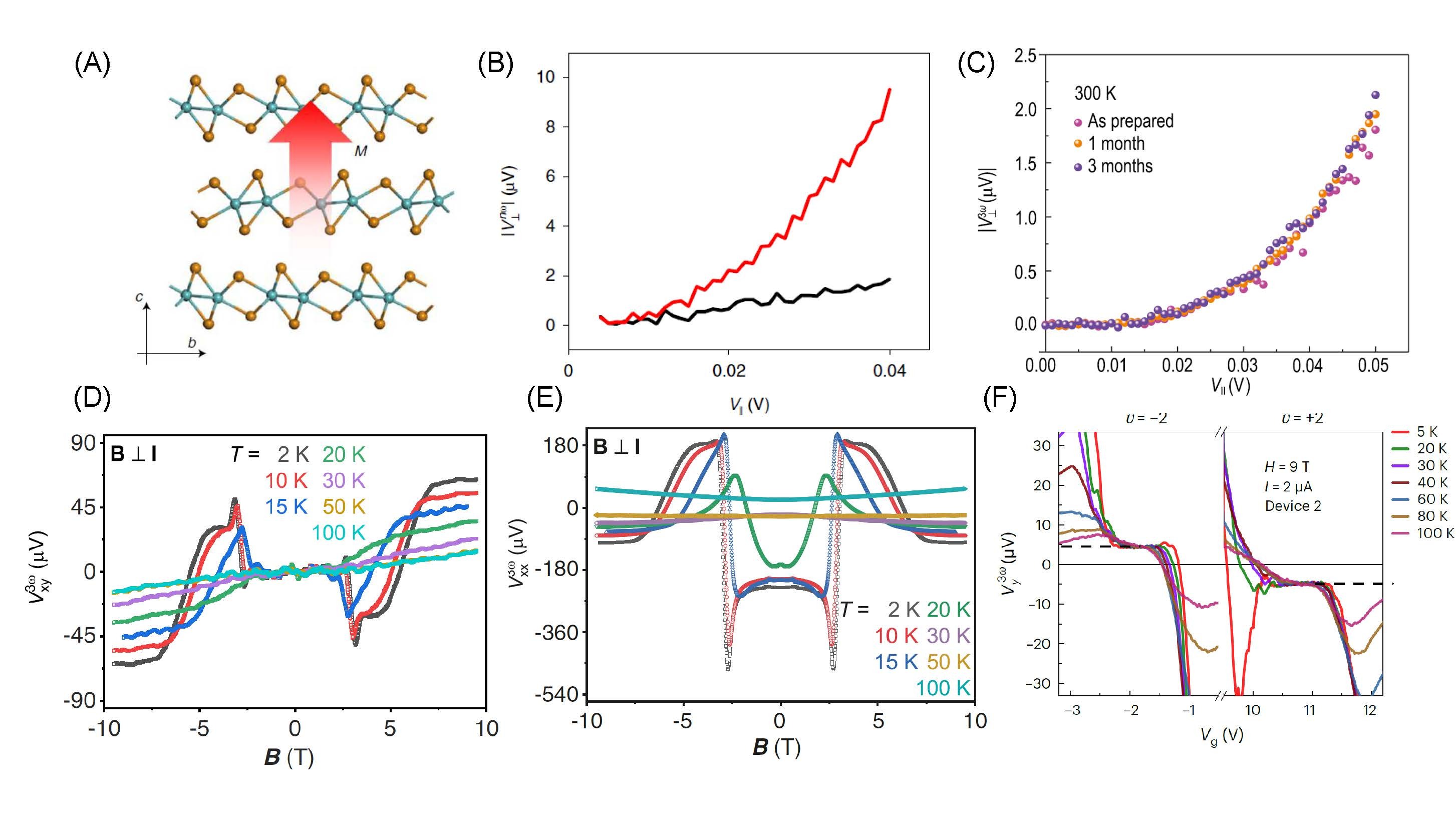}
    \caption{
    The third-order nonlinear Hall effect in several materials. (A) b–c plane of few-layer $T_{d}$~\ce{MoTe2}. (B) The $V_{\bot}^{n\omega}$ of \ce{MoTe2} as a function of $V_{\parallel}$, with red demoting the third-order response and black representing the second-order response. (C) The $V_{\bot}^{3\omega}$ in~\ce{TaIrTe4} as a function of $V_{\parallel}$. (D, E) The $V_{xy}^{3\omega}$ and $V_{xx}^{3\omega}$ of \ce{MnBi2Te4} as functions of magnetic field at different temperatures. (F) The $V_{y}^{3\omega}$ of graphene under quantum Hall states as a function of $V_{g}$ at varying temperatures.   Panels A-B reprinted with permission\cite{lai2021third}, 2021, CC BY license. Panel C reprinted with permission\cite{wang2022room}, 2022, CC BY license. Panels D-E reprinted with permission\cite{li2024quantum}, 2024, CC BY license. Panel F reprinted with permission\cite{he2024third}, 2024, CC BY license.
}
    \label{fig:third order nonlinear hall effect}
\end{figure}
Distinct from the linear and second-order nonlinear Hall effects, the third-order Hall effect follows unique symmetry constraints and does not require broken \textit{T} or \textit{P} symmetry, making it a valuable tool for characterizing materials that preserve both symmetries\cite{ye2022orbital}. 

The first observation of the third-order nonlinear Hall effect was made in \ce{MoTe2}\cite{lai2021third}. As depicted in Figure \ref{fig:third order nonlinear hall effect}A. In the $T_{d}$ phase of \ce{MoTe2}, symmetry broken arises from the material's surface. With increasing thickness, the third-order nonlinear Hall effect becomes increasingly distinct, as illustrated in Figure \ref{fig:third order nonlinear hall effect}B. Furthermore, Wang~\etal~reported a pronounced third-order nonlinear Hall effect at room temperature\cite{wang2022room}, as shown in Figure \ref{fig:third order nonlinear hall effect}C. Additionally, the third-order nonlinear Hall effect has been observed in magnetic systems. Li~\etal~identified a third-order response in \ce{MnBi2Te4}\cite{li2024quantum}. Notably, Figures \ref{fig:third order nonlinear hall effect}D and E reveal that the measured $V^{3\omega}_{xx}$ exhibits an even dependence on the magnetic field, whereas $V^{3\omega}_{xy}$ displays an odd dependence. The field dependence of these responses suggests that the longitudinal third-order response is driven by the Berry connection polarizability, while the transverse response is governed by the Berry curvature quadrupole. Similarly, the quantum Hall effect also manifests a third-order response. He \etal~observed the third-order nonlinear Hall effect in graphene within quantum Hall phases\cite{he2024third}. Figure \ref{fig:third order nonlinear hall effect}F illustrates that the third-order Hall response is independent of temperature. The extrinsic effects, including skew and side-jump scattering of electrons, which are dependent on scattering time, are supposed to be suppressed in quantum Hall states. As a result, the nonlinear Hall response observed in quantum states provides an ideal platform for exploring the intrinsic nonlinear Hall effect. In addition to the materials discussed above, the third-order nonlinear Hall effect has also been observed in Weyl semimetals \ce{WTe2}\cite{ye2022orbital}, in transition metal dichalcogenides \ce{VSe2}\cite{chen2024charge}, which is characterized by charge density wave modulation, and in the misfit layer compound ${\ce{(SnS)}}_{1.17}\ce{(NbS2)3}$\cite{li2024giant}, which exhibits a giant third-order response.

\begin{figure}[htbp]
    \centering
    \includegraphics[width=0.95\linewidth]{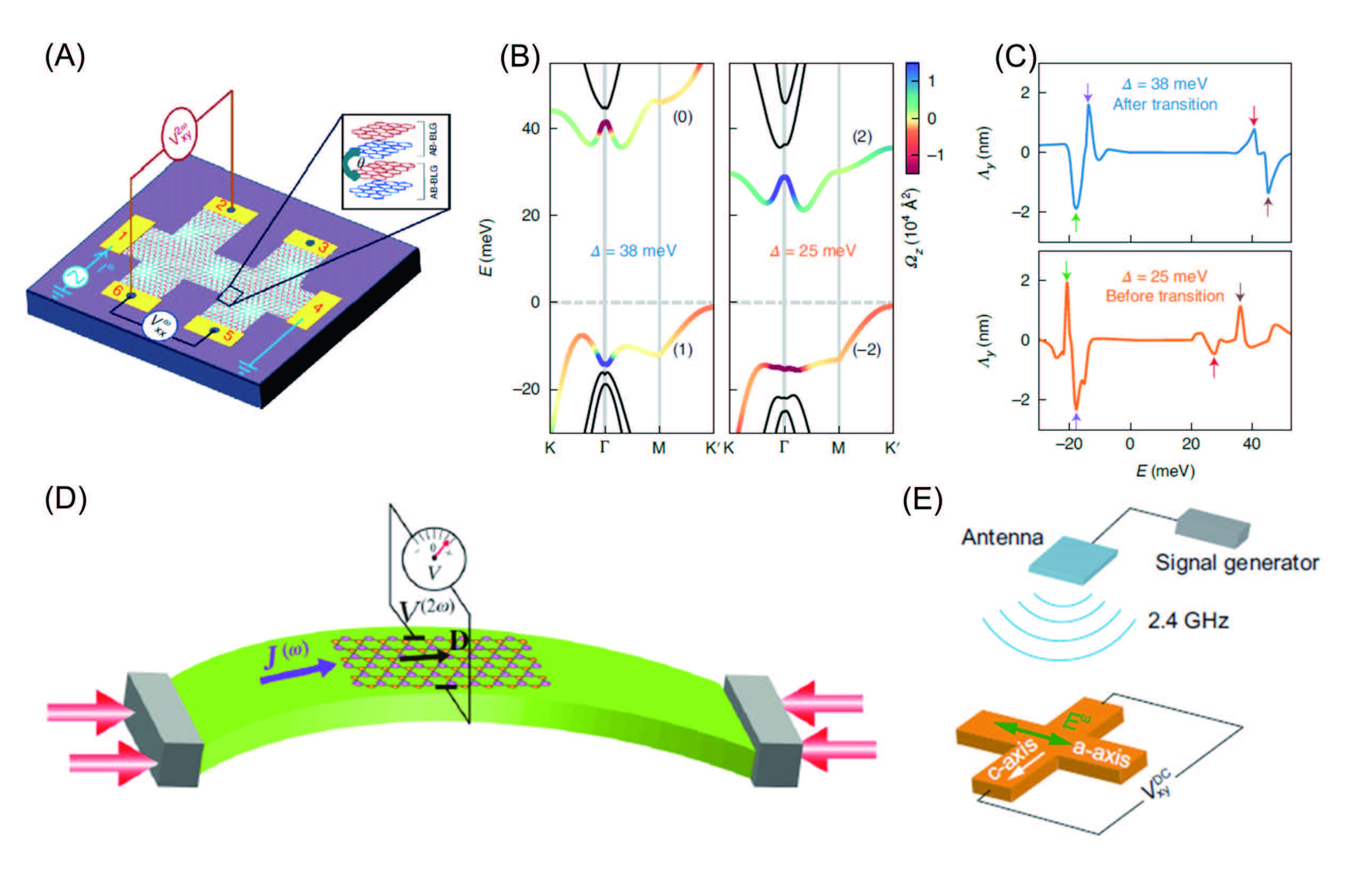}
    \caption{Device applications of the nonlinear Hall effect span diverse fields, including probing topological transitions, developing strain sensors, and creating wireless radiofrequency rectifiers. (A) Schematic of the nonlinear Hall measurement. (B) Band structures of the \textit{K} valley for two different interlayer potential values, with the overlaid color representing the Berry curvature of the corresponding flat bands. (C) The variation in the Berry curvature dipole as a function of energy $\textit{E}$ at two different Fermi energy. (D) A strain sensor leveraging the nonlinear Hall response. (E) Wireless radiofrequency rectifier via the nonlinear Hall response. Panels A-C reprinted with permission\cite{sinha2022berry}, 2022, CC BY license. Panel D reprinted with permission\cite{xiao2020two}, Copyright 2020, The American Physical Society. Panel E reprinted with permission\cite{cheng2024giant}, 2024, CC BY license.}
    \label{fig:Device applications}
\end{figure}
In addition to the higher order nonlinear Hall effect, the nonlinear spin Hall response can be realized through charge-to-spin conversion\cite{hamamoto2017nonlinear,araki2018strain}. In thermally driven systems, the nonlinear response includes effects such as the nonlinear Nernst effect, the nonlinear Seebeck effect, and the nonlinear anomalous thermal Hall effect\cite{zeng2019nonlinear,zeng2020fundamental,nakai2019nonreciprocal,yu2019topological}. The planar Hall effect, unlike the traditional Hall effect, occurs when the Hall voltage, electric field, and magnetic field are coplanar\cite{kumar2018planar,burkov2017giant,tang2003giant}. The nonlinear planar Hall effect further introduces a higher-order response, where the transverse voltage scales nonlinearly with the applied current or magnetic field\cite{he2019nonlinearl,huang2023intrinsic,rao2021theory}. Collectively, these nonlinear transport phenomena offer rich opportunities for exploring new physics and developing innovative applications in science and technology.

\subsection{5.2 Device applications}
Due to its unique sensitivity to symmetry breaking and material properties, the nonlinear Hall effect offers applications in phase probing and the investigation of topological phenomena in condensed matter systems\cite{xiao2020berry,shao2020nonlinear}. In twisted double bilayer graphene, the application of a perpendicular electric field can concurrently adjust both the valley Chern number and the Berry curvature dipole, offering a tunable platform to investigate topological transitions\cite{sinha2022berry}, as shown in Figure \ref{fig:Device applications}A - C. The Berry curvature dipole can also be tuned by strain\cite{qin2021strain,xiao2020two} (Figure \ref{fig:Device applications}D). This provides new ideas and methods for designing piezoelectric-like devices, such as strain sensors.

Beyond its scientific significance in probing quantum geometry and crystalline symmetry, the nonlinear Hall effect holds substantial potential for practical applications, particularly in energy harvesting and rectifying devices\cite{kumar2021room,isobe2020high,zhang2021terahertz}, as shown in Figure \ref{fig:Device applications}E. Materials exhibiting the nonlinear Hall effect can function  as wireless radiofrequency (RF) rectifiers operating without an external bias (battery-free), and in the absence of a magnetic field by utilizing a driving alternating current in place of an oscillating electromagnetic field\cite{suarez2024microscale}. For example, the use of materials such as \ce{TaIrTe4}\cite{kumar2021room} and \ce{MnBi2Te4}\cite{gao2023quantum} in radiofrequency rectification has demonstrated cutoff frequencies reaching up to 5 GHz, which is sufficient to encompass the widely utilized 2.4 GHz Wi-Fi channel. Furthermore, BiTeBr-based rectifiers have exhibited radiofrequency rectification initiating at power levels as low as -15 dBm (~0.03 mW), aligning closely with ambient RF power levels ranging from -20 to -10 dBm\cite{lu2024nonlinear,muhammad2022harvesting}. In addition, the rectified output in tellurium thin flakes can be significantly enhanced through the application of a gate voltage\cite{cheng2024giant}. This class of Hall rectifiers, relying on the intrinsic material properties, effectively bypasses the limitations imposed by transition time and thermal voltage thresholds, presenting a promising solution for efficient, low-power energy conversion technologies~\cite{Hueso-review-arxiv2024}.

\section{6. PERSPECTIVE}
The study of the nonlinear Hall response is rapidly advancing the condensed matter physics has opened new frontiers within material science research, particularly in the realm of 2D systems. By leveraging the principles of symmetry, researchers have uncovered novel mechanisms that drive this effect, distinguishing it from traditional Hall phenomena\cite{sodemann2015quantum,du2019disorder,wang2021intrinsic}. The insights gained from these research not only deepen our understanding of the interplay between symmetry and electronic properties, but also pave the way for innovative applications in electronic devices and sensors.

However, realizing practical devices based on the nonlinear Hall effect presents several challenges. First, the scale of most 2D materials exhibiting the nonlinear Hall effect is limited, with sizes typically restricted to a few tens of micrometers. This constraint poses a significant challenge to the development of large-scale, integrated applications leveraging the nonlinear Hall effect. Furthermore, although wireless radiofrequency rectifiers based on this effect can function without an external bias, their conversion efficiency remains low. Moreover, the nonlinear Hall effect weakens at higher frequencies due to the interplay with material relaxation times, further complicating its integration into existing technological frameworks. 

Future research on the nonlinear Hall effect should strategically focus on advancing theoretical frameworks, driving material discovery, and addressing device scalability to fully realize its potential. Refining theoretical models to encompass higher-order nonlinear responses, quantum geometric effects, and the role of disorder is imperative for deepening our understanding of the underlying mechanisms. In parallel, identifying materials with pronounced nonlinear Hall effect responses, particularly those operational at high temperatures, is essential. Promising candidates include non-centrosymmetric two-dimensional materials, topological semimetals, and moiré superlattices. Moreover, dynamic control of the nonlinear Hall effect through effective gating strategies, such as the modulation of the band structure through electric and optical fields\cite{huang2023intrinsic,qin2024light,qin2024nonlinear}, strain engineering\cite{qin2021strain}, and chemical doping\cite{wang2024nonlinear}, will be key to tailoring device performance. Furthermore, achieving scalable fabrication of high-quality 2D materials is vital for the development of energy-harvesting devices that leverage the nonlinear Hall effect, propelling it from fundamental exploration to transformative applications.

\section{7. CONCLUSION}
In conclusion, the nonlinear Hall effect arises from diverse mechanisms, including contributions from the quantum metric and Berry curvature dipole, alongside skew scattering and side-jump effects, each uniquely shaping the nonlinear response. Experimentally, this effect has been extensively explored in 2D materials, encompassing intrinsic symmetry breaking in specific systems as well as engineered configurations, such as heterostructures and moiré superlattices. The ability to manipulate electronic responses without requiring an external magnetic field positions the nonlinear Hall effect as a transformative platform for next-generation electronic devices. Potential applications span highly efficient rectifiers, sensitive detectors, and innovative logic devices. Looking ahead, the nonlinear Hall effect is poised to serve as a cornerstone for the exploration of other nonlinear transport phenomena, offering vast opportunities for groundbreaking discoveries. The continued study of this phenomenon promises to deepen insights into condensed matter physics while paving the way for advanced technologies that capitalize on the unique electronic properties of two-dimensional materials.

\section*{Declarations}

\subsection*{Acknowledgments}
The work carried out by Shuo Wang and Wei Niu was supported by the Natural Science Foundation of Nanjing University of Posts and Telecommunications (NY222170).
\subsection*{Authors’ contributions}
W.N. \& Y.-W.F. conceived this project upon the invitation from OAE to Y.W.-F. All users contributed to the writing and interpretation.
\subsection*{Conflicts of interest}
All authors declared that there are no conflicts of interest.



\begin{thebibliography}{0}%
\makeatletter
\providecommand \@ifxundefined [1]{%
 \@ifx{#1\undefined}
}%
\providecommand \@ifnum [1]{%
 \ifnum #1\expandafter \@firstoftwo
 \else \expandafter \@secondoftwo
 \fi
}%
\providecommand \@ifx [1]{%
 \ifx #1\expandafter \@firstoftwo
 \else \expandafter \@secondoftwo
 \fi
}%
\providecommand \natexlab [1]{#1}%
\providecommand \enquote  [1]{``#1''}%
\providecommand \bibnamefont  [1]{#1}%
\providecommand \bibfnamefont [1]{#1}%
\providecommand \citenamefont [1]{#1}%
\providecommand \href@noop [0]{\@secondoftwo}%
\providecommand \href [0]{\begingroup \@sanitize@url \@href}%
\providecommand \@href[1]{\@@startlink{#1}\@@href}%
\providecommand \@@href[1]{\endgroup#1\@@endlink}%
\providecommand \@sanitize@url [0]{\catcode `\\12\catcode `\$12\catcode
  `\&12\catcode `\#12\catcode `\^12\catcode `\_12\catcode `\%12\relax}%
\providecommand \@@startlink[1]{}%
\providecommand \@@endlink[0]{}%
\providecommand \url  [0]{\begingroup\@sanitize@url \@url }%
\providecommand \@url [1]{\endgroup\@href {#1}{\urlprefix }}%
\providecommand \urlprefix  [0]{URL }%
\providecommand \Eprint [0]{\href }%
\providecommand \doibase [0]{https://doi.org/}%
\providecommand \selectlanguage [0]{\@gobble}%
\providecommand \bibinfo  [0]{\@secondoftwo}%
\providecommand \bibfield  [0]{\@secondoftwo}%
\providecommand \translation [1]{[#1]}%
\providecommand \BibitemOpen [0]{}%
\providecommand \bibitemStop [0]{}%
\providecommand \bibitemNoStop [0]{.\EOS\space}%
\providecommand \EOS [0]{\spacefactor3000\relax}%
\providecommand \BibitemShut  [1]{\csname bibitem#1\endcsname}%
\let\auto@bib@innerbib\@empty
\end{thebibliography}%


\begin{thebibliography}{139}%
\makeatletter
\providecommand \@ifxundefined [1]{%
 \@ifx{#1\undefined}
}%
\providecommand \@ifnum [1]{%
 \ifnum #1\expandafter \@firstoftwo
 \else \expandafter \@secondoftwo
 \fi
}%
\providecommand \@ifx [1]{%
 \ifx #1\expandafter \@firstoftwo
 \else \expandafter \@secondoftwo
 \fi
}%
\providecommand \natexlab [1]{#1}%
\providecommand \enquote  [1]{``#1''}%
\providecommand \bibnamefont  [1]{#1}%
\providecommand \bibfnamefont [1]{#1}%
\providecommand \citenamefont [1]{#1}%
\providecommand \href@noop [0]{\@secondoftwo}%
\providecommand \href [0]{\begingroup \@sanitize@url \@href}%
\providecommand \@href[1]{\@@startlink{#1}\@@href}%
\providecommand \@@href[1]{\endgroup#1\@@endlink}%
\providecommand \@sanitize@url [0]{\catcode `\\12\catcode `\$12\catcode
  `\&12\catcode `\#12\catcode `\^12\catcode `\_12\catcode `\%12\relax}%
\providecommand \@@startlink[1]{}%
\providecommand \@@endlink[0]{}%
\providecommand \url  [0]{\begingroup\@sanitize@url \@url }%
\providecommand \@url [1]{\endgroup\@href {#1}{\urlprefix }}%
\providecommand \urlprefix  [0]{URL }%
\providecommand \Eprint [0]{\href }%
\providecommand \doibase [0]{https://doi.org/}%
\providecommand \selectlanguage [0]{\@gobble}%
\providecommand \bibinfo  [0]{\@secondoftwo}%
\providecommand \bibfield  [0]{\@secondoftwo}%
\providecommand \translation [1]{[#1]}%
\providecommand \BibitemOpen [0]{}%
\providecommand \bibitemStop [0]{}%
\providecommand \bibitemNoStop [0]{.\EOS\space}%
\providecommand \EOS [0]{\spacefactor3000\relax}%
\providecommand \BibitemShut  [1]{\csname bibitem#1\endcsname}%
\let\auto@bib@innerbib\@empty
\bibitem [{\citenamefont {R{\"u}ter}\ \emph {et~al.}(2010)\citenamefont
  {R{\"u}ter}, \citenamefont {Makris}, \citenamefont {El-Ganainy},
  \citenamefont {Christodoulides}, \citenamefont {Segev},\ and\ \citenamefont
  {Kip}}]{ruter2010observation}%
  \BibitemOpen
  \bibfield  {author} {\bibinfo {author} {\bibfnamefont {C.~E.}\ \bibnamefont
  {R{\"u}ter}}, \bibinfo {author} {\bibfnamefont {K.~G.}\ \bibnamefont
  {Makris}}, \bibinfo {author} {\bibfnamefont {R.}~\bibnamefont {El-Ganainy}},
  \bibinfo {author} {\bibfnamefont {D.~N.}\ \bibnamefont {Christodoulides}},
  \bibinfo {author} {\bibfnamefont {M.}~\bibnamefont {Segev}},\ and\ \bibinfo
  {author} {\bibfnamefont {D.}~\bibnamefont {Kip}},\ }\bibfield  {title}
  {\bibinfo {title} {Observation of parity--time symmetry in optics},\ }\href
  {https://doi.org/10.1038/nphys1515} {\bibfield  {journal} {\bibinfo
  {journal} {Nature physics}\ }\textbf {\bibinfo {volume} {6}},\ \bibinfo
  {pages} {192} (\bibinfo {year} {2010})}\BibitemShut {NoStop}%
\bibitem [{\citenamefont {Liu}\ \emph {et~al.}(2023)\citenamefont {Liu},
  \citenamefont {Zhao}, \citenamefont {Lin}, \citenamefont {Shu}, \citenamefont
  {Zhou}, \citenamefont {Zheng}, \citenamefont {Chen}, \citenamefont {Jia},\
  and\ \citenamefont {Chen}}]{liu2023symmetry}%
  \BibitemOpen
  \bibfield  {author} {\bibinfo {author} {\bibfnamefont {L.}~\bibnamefont
  {Liu}}, \bibinfo {author} {\bibfnamefont {T.}~\bibnamefont {Zhao}}, \bibinfo
  {author} {\bibfnamefont {W.}~\bibnamefont {Lin}}, \bibinfo {author}
  {\bibfnamefont {X.}~\bibnamefont {Shu}}, \bibinfo {author} {\bibfnamefont
  {J.}~\bibnamefont {Zhou}}, \bibinfo {author} {\bibfnamefont {Z.}~\bibnamefont
  {Zheng}}, \bibinfo {author} {\bibfnamefont {H.}~\bibnamefont {Chen}},
  \bibinfo {author} {\bibfnamefont {L.}~\bibnamefont {Jia}},\ and\ \bibinfo
  {author} {\bibfnamefont {J.}~\bibnamefont {Chen}},\ }\bibfield  {title}
  {\bibinfo {title} {Symmetry breaking for current-induced magnetization
  switching},\ }\bibfield  {journal} {\bibinfo  {journal} {Applied Physics
  Reviews}\ }\textbf {\bibinfo {volume} {10}},\ \href
  {https://doi.org/10.1063/5.0149290} {10.1063/5.0149290} (\bibinfo {year}
  {2023})\BibitemShut {NoStop}%
\bibitem [{\citenamefont {Dembowski}\ \emph {et~al.}(2003)\citenamefont
  {Dembowski}, \citenamefont {Dietz}, \citenamefont {Gr{\"a}f}, \citenamefont
  {Harney}, \citenamefont {Heine}, \citenamefont {Heiss},\ and\ \citenamefont
  {Richter}}]{dembowski2003observation}%
  \BibitemOpen
  \bibfield  {author} {\bibinfo {author} {\bibfnamefont {C.}~\bibnamefont
  {Dembowski}}, \bibinfo {author} {\bibfnamefont {B.}~\bibnamefont {Dietz}},
  \bibinfo {author} {\bibfnamefont {H.-D.}\ \bibnamefont {Gr{\"a}f}}, \bibinfo
  {author} {\bibfnamefont {H.}~\bibnamefont {Harney}}, \bibinfo {author}
  {\bibfnamefont {A.}~\bibnamefont {Heine}}, \bibinfo {author} {\bibfnamefont
  {W.}~\bibnamefont {Heiss}},\ and\ \bibinfo {author} {\bibfnamefont
  {A.}~\bibnamefont {Richter}},\ }\bibfield  {title} {\bibinfo {title}
  {Observation of a chiral state in a microwave cavity},\ }\href
  {https://doi.org/10.1103/PhysRevLett.90.034101} {\bibfield  {journal}
  {\bibinfo  {journal} {Physical Review Letters}\ }\textbf {\bibinfo {volume}
  {90}},\ \bibinfo {pages} {034101} (\bibinfo {year} {2003})}\BibitemShut
  {NoStop}%
\bibitem [{\citenamefont {Zhang}\ \emph
  {et~al.}(2024{\natexlab{a}})\citenamefont {Zhang}, \citenamefont {Zhu},
  \citenamefont {Zhang}, \citenamefont {Chen}, \citenamefont {Song},
  \citenamefont {Zhang}, \citenamefont {Gao}, \citenamefont {Niu},
  \citenamefont {Chen}, \citenamefont {Fei} \emph {et~al.}}]{zhang2024light}%
  \BibitemOpen
  \bibfield  {author} {\bibinfo {author} {\bibfnamefont {X.}~\bibnamefont
  {Zhang}}, \bibinfo {author} {\bibfnamefont {T.}~\bibnamefont {Zhu}}, \bibinfo
  {author} {\bibfnamefont {S.}~\bibnamefont {Zhang}}, \bibinfo {author}
  {\bibfnamefont {Z.}~\bibnamefont {Chen}}, \bibinfo {author} {\bibfnamefont
  {A.}~\bibnamefont {Song}}, \bibinfo {author} {\bibfnamefont {C.}~\bibnamefont
  {Zhang}}, \bibinfo {author} {\bibfnamefont {R.}~\bibnamefont {Gao}}, \bibinfo
  {author} {\bibfnamefont {W.}~\bibnamefont {Niu}}, \bibinfo {author}
  {\bibfnamefont {Y.}~\bibnamefont {Chen}}, \bibinfo {author} {\bibfnamefont
  {F.}~\bibnamefont {Fei}}, \emph {et~al.},\ }\bibfield  {title} {\bibinfo
  {title} {Light-induced giant enhancement of nonreciprocal transport at
  ktao3-based interfaces},\ }\href {https://doi.org/10.1038/s41467-024-47231-6}
  {\bibfield  {journal} {\bibinfo  {journal} {Nature Communications}\ }\textbf
  {\bibinfo {volume} {15}},\ \bibinfo {pages} {2992} (\bibinfo {year}
  {2024}{\natexlab{a}})}\BibitemShut {NoStop}%
\bibitem [{\citenamefont {Li}\ \emph {et~al.}(2024{\natexlab{a}})\citenamefont
  {Li}, \citenamefont {Liang}, \citenamefont {Zhao}, \citenamefont {Wei},\ and\
  \citenamefont {Zhang}}]{structural-symmetry2024JMI-Zhang}%
  \BibitemOpen
  \bibfield  {author} {\bibinfo {author} {\bibfnamefont {C.-N.}\ \bibnamefont
  {Li}}, \bibinfo {author} {\bibfnamefont {H.-P.}\ \bibnamefont {Liang}},
  \bibinfo {author} {\bibfnamefont {B.-Q.}\ \bibnamefont {Zhao}}, \bibinfo
  {author} {\bibfnamefont {S.-H.}\ \bibnamefont {Wei}},\ and\ \bibinfo {author}
  {\bibfnamefont {X.}~\bibnamefont {Zhang}},\ }\bibfield  {title} {\bibinfo
  {title} {Machine learning assisted crystal structure prediction made
  simple},\ }\bibfield  {journal} {\bibinfo  {journal} {Journal of Materials
  Informatics}\ }\textbf {\bibinfo {volume} {4}},\ \href
  {https://doi.org/10.20517/jmi.2024.18} {10.20517/jmi.2024.18} (\bibinfo
  {year} {2024}{\natexlab{a}})\BibitemShut {NoStop}%
\bibitem [{\citenamefont {Ji}\ and\ \citenamefont
  {Wen}(2020)}]{ji2020categorical}%
  \BibitemOpen
  \bibfield  {author} {\bibinfo {author} {\bibfnamefont {W.}~\bibnamefont
  {Ji}}\ and\ \bibinfo {author} {\bibfnamefont {X.-G.}\ \bibnamefont {Wen}},\
  }\bibfield  {title} {\bibinfo {title} {Categorical symmetry and noninvertible
  anomaly in symmetry-breaking and topological phase transitions},\ }\href
  {https://doi.org/10.1103/PhysRevResearch.2.033417} {\bibfield  {journal}
  {\bibinfo  {journal} {Physical Review Research}\ }\textbf {\bibinfo {volume}
  {2}},\ \bibinfo {pages} {033417} (\bibinfo {year} {2020})}\BibitemShut
  {NoStop}%
\bibitem [{\citenamefont {Bender}(2005)}]{bender2005introduction}%
  \BibitemOpen
  \bibfield  {author} {\bibinfo {author} {\bibfnamefont {C.~M.}\ \bibnamefont
  {Bender}},\ }\bibfield  {title} {\bibinfo {title} {Introduction to
  pt-symmetric quantum theory},\ }\href
  {https://doi.org/10.1080/00107500072632} {\bibfield  {journal} {\bibinfo
  {journal} {Contemporary physics}\ }\textbf {\bibinfo {volume} {46}},\
  \bibinfo {pages} {277} (\bibinfo {year} {2005})}\BibitemShut {NoStop}%
\bibitem [{\citenamefont {Li}\ \emph {et~al.}(2024{\natexlab{b}})\citenamefont
  {Li}, \citenamefont {Wang}, \citenamefont {Zhang}, \citenamefont {Qin},
  \citenamefont {Ying}, \citenamefont {Wei}, \citenamefont {Dai}, \citenamefont
  {Guo}, \citenamefont {Chen}, \citenamefont {Zhang} \emph
  {et~al.}}]{li2024observation}%
  \BibitemOpen
  \bibfield  {author} {\bibinfo {author} {\bibfnamefont {C.}~\bibnamefont
  {Li}}, \bibinfo {author} {\bibfnamefont {R.}~\bibnamefont {Wang}}, \bibinfo
  {author} {\bibfnamefont {S.}~\bibnamefont {Zhang}}, \bibinfo {author}
  {\bibfnamefont {Y.}~\bibnamefont {Qin}}, \bibinfo {author} {\bibfnamefont
  {Z.}~\bibnamefont {Ying}}, \bibinfo {author} {\bibfnamefont {B.}~\bibnamefont
  {Wei}}, \bibinfo {author} {\bibfnamefont {Z.}~\bibnamefont {Dai}}, \bibinfo
  {author} {\bibfnamefont {F.}~\bibnamefont {Guo}}, \bibinfo {author}
  {\bibfnamefont {W.}~\bibnamefont {Chen}}, \bibinfo {author} {\bibfnamefont
  {R.}~\bibnamefont {Zhang}}, \emph {et~al.},\ }\bibfield  {title} {\bibinfo
  {title} {Observation of giant non-reciprocal charge transport from quantum
  hall states in a topological insulator},\ }\href
  {https://doi.org/10.1038/s41563-024-01874-4} {\bibfield  {journal} {\bibinfo
  {journal} {Nature Materials}\ ,\ \bibinfo {pages} {1}} (\bibinfo {year}
  {2024}{\natexlab{b}})}\BibitemShut {NoStop}%
\bibitem [{\citenamefont {Zhao}\ \emph {et~al.}(2020)\citenamefont {Zhao},
  \citenamefont {Chen}, \citenamefont {Paillard}, \citenamefont {Arras},
  \citenamefont {Fang}, \citenamefont {Li}, \citenamefont {Gosteau},
  \citenamefont {Yang},\ and\ \citenamefont
  {Bellaiche}}]{Zhao-PRB-symmetry-Rashba2020}%
  \BibitemOpen
  \bibfield  {author} {\bibinfo {author} {\bibfnamefont {H.~J.}\ \bibnamefont
  {Zhao}}, \bibinfo {author} {\bibfnamefont {P.}~\bibnamefont {Chen}}, \bibinfo
  {author} {\bibfnamefont {C.}~\bibnamefont {Paillard}}, \bibinfo {author}
  {\bibfnamefont {R.}~\bibnamefont {Arras}}, \bibinfo {author} {\bibfnamefont
  {Y.-W.}\ \bibnamefont {Fang}}, \bibinfo {author} {\bibfnamefont
  {X.}~\bibnamefont {Li}}, \bibinfo {author} {\bibfnamefont {J.}~\bibnamefont
  {Gosteau}}, \bibinfo {author} {\bibfnamefont {Y.}~\bibnamefont {Yang}},\ and\
  \bibinfo {author} {\bibfnamefont {L.}~\bibnamefont {Bellaiche}},\ }\bibfield
  {title} {\bibinfo {title} {Large spin splittings due to the orbital degree of
  freedom and spin textures in a ferroelectric nitride perovskite},\ }\href
  {https://doi.org/10.1103/PhysRevB.102.041203} {\bibfield  {journal} {\bibinfo
   {journal} {Phys. Rev. B}\ }\textbf {\bibinfo {volume} {102}},\ \bibinfo
  {pages} {041203} (\bibinfo {year} {2020})}\BibitemShut {NoStop}%
\bibitem [{\citenamefont {Brunschwig}\ \emph {et~al.}(2002)\citenamefont
  {Brunschwig}, \citenamefont {Creutz},\ and\ \citenamefont
  {Sutin}}]{brunschwig2002optical}%
  \BibitemOpen
  \bibfield  {author} {\bibinfo {author} {\bibfnamefont {B.~S.}\ \bibnamefont
  {Brunschwig}}, \bibinfo {author} {\bibfnamefont {C.}~\bibnamefont {Creutz}},\
  and\ \bibinfo {author} {\bibfnamefont {N.}~\bibnamefont {Sutin}},\ }\bibfield
   {title} {\bibinfo {title} {Optical transitions of symmetrical mixed-valence
  systems in the class ii--iii transition regime},\ }\href
  {https://doi.org/10.1039/B008034I} {\bibfield  {journal} {\bibinfo  {journal}
  {Chemical Society Reviews}\ }\textbf {\bibinfo {volume} {31}},\ \bibinfo
  {pages} {168} (\bibinfo {year} {2002})}\BibitemShut {NoStop}%
\bibitem [{\citenamefont {Chen}\ \emph
  {et~al.}(2024{\natexlab{a}})\citenamefont {Chen}, \citenamefont {Qiu},
  \citenamefont {Cheng}, \citenamefont {Cui}, \citenamefont {Jin},
  \citenamefont {Tian}, \citenamefont {Zhang}, \citenamefont {Xu},
  \citenamefont {Liu}, \citenamefont {Niu} \emph {et~al.}}]{chen2024defect}%
  \BibitemOpen
  \bibfield  {author} {\bibinfo {author} {\bibfnamefont {Z.}~\bibnamefont
  {Chen}}, \bibinfo {author} {\bibfnamefont {H.}~\bibnamefont {Qiu}}, \bibinfo
  {author} {\bibfnamefont {X.}~\bibnamefont {Cheng}}, \bibinfo {author}
  {\bibfnamefont {J.}~\bibnamefont {Cui}}, \bibinfo {author} {\bibfnamefont
  {Z.}~\bibnamefont {Jin}}, \bibinfo {author} {\bibfnamefont {D.}~\bibnamefont
  {Tian}}, \bibinfo {author} {\bibfnamefont {X.}~\bibnamefont {Zhang}},
  \bibinfo {author} {\bibfnamefont {K.}~\bibnamefont {Xu}}, \bibinfo {author}
  {\bibfnamefont {R.}~\bibnamefont {Liu}}, \bibinfo {author} {\bibfnamefont
  {W.}~\bibnamefont {Niu}}, \emph {et~al.},\ }\bibfield  {title} {\bibinfo
  {title} {Defect-induced helicity dependent terahertz emission in dirac
  semimetal ptte2 thin films},\ }\href
  {https://doi.org/10.1038/s41467-024-46821-8} {\bibfield  {journal} {\bibinfo
  {journal} {Nature Communications}\ }\textbf {\bibinfo {volume} {15}},\
  \bibinfo {pages} {2605} (\bibinfo {year} {2024}{\natexlab{a}})}\BibitemShut
  {NoStop}%
\bibitem [{\citenamefont {Niu}\ \emph {et~al.}(2022)\citenamefont {Niu},
  \citenamefont {Fang}, \citenamefont {Liu}, \citenamefont {Wu}, \citenamefont
  {Chen}, \citenamefont {Gan}, \citenamefont {Zhang}, \citenamefont {Zhu},
  \citenamefont {Wang}, \citenamefont {Xu}, \citenamefont {Pu}, \citenamefont
  {Chen},\ and\ \citenamefont {Wang}}]{Niu-jpclett2022}%
  \BibitemOpen
  \bibfield  {author} {\bibinfo {author} {\bibfnamefont {W.}~\bibnamefont
  {Niu}}, \bibinfo {author} {\bibfnamefont {Y.-W.}\ \bibnamefont {Fang}},
  \bibinfo {author} {\bibfnamefont {R.}~\bibnamefont {Liu}}, \bibinfo {author}
  {\bibfnamefont {Z.}~\bibnamefont {Wu}}, \bibinfo {author} {\bibfnamefont
  {Y.}~\bibnamefont {Chen}}, \bibinfo {author} {\bibfnamefont {Y.}~\bibnamefont
  {Gan}}, \bibinfo {author} {\bibfnamefont {X.}~\bibnamefont {Zhang}}, \bibinfo
  {author} {\bibfnamefont {C.}~\bibnamefont {Zhu}}, \bibinfo {author}
  {\bibfnamefont {L.}~\bibnamefont {Wang}}, \bibinfo {author} {\bibfnamefont
  {Y.}~\bibnamefont {Xu}}, \bibinfo {author} {\bibfnamefont {Y.}~\bibnamefont
  {Pu}}, \bibinfo {author} {\bibfnamefont {Y.}~\bibnamefont {Chen}},\ and\
  \bibinfo {author} {\bibfnamefont {X.}~\bibnamefont {Wang}},\ }\bibfield
  {title} {\bibinfo {title} {Fully optical modulation of the two-dimensional
  electron gas at the $\gamma$-al2o3/srtio3 interface},\ }\href
  {https://doi.org/10.1021/acs.jpclett.2c00384} {\bibfield  {journal} {\bibinfo
   {journal} {The Journal of Physical Chemistry Letters}\ }\textbf {\bibinfo
  {volume} {13}},\ \bibinfo {pages} {2976} (\bibinfo {year} {2022})},\ \bibinfo
  {note} {pMID: 35343699}\BibitemShut {NoStop}%
\bibitem [{\citenamefont {Niu}\ \emph {et~al.}(2017)\citenamefont {Niu},
  \citenamefont {Zhang}, \citenamefont {Gan}, \citenamefont {Christensen},
  \citenamefont {Soosten}, \citenamefont {Garcia-Suarez}, \citenamefont
  {Riisager}, \citenamefont {Wang}, \citenamefont {Xu}, \citenamefont {Zhang},
  \citenamefont {Pryds},\ and\ \citenamefont
  {Chen}}]{Niu-acs.nanolett2017-GAOSTO}%
  \BibitemOpen
  \bibfield  {author} {\bibinfo {author} {\bibfnamefont {W.}~\bibnamefont
  {Niu}}, \bibinfo {author} {\bibfnamefont {Y.}~\bibnamefont {Zhang}}, \bibinfo
  {author} {\bibfnamefont {Y.}~\bibnamefont {Gan}}, \bibinfo {author}
  {\bibfnamefont {D.~V.}\ \bibnamefont {Christensen}}, \bibinfo {author}
  {\bibfnamefont {M.~V.}\ \bibnamefont {Soosten}}, \bibinfo {author}
  {\bibfnamefont {E.~J.}\ \bibnamefont {Garcia-Suarez}}, \bibinfo {author}
  {\bibfnamefont {A.}~\bibnamefont {Riisager}}, \bibinfo {author}
  {\bibfnamefont {X.}~\bibnamefont {Wang}}, \bibinfo {author} {\bibfnamefont
  {Y.}~\bibnamefont {Xu}}, \bibinfo {author} {\bibfnamefont {R.}~\bibnamefont
  {Zhang}}, \bibinfo {author} {\bibfnamefont {N.}~\bibnamefont {Pryds}},\ and\
  \bibinfo {author} {\bibfnamefont {Y.}~\bibnamefont {Chen}},\ }\bibfield
  {title} {\bibinfo {title} {Giant tunability of the two-dimensional electron
  gas at the interface of $\gamma$-al2o3/srtio3},\ }\href
  {https://doi.org/10.1021/acs.nanolett.7b03209} {\bibfield  {journal}
  {\bibinfo  {journal} {Nano Letters}\ }\textbf {\bibinfo {volume} {17}},\
  \bibinfo {pages} {6878} (\bibinfo {year} {2017})},\ \bibinfo {note} {pMID:
  28968124}\BibitemShut {NoStop}%
\bibitem [{\citenamefont {Hu}\ \emph {et~al.}(2021{\natexlab{a}})\citenamefont
  {Hu}, \citenamefont {Luo}, \citenamefont {Fang}, \citenamefont {Qin},
  \citenamefont {Cao}, \citenamefont {Xie}, \citenamefont {Liu}, \citenamefont
  {Dong}, \citenamefont {Sanson}, \citenamefont {Giarola}, \citenamefont {Tan},
  \citenamefont {Zheng}, \citenamefont {Suwardi}, \citenamefont {Huang},
  \citenamefont {Hippalgaonkar}, \citenamefont {He}, \citenamefont {Zhang},
  \citenamefont {Xu}, \citenamefont {Yan},\ and\ \citenamefont
  {Kanatzidis}}]{Hu2021-AdvEnergyMater}%
  \BibitemOpen
  \bibfield  {author} {\bibinfo {author} {\bibfnamefont {L.}~\bibnamefont
  {Hu}}, \bibinfo {author} {\bibfnamefont {Y.}~\bibnamefont {Luo}}, \bibinfo
  {author} {\bibfnamefont {Y.}~\bibnamefont {Fang}}, \bibinfo {author}
  {\bibfnamefont {F.}~\bibnamefont {Qin}}, \bibinfo {author} {\bibfnamefont
  {X.}~\bibnamefont {Cao}}, \bibinfo {author} {\bibfnamefont {H.}~\bibnamefont
  {Xie}}, \bibinfo {author} {\bibfnamefont {J.}~\bibnamefont {Liu}}, \bibinfo
  {author} {\bibfnamefont {J.}~\bibnamefont {Dong}}, \bibinfo {author}
  {\bibfnamefont {A.}~\bibnamefont {Sanson}}, \bibinfo {author} {\bibfnamefont
  {M.}~\bibnamefont {Giarola}}, \bibinfo {author} {\bibfnamefont
  {X.}~\bibnamefont {Tan}}, \bibinfo {author} {\bibfnamefont {Y.}~\bibnamefont
  {Zheng}}, \bibinfo {author} {\bibfnamefont {A.}~\bibnamefont {Suwardi}},
  \bibinfo {author} {\bibfnamefont {Y.}~\bibnamefont {Huang}}, \bibinfo
  {author} {\bibfnamefont {K.}~\bibnamefont {Hippalgaonkar}}, \bibinfo {author}
  {\bibfnamefont {J.}~\bibnamefont {He}}, \bibinfo {author} {\bibfnamefont
  {W.}~\bibnamefont {Zhang}}, \bibinfo {author} {\bibfnamefont
  {J.}~\bibnamefont {Xu}}, \bibinfo {author} {\bibfnamefont {Q.}~\bibnamefont
  {Yan}},\ and\ \bibinfo {author} {\bibfnamefont {M.~G.}\ \bibnamefont
  {Kanatzidis}},\ }\bibfield  {title} {\bibinfo {title} {High thermoelectric
  performance through crystal symmetry enhancement in triply doped diamondoid
  compound cu2snse3},\ }\href {https://doi.org/10.1002/aenm.202100661}
  {\bibfield  {journal} {\bibinfo  {journal} {Advanced Energy Materials}\
  }\textbf {\bibinfo {volume} {11}},\ \bibinfo {pages} {51} (\bibinfo {year}
  {2021}{\natexlab{a}})}\BibitemShut {NoStop}%
\bibitem [{\citenamefont {Hu}\ \emph {et~al.}(2021{\natexlab{b}})\citenamefont
  {Hu}, \citenamefont {Fang}, \citenamefont {Qin}, \citenamefont {Cao},
  \citenamefont {Zhao}, \citenamefont {Luo}, \citenamefont {Repaka},
  \citenamefont {Luo}, \citenamefont {Suwardi}, \citenamefont {Soldi},
  \citenamefont {Aydemir}, \citenamefont {Huang}, \citenamefont {Liu},
  \citenamefont {Hippalgaonkar}, \citenamefont {Snyder}, \citenamefont {Xu},\
  and\ \citenamefont {Yan}}]{Hu2021-NatCommun-Pb7Bi4Se13}%
  \BibitemOpen
  \bibfield  {author} {\bibinfo {author} {\bibfnamefont {L.}~\bibnamefont
  {Hu}}, \bibinfo {author} {\bibfnamefont {Y.-W.}\ \bibnamefont {Fang}},
  \bibinfo {author} {\bibfnamefont {F.}~\bibnamefont {Qin}}, \bibinfo {author}
  {\bibfnamefont {X.}~\bibnamefont {Cao}}, \bibinfo {author} {\bibfnamefont
  {X.}~\bibnamefont {Zhao}}, \bibinfo {author} {\bibfnamefont {Y.}~\bibnamefont
  {Luo}}, \bibinfo {author} {\bibfnamefont {D.~V.~M.}\ \bibnamefont {Repaka}},
  \bibinfo {author} {\bibfnamefont {W.}~\bibnamefont {Luo}}, \bibinfo {author}
  {\bibfnamefont {A.}~\bibnamefont {Suwardi}}, \bibinfo {author} {\bibfnamefont
  {T.}~\bibnamefont {Soldi}}, \bibinfo {author} {\bibfnamefont
  {U.}~\bibnamefont {Aydemir}}, \bibinfo {author} {\bibfnamefont
  {Y.}~\bibnamefont {Huang}}, \bibinfo {author} {\bibfnamefont
  {Z.}~\bibnamefont {Liu}}, \bibinfo {author} {\bibfnamefont {K.}~\bibnamefont
  {Hippalgaonkar}}, \bibinfo {author} {\bibfnamefont {G.~J.}\ \bibnamefont
  {Snyder}}, \bibinfo {author} {\bibfnamefont {J.}~\bibnamefont {Xu}},\ and\
  \bibinfo {author} {\bibfnamefont {Q.}~\bibnamefont {Yan}},\ }\bibfield
  {title} {\bibinfo {title} {High thermoelectric performance enabled by
  convergence of nested conduction bands in pb7bi4se13 with low thermal
  conductivity},\ }\href {https://doi.org/10.1038/s41467-021-25119-z}
  {\bibfield  {journal} {\bibinfo  {journal} {Nat Commun}\ }\textbf {\bibinfo
  {volume} {12}},\ \bibinfo {pages} {105} (\bibinfo {year}
  {2021}{\natexlab{b}})}\BibitemShut {NoStop}%
\bibitem [{\citenamefont {Zheng}\ \emph {et~al.}(2021)\citenamefont {Zheng},
  \citenamefont {Fang}, \citenamefont {Zhang}, \citenamefont {Li},
  \citenamefont {Wen}, \citenamefont {Fang}, \citenamefont {He}, \citenamefont
  {Li}, \citenamefont {Zhang}, \citenamefont {Tong}, \citenamefont {Mi},
  \citenamefont {Bai}, \citenamefont {Alshareef}, \citenamefont {Qiu},\ and\
  \citenamefont {Zhang}}]{DXZheng-acsnano2021-halleffect}%
  \BibitemOpen
  \bibfield  {author} {\bibinfo {author} {\bibfnamefont {D.}~\bibnamefont
  {Zheng}}, \bibinfo {author} {\bibfnamefont {Y.-W.}\ \bibnamefont {Fang}},
  \bibinfo {author} {\bibfnamefont {S.}~\bibnamefont {Zhang}}, \bibinfo
  {author} {\bibfnamefont {P.}~\bibnamefont {Li}}, \bibinfo {author}
  {\bibfnamefont {Y.}~\bibnamefont {Wen}}, \bibinfo {author} {\bibfnamefont
  {B.}~\bibnamefont {Fang}}, \bibinfo {author} {\bibfnamefont {X.}~\bibnamefont
  {He}}, \bibinfo {author} {\bibfnamefont {Y.}~\bibnamefont {Li}}, \bibinfo
  {author} {\bibfnamefont {C.}~\bibnamefont {Zhang}}, \bibinfo {author}
  {\bibfnamefont {W.}~\bibnamefont {Tong}}, \bibinfo {author} {\bibfnamefont
  {W.}~\bibnamefont {Mi}}, \bibinfo {author} {\bibfnamefont {H.}~\bibnamefont
  {Bai}}, \bibinfo {author} {\bibfnamefont {H.~N.}\ \bibnamefont {Alshareef}},
  \bibinfo {author} {\bibfnamefont {Z.~Q.}\ \bibnamefont {Qiu}},\ and\ \bibinfo
  {author} {\bibfnamefont {X.}~\bibnamefont {Zhang}},\ }\bibfield  {title}
  {\bibinfo {title} {Berry phase engineering in srruo3/sriro3/srtio3
  superlattices induced by band structure reconstruction},\ }\href
  {https://doi.org/10.1021/acsnano.0c10200} {\bibfield  {journal} {\bibinfo
  {journal} {ACS Nano}\ }\textbf {\bibinfo {volume} {15}},\ \bibinfo {pages}
  {5086} (\bibinfo {year} {2021})},\ \bibinfo {note} {pMID:
  33606942}\BibitemShut {NoStop}%
\bibitem [{\citenamefont {Stormer}(1999)}]{stormer1999nobel}%
  \BibitemOpen
  \bibfield  {author} {\bibinfo {author} {\bibfnamefont {H.~L.}\ \bibnamefont
  {Stormer}},\ }\bibfield  {title} {\bibinfo {title} {Nobel lecture: the
  fractional quantum hall effect},\ }\href
  {https://doi.org/10.1103/RevModPhys.71.875} {\bibfield  {journal} {\bibinfo
  {journal} {Reviews of Modern Physics}\ }\textbf {\bibinfo {volume} {71}},\
  \bibinfo {pages} {875} (\bibinfo {year} {1999})}\BibitemShut {NoStop}%
\bibitem [{\citenamefont {Popovi{\'c}}(1989)}]{popovic1989hall}%
  \BibitemOpen
  \bibfield  {author} {\bibinfo {author} {\bibfnamefont {R.}~\bibnamefont
  {Popovi{\'c}}},\ }\bibfield  {title} {\bibinfo {title} {Hall-effect
  devices},\ }\href {https://doi.org/10.1016/0250-6874(89)80063-0} {\bibfield
  {journal} {\bibinfo  {journal} {Sensors and Actuators}\ }\textbf {\bibinfo
  {volume} {17}},\ \bibinfo {pages} {39} (\bibinfo {year} {1989})}\BibitemShut
  {NoStop}%
\bibitem [{\citenamefont {Nagaosa}\ \emph {et~al.}(2010)\citenamefont
  {Nagaosa}, \citenamefont {Sinova}, \citenamefont {Onoda}, \citenamefont
  {MacDonald},\ and\ \citenamefont {Ong}}]{nagaosa2010anomalous}%
  \BibitemOpen
  \bibfield  {author} {\bibinfo {author} {\bibfnamefont {N.}~\bibnamefont
  {Nagaosa}}, \bibinfo {author} {\bibfnamefont {J.}~\bibnamefont {Sinova}},
  \bibinfo {author} {\bibfnamefont {S.}~\bibnamefont {Onoda}}, \bibinfo
  {author} {\bibfnamefont {A.~H.}\ \bibnamefont {MacDonald}},\ and\ \bibinfo
  {author} {\bibfnamefont {N.~P.}\ \bibnamefont {Ong}},\ }\bibfield  {title}
  {\bibinfo {title} {Anomalous hall effect},\ }\href
  {https://doi.org/10.1103/RevModPhys.82.1539} {\bibfield  {journal} {\bibinfo
  {journal} {Reviews of modern physics}\ }\textbf {\bibinfo {volume} {82}},\
  \bibinfo {pages} {1539} (\bibinfo {year} {2010})}\BibitemShut {NoStop}%
\bibitem [{\citenamefont {Sinova}\ \emph {et~al.}(2015)\citenamefont {Sinova},
  \citenamefont {Valenzuela}, \citenamefont {Wunderlich}, \citenamefont
  {Back},\ and\ \citenamefont {Jungwirth}}]{sinova2015spin}%
  \BibitemOpen
  \bibfield  {author} {\bibinfo {author} {\bibfnamefont {J.}~\bibnamefont
  {Sinova}}, \bibinfo {author} {\bibfnamefont {S.~O.}\ \bibnamefont
  {Valenzuela}}, \bibinfo {author} {\bibfnamefont {J.}~\bibnamefont
  {Wunderlich}}, \bibinfo {author} {\bibfnamefont {C.}~\bibnamefont {Back}},\
  and\ \bibinfo {author} {\bibfnamefont {T.}~\bibnamefont {Jungwirth}},\
  }\bibfield  {title} {\bibinfo {title} {Spin hall effects},\ }\href
  {https://doi.org/10.1103/RevModPhys.87.1213} {\bibfield  {journal} {\bibinfo
  {journal} {Reviews of modern physics}\ }\textbf {\bibinfo {volume} {87}},\
  \bibinfo {pages} {1213} (\bibinfo {year} {2015})}\BibitemShut {NoStop}%
\bibitem [{\citenamefont {Sodemann}\ and\ \citenamefont
  {Fu}(2015)}]{sodemann2015quantum}%
  \BibitemOpen
  \bibfield  {author} {\bibinfo {author} {\bibfnamefont {I.}~\bibnamefont
  {Sodemann}}\ and\ \bibinfo {author} {\bibfnamefont {L.}~\bibnamefont {Fu}},\
  }\bibfield  {title} {\bibinfo {title} {Quantum nonlinear hall effect induced
  by berry curvature dipole in time-reversal invariant materials},\ }\href
  {https://doi.org/10.1103/PhysRevLett.115.216806} {\bibfield  {journal}
  {\bibinfo  {journal} {Physical review letters}\ }\textbf {\bibinfo {volume}
  {115}},\ \bibinfo {pages} {216806} (\bibinfo {year} {2015})}\BibitemShut
  {NoStop}%
\bibitem [{\citenamefont {Kang}\ \emph {et~al.}(2019)\citenamefont {Kang},
  \citenamefont {Li}, \citenamefont {Sohn}, \citenamefont {Shan},\ and\
  \citenamefont {Mak}}]{kang2019nonlinear}%
  \BibitemOpen
  \bibfield  {author} {\bibinfo {author} {\bibfnamefont {K.}~\bibnamefont
  {Kang}}, \bibinfo {author} {\bibfnamefont {T.}~\bibnamefont {Li}}, \bibinfo
  {author} {\bibfnamefont {E.}~\bibnamefont {Sohn}}, \bibinfo {author}
  {\bibfnamefont {J.}~\bibnamefont {Shan}},\ and\ \bibinfo {author}
  {\bibfnamefont {K.~F.}\ \bibnamefont {Mak}},\ }\bibfield  {title} {\bibinfo
  {title} {Nonlinear anomalous hall effect in few-layer wte2},\ }\href
  {https://doi.org/10.1038/s41563-019-0294-7} {\bibfield  {journal} {\bibinfo
  {journal} {Nature materials}\ }\textbf {\bibinfo {volume} {18}},\ \bibinfo
  {pages} {324} (\bibinfo {year} {2019})}\BibitemShut {NoStop}%
\bibitem [{\citenamefont {Ma}\ \emph {et~al.}(2019)\citenamefont {Ma},
  \citenamefont {Xu}, \citenamefont {Shen}, \citenamefont {MacNeill},
  \citenamefont {Fatemi}, \citenamefont {Chang}, \citenamefont {Mier~Valdivia},
  \citenamefont {Wu}, \citenamefont {Du}, \citenamefont {Hsu} \emph
  {et~al.}}]{ma2019observation}%
  \BibitemOpen
  \bibfield  {author} {\bibinfo {author} {\bibfnamefont {Q.}~\bibnamefont
  {Ma}}, \bibinfo {author} {\bibfnamefont {S.-Y.}\ \bibnamefont {Xu}}, \bibinfo
  {author} {\bibfnamefont {H.}~\bibnamefont {Shen}}, \bibinfo {author}
  {\bibfnamefont {D.}~\bibnamefont {MacNeill}}, \bibinfo {author}
  {\bibfnamefont {V.}~\bibnamefont {Fatemi}}, \bibinfo {author} {\bibfnamefont
  {T.-R.}\ \bibnamefont {Chang}}, \bibinfo {author} {\bibfnamefont {A.~M.}\
  \bibnamefont {Mier~Valdivia}}, \bibinfo {author} {\bibfnamefont
  {S.}~\bibnamefont {Wu}}, \bibinfo {author} {\bibfnamefont {Z.}~\bibnamefont
  {Du}}, \bibinfo {author} {\bibfnamefont {C.-H.}\ \bibnamefont {Hsu}}, \emph
  {et~al.},\ }\bibfield  {title} {\bibinfo {title} {Observation of the
  nonlinear hall effect under time-reversal-symmetric conditions},\ }\href
  {https://doi.org/10.1038/s41586-018-0807-6} {\bibfield  {journal} {\bibinfo
  {journal} {Nature}\ }\textbf {\bibinfo {volume} {565}},\ \bibinfo {pages}
  {337} (\bibinfo {year} {2019})}\BibitemShut {NoStop}%
\bibitem [{\citenamefont {Tiwari}\ \emph {et~al.}(2021)\citenamefont {Tiwari},
  \citenamefont {Chen}, \citenamefont {Zhong}, \citenamefont {Drueke},
  \citenamefont {Koo}, \citenamefont {Kaczmarek}, \citenamefont {Xiao},
  \citenamefont {Gao}, \citenamefont {Luo}, \citenamefont {Niu} \emph
  {et~al.}}]{tiwari2021giant}%
  \BibitemOpen
  \bibfield  {author} {\bibinfo {author} {\bibfnamefont {A.}~\bibnamefont
  {Tiwari}}, \bibinfo {author} {\bibfnamefont {F.}~\bibnamefont {Chen}},
  \bibinfo {author} {\bibfnamefont {S.}~\bibnamefont {Zhong}}, \bibinfo
  {author} {\bibfnamefont {E.}~\bibnamefont {Drueke}}, \bibinfo {author}
  {\bibfnamefont {J.}~\bibnamefont {Koo}}, \bibinfo {author} {\bibfnamefont
  {A.}~\bibnamefont {Kaczmarek}}, \bibinfo {author} {\bibfnamefont
  {C.}~\bibnamefont {Xiao}}, \bibinfo {author} {\bibfnamefont {J.}~\bibnamefont
  {Gao}}, \bibinfo {author} {\bibfnamefont {X.}~\bibnamefont {Luo}}, \bibinfo
  {author} {\bibfnamefont {Q.}~\bibnamefont {Niu}}, \emph {et~al.},\ }\bibfield
   {title} {\bibinfo {title} {Giant c-axis nonlinear anomalous hall effect in
  td-mote2 and wte2},\ }\href {https://doi.org/10.1038/s41467-021-22343-5}
  {\bibfield  {journal} {\bibinfo  {journal} {Nature communications}\ }\textbf
  {\bibinfo {volume} {12}},\ \bibinfo {pages} {2049} (\bibinfo {year}
  {2021})}\BibitemShut {NoStop}%
\bibitem [{\citenamefont {Kumar}\ \emph {et~al.}(2021)\citenamefont {Kumar},
  \citenamefont {Hsu}, \citenamefont {Sharma}, \citenamefont {Chang},
  \citenamefont {Yu}, \citenamefont {Wang}, \citenamefont {Eda}, \citenamefont
  {Liang},\ and\ \citenamefont {Yang}}]{kumar2021room}%
  \BibitemOpen
  \bibfield  {author} {\bibinfo {author} {\bibfnamefont {D.}~\bibnamefont
  {Kumar}}, \bibinfo {author} {\bibfnamefont {C.-H.}\ \bibnamefont {Hsu}},
  \bibinfo {author} {\bibfnamefont {R.}~\bibnamefont {Sharma}}, \bibinfo
  {author} {\bibfnamefont {T.-R.}\ \bibnamefont {Chang}}, \bibinfo {author}
  {\bibfnamefont {P.}~\bibnamefont {Yu}}, \bibinfo {author} {\bibfnamefont
  {J.}~\bibnamefont {Wang}}, \bibinfo {author} {\bibfnamefont {G.}~\bibnamefont
  {Eda}}, \bibinfo {author} {\bibfnamefont {G.}~\bibnamefont {Liang}},\ and\
  \bibinfo {author} {\bibfnamefont {H.}~\bibnamefont {Yang}},\ }\bibfield
  {title} {\bibinfo {title} {Room-temperature nonlinear hall effect and
  wireless radiofrequency rectification in weyl semimetal tairte4},\ }\href
  {https://doi.org/10.1038/s41565-020-00839-3} {\bibfield  {journal} {\bibinfo
  {journal} {Nature Nanotechnology}\ }\textbf {\bibinfo {volume} {16}},\
  \bibinfo {pages} {421} (\bibinfo {year} {2021})}\BibitemShut {NoStop}%
\bibitem [{\citenamefont {Wang}\ \emph {et~al.}(2023)\citenamefont {Wang},
  \citenamefont {Kaplan}, \citenamefont {Zhang}, \citenamefont {Holder},
  \citenamefont {Cao}, \citenamefont {Wang}, \citenamefont {Zhou},
  \citenamefont {Zhou}, \citenamefont {Jiang}, \citenamefont {Zhang} \emph
  {et~al.}}]{wang2023quantum}%
  \BibitemOpen
  \bibfield  {author} {\bibinfo {author} {\bibfnamefont {N.}~\bibnamefont
  {Wang}}, \bibinfo {author} {\bibfnamefont {D.}~\bibnamefont {Kaplan}},
  \bibinfo {author} {\bibfnamefont {Z.}~\bibnamefont {Zhang}}, \bibinfo
  {author} {\bibfnamefont {T.}~\bibnamefont {Holder}}, \bibinfo {author}
  {\bibfnamefont {N.}~\bibnamefont {Cao}}, \bibinfo {author} {\bibfnamefont
  {A.}~\bibnamefont {Wang}}, \bibinfo {author} {\bibfnamefont {X.}~\bibnamefont
  {Zhou}}, \bibinfo {author} {\bibfnamefont {F.}~\bibnamefont {Zhou}}, \bibinfo
  {author} {\bibfnamefont {Z.}~\bibnamefont {Jiang}}, \bibinfo {author}
  {\bibfnamefont {C.}~\bibnamefont {Zhang}}, \emph {et~al.},\ }\bibfield
  {title} {\bibinfo {title} {Quantum-metric-induced nonlinear transport in a
  topological antiferromagnet},\ }\href
  {https://doi.org/10.1038/s41586-023-06363-3} {\bibfield  {journal} {\bibinfo
  {journal} {Nature}\ }\textbf {\bibinfo {volume} {621}},\ \bibinfo {pages}
  {487} (\bibinfo {year} {2023})}\BibitemShut {NoStop}%
\bibitem [{\citenamefont {Gao}\ \emph {et~al.}(2023)\citenamefont {Gao},
  \citenamefont {Liu}, \citenamefont {Qiu}, \citenamefont {Ghosh},
  \citenamefont {V.~Trevisan}, \citenamefont {Onishi}, \citenamefont {Hu},
  \citenamefont {Qian}, \citenamefont {Tien}, \citenamefont {Chen} \emph
  {et~al.}}]{gao2023quantum}%
  \BibitemOpen
  \bibfield  {author} {\bibinfo {author} {\bibfnamefont {A.}~\bibnamefont
  {Gao}}, \bibinfo {author} {\bibfnamefont {Y.-F.}\ \bibnamefont {Liu}},
  \bibinfo {author} {\bibfnamefont {J.-X.}\ \bibnamefont {Qiu}}, \bibinfo
  {author} {\bibfnamefont {B.}~\bibnamefont {Ghosh}}, \bibinfo {author}
  {\bibfnamefont {T.}~\bibnamefont {V.~Trevisan}}, \bibinfo {author}
  {\bibfnamefont {Y.}~\bibnamefont {Onishi}}, \bibinfo {author} {\bibfnamefont
  {C.}~\bibnamefont {Hu}}, \bibinfo {author} {\bibfnamefont {T.}~\bibnamefont
  {Qian}}, \bibinfo {author} {\bibfnamefont {H.-J.}\ \bibnamefont {Tien}},
  \bibinfo {author} {\bibfnamefont {S.-W.}\ \bibnamefont {Chen}}, \emph
  {et~al.},\ }\bibfield  {title} {\bibinfo {title} {Quantum metric nonlinear
  hall effect in a topological antiferromagnetic heterostructure},\ }\href
  {https://doi.org/10.1126/science.adf1506} {\bibfield  {journal} {\bibinfo
  {journal} {Science}\ }\textbf {\bibinfo {volume} {381}},\ \bibinfo {pages}
  {181} (\bibinfo {year} {2023})}\BibitemShut {NoStop}%
\bibitem [{\citenamefont {Rostami}\ and\ \citenamefont
  {Juri{\v{c}}i{\'c}}(2020)}]{rostami2020probing}%
  \BibitemOpen
  \bibfield  {author} {\bibinfo {author} {\bibfnamefont {H.}~\bibnamefont
  {Rostami}}\ and\ \bibinfo {author} {\bibfnamefont {V.}~\bibnamefont
  {Juri{\v{c}}i{\'c}}},\ }\bibfield  {title} {\bibinfo {title} {Probing quantum
  criticality using nonlinear hall effect in a metallic dirac system},\ }\href
  {https://doi.org/10.1103/PhysRevResearch.2.013069} {\bibfield  {journal}
  {\bibinfo  {journal} {Physical Review Research}\ }\textbf {\bibinfo {volume}
  {2}},\ \bibinfo {pages} {013069} (\bibinfo {year} {2020})}\BibitemShut
  {NoStop}%
\bibitem [{\citenamefont {He}\ and\ \citenamefont
  {Law}(2024)}]{He2024_nonlinearhalleffect_insulators}%
  \BibitemOpen
  \bibfield  {author} {\bibinfo {author} {\bibfnamefont {W.-Y.}\ \bibnamefont
  {He}}\ and\ \bibinfo {author} {\bibfnamefont {K.~T.}\ \bibnamefont {Law}},\
  }\href {https://arxiv.org/abs/2411.07456} {\bibinfo {title} {Nonlinear hall
  effect in insulators}} (\bibinfo {year} {2024}),\ \Eprint
  {https://arxiv.org/abs/2411.07456} {arXiv:2411.07456 [cond-mat.mes-hall]}
  \BibitemShut {NoStop}%
\bibitem [{\citenamefont {Lee}\ \emph {et~al.}(2024)\citenamefont {Lee},
  \citenamefont {Wang}, \citenamefont {Chen}, \citenamefont {Kwon},
  \citenamefont {Hwang}, \citenamefont {Cho}, \citenamefont {Son},
  \citenamefont {Han}, \citenamefont {Choi}, \citenamefont {Kim} \emph
  {et~al.}}]{lee2024spin}%
  \BibitemOpen
  \bibfield  {author} {\bibinfo {author} {\bibfnamefont {J.-E.}\ \bibnamefont
  {Lee}}, \bibinfo {author} {\bibfnamefont {A.}~\bibnamefont {Wang}}, \bibinfo
  {author} {\bibfnamefont {S.}~\bibnamefont {Chen}}, \bibinfo {author}
  {\bibfnamefont {M.}~\bibnamefont {Kwon}}, \bibinfo {author} {\bibfnamefont
  {J.}~\bibnamefont {Hwang}}, \bibinfo {author} {\bibfnamefont
  {M.}~\bibnamefont {Cho}}, \bibinfo {author} {\bibfnamefont {K.-H.}\
  \bibnamefont {Son}}, \bibinfo {author} {\bibfnamefont {D.-S.}\ \bibnamefont
  {Han}}, \bibinfo {author} {\bibfnamefont {J.~W.}\ \bibnamefont {Choi}},
  \bibinfo {author} {\bibfnamefont {Y.~D.}\ \bibnamefont {Kim}}, \emph
  {et~al.},\ }\bibfield  {title} {\bibinfo {title} {Spin-orbit-splitting-driven
  nonlinear hall effect in nbirte4},\ }\href
  {https://doi.org/10.1038/s41467-024-47643-4} {\bibfield  {journal} {\bibinfo
  {journal} {Nature Communications}\ }\textbf {\bibinfo {volume} {15}},\
  \bibinfo {pages} {3971} (\bibinfo {year} {2024})}\BibitemShut {NoStop}%
\bibitem [{\citenamefont {Duan}\ \emph {et~al.}(2023)\citenamefont {Duan},
  \citenamefont {Qin}, \citenamefont {Chen}, \citenamefont {Yang},
  \citenamefont {Qiu}, \citenamefont {Huang}, \citenamefont {Liu},
  \citenamefont {Li}, \citenamefont {Bi}, \citenamefont {Meng} \emph
  {et~al.}}]{duan2023berry}%
  \BibitemOpen
  \bibfield  {author} {\bibinfo {author} {\bibfnamefont {S.}~\bibnamefont
  {Duan}}, \bibinfo {author} {\bibfnamefont {F.}~\bibnamefont {Qin}}, \bibinfo
  {author} {\bibfnamefont {P.}~\bibnamefont {Chen}}, \bibinfo {author}
  {\bibfnamefont {X.}~\bibnamefont {Yang}}, \bibinfo {author} {\bibfnamefont
  {C.}~\bibnamefont {Qiu}}, \bibinfo {author} {\bibfnamefont {J.}~\bibnamefont
  {Huang}}, \bibinfo {author} {\bibfnamefont {G.}~\bibnamefont {Liu}}, \bibinfo
  {author} {\bibfnamefont {Z.}~\bibnamefont {Li}}, \bibinfo {author}
  {\bibfnamefont {X.}~\bibnamefont {Bi}}, \bibinfo {author} {\bibfnamefont
  {F.}~\bibnamefont {Meng}}, \emph {et~al.},\ }\bibfield  {title} {\bibinfo
  {title} {Berry curvature dipole generation and helicity-to-spin conversion at
  symmetry-mismatched heterointerfaces},\ }\href
  {https://doi.org/10.1038/s41565-023-01417-z} {\bibfield  {journal} {\bibinfo
  {journal} {Nature Nanotechnology}\ }\textbf {\bibinfo {volume} {18}},\
  \bibinfo {pages} {867} (\bibinfo {year} {2023})}\BibitemShut {NoStop}%
\bibitem [{\citenamefont {He}\ and\ \citenamefont {Weng}(2021)}]{he2021giant}%
  \BibitemOpen
  \bibfield  {author} {\bibinfo {author} {\bibfnamefont {Z.}~\bibnamefont
  {He}}\ and\ \bibinfo {author} {\bibfnamefont {H.}~\bibnamefont {Weng}},\
  }\bibfield  {title} {\bibinfo {title} {Giant nonlinear hall effect in twisted
  bilayer wte2},\ }\href {https://doi.org/10.1038/s41535-021-00403-9}
  {\bibfield  {journal} {\bibinfo  {journal} {npj Quantum Materials}\ }\textbf
  {\bibinfo {volume} {6}},\ \bibinfo {pages} {101} (\bibinfo {year}
  {2021})}\BibitemShut {NoStop}%
\bibitem [{\citenamefont {Hu}\ \emph {et~al.}(2022)\citenamefont {Hu},
  \citenamefont {Zhang}, \citenamefont {Xie},\ and\ \citenamefont
  {Law}}]{hu2022nonlinear}%
  \BibitemOpen
  \bibfield  {author} {\bibinfo {author} {\bibfnamefont {J.-X.}\ \bibnamefont
  {Hu}}, \bibinfo {author} {\bibfnamefont {C.-P.}\ \bibnamefont {Zhang}},
  \bibinfo {author} {\bibfnamefont {Y.-M.}\ \bibnamefont {Xie}},\ and\ \bibinfo
  {author} {\bibfnamefont {K.}~\bibnamefont {Law}},\ }\bibfield  {title}
  {\bibinfo {title} {Nonlinear hall effects in strained twisted bilayer wse2},\
  }\href {https://doi.org/10.1038/s42005-022-01034-7} {\bibfield  {journal}
  {\bibinfo  {journal} {Communications Physics}\ }\textbf {\bibinfo {volume}
  {5}},\ \bibinfo {pages} {255} (\bibinfo {year} {2022})}\BibitemShut {NoStop}%
\bibitem [{\citenamefont {Gao}\ \emph {et~al.}(2021)\citenamefont {Gao},
  \citenamefont {Liu}, \citenamefont {Hu}, \citenamefont {Qiu}, \citenamefont
  {Tzschaschel}, \citenamefont {Ghosh}, \citenamefont {Ho}, \citenamefont
  {B{\'e}rub{\'e}}, \citenamefont {Chen}, \citenamefont {Sun} \emph
  {et~al.}}]{gao2021layer}%
  \BibitemOpen
  \bibfield  {author} {\bibinfo {author} {\bibfnamefont {A.}~\bibnamefont
  {Gao}}, \bibinfo {author} {\bibfnamefont {Y.-F.}\ \bibnamefont {Liu}},
  \bibinfo {author} {\bibfnamefont {C.}~\bibnamefont {Hu}}, \bibinfo {author}
  {\bibfnamefont {J.-X.}\ \bibnamefont {Qiu}}, \bibinfo {author} {\bibfnamefont
  {C.}~\bibnamefont {Tzschaschel}}, \bibinfo {author} {\bibfnamefont
  {B.}~\bibnamefont {Ghosh}}, \bibinfo {author} {\bibfnamefont {S.-C.}\
  \bibnamefont {Ho}}, \bibinfo {author} {\bibfnamefont {D.}~\bibnamefont
  {B{\'e}rub{\'e}}}, \bibinfo {author} {\bibfnamefont {R.}~\bibnamefont
  {Chen}}, \bibinfo {author} {\bibfnamefont {H.}~\bibnamefont {Sun}}, \emph
  {et~al.},\ }\bibfield  {title} {\bibinfo {title} {Layer hall effect in a 2d
  topological axion antiferromagnet},\ }\href
  {https://doi.org/10.1038/s41586-021-03679-w} {\bibfield  {journal} {\bibinfo
  {journal} {Nature}\ }\textbf {\bibinfo {volume} {595}},\ \bibinfo {pages}
  {521} (\bibinfo {year} {2021})}\BibitemShut {NoStop}%
\bibitem [{\citenamefont {Du}\ \emph {et~al.}(2021{\natexlab{a}})\citenamefont
  {Du}, \citenamefont {Lu},\ and\ \citenamefont {Xie}}]{du2021nonlinear}%
  \BibitemOpen
  \bibfield  {author} {\bibinfo {author} {\bibfnamefont {Z.}~\bibnamefont
  {Du}}, \bibinfo {author} {\bibfnamefont {H.-Z.}\ \bibnamefont {Lu}},\ and\
  \bibinfo {author} {\bibfnamefont {X.}~\bibnamefont {Xie}},\ }\bibfield
  {title} {\bibinfo {title} {Nonlinear hall effects},\ }\href
  {https://doi.org/10.1038/s42254-021-00359-6} {\bibfield  {journal} {\bibinfo
  {journal} {Nature Reviews Physics}\ }\textbf {\bibinfo {volume} {3}},\
  \bibinfo {pages} {744} (\bibinfo {year} {2021}{\natexlab{a}})}\BibitemShut
  {NoStop}%
\bibitem [{\citenamefont {Klitzing}\ \emph {et~al.}(1980)\citenamefont
  {Klitzing}, \citenamefont {Dorda},\ and\ \citenamefont
  {Pepper}}]{klitzing1980new}%
  \BibitemOpen
  \bibfield  {author} {\bibinfo {author} {\bibfnamefont {K.~v.}\ \bibnamefont
  {Klitzing}}, \bibinfo {author} {\bibfnamefont {G.}~\bibnamefont {Dorda}},\
  and\ \bibinfo {author} {\bibfnamefont {M.}~\bibnamefont {Pepper}},\
  }\bibfield  {title} {\bibinfo {title} {New method for high-accuracy
  determination of the fine-structure constant based on quantized hall
  resistance},\ }\href {https://doi.org/10.1103/PhysRevLett.45.494} {\bibfield
  {journal} {\bibinfo  {journal} {Physical review letters}\ }\textbf {\bibinfo
  {volume} {45}},\ \bibinfo {pages} {494} (\bibinfo {year} {1980})}\BibitemShut
  {NoStop}%
\bibitem [{\citenamefont {Yasuda}\ \emph {et~al.}(2016)\citenamefont {Yasuda},
  \citenamefont {Wakatsuki}, \citenamefont {Morimoto}, \citenamefont {Yoshimi},
  \citenamefont {Tsukazaki}, \citenamefont {Takahashi}, \citenamefont {Ezawa},
  \citenamefont {Kawasaki}, \citenamefont {Nagaosa},\ and\ \citenamefont
  {Tokura}}]{yasuda2016geometric}%
  \BibitemOpen
  \bibfield  {author} {\bibinfo {author} {\bibfnamefont {K.}~\bibnamefont
  {Yasuda}}, \bibinfo {author} {\bibfnamefont {R.}~\bibnamefont {Wakatsuki}},
  \bibinfo {author} {\bibfnamefont {T.}~\bibnamefont {Morimoto}}, \bibinfo
  {author} {\bibfnamefont {R.}~\bibnamefont {Yoshimi}}, \bibinfo {author}
  {\bibfnamefont {A.}~\bibnamefont {Tsukazaki}}, \bibinfo {author}
  {\bibfnamefont {K.}~\bibnamefont {Takahashi}}, \bibinfo {author}
  {\bibfnamefont {M.}~\bibnamefont {Ezawa}}, \bibinfo {author} {\bibfnamefont
  {M.}~\bibnamefont {Kawasaki}}, \bibinfo {author} {\bibfnamefont
  {N.}~\bibnamefont {Nagaosa}},\ and\ \bibinfo {author} {\bibfnamefont
  {Y.}~\bibnamefont {Tokura}},\ }\bibfield  {title} {\bibinfo {title}
  {Geometric hall effects in topological insulator heterostructures},\ }\href
  {https://doi.org/10.1038/nphys3671} {\bibfield  {journal} {\bibinfo
  {journal} {Nature Physics}\ }\textbf {\bibinfo {volume} {12}},\ \bibinfo
  {pages} {555} (\bibinfo {year} {2016})}\BibitemShut {NoStop}%
\bibitem [{\citenamefont {Gao}\ \emph {et~al.}(2014)\citenamefont {Gao},
  \citenamefont {Yang},\ and\ \citenamefont {Niu}}]{gao2014field}%
  \BibitemOpen
  \bibfield  {author} {\bibinfo {author} {\bibfnamefont {Y.}~\bibnamefont
  {Gao}}, \bibinfo {author} {\bibfnamefont {S.~A.}\ \bibnamefont {Yang}},\ and\
  \bibinfo {author} {\bibfnamefont {Q.}~\bibnamefont {Niu}},\ }\bibfield
  {title} {\bibinfo {title} {Field induced positional shift of bloch electrons
  and its dynamical implications},\ }\href
  {https://doi.org/10.1103/PhysRevLett.112.166601} {\bibfield  {journal}
  {\bibinfo  {journal} {Physical review letters}\ }\textbf {\bibinfo {volume}
  {112}},\ \bibinfo {pages} {166601} (\bibinfo {year} {2014})}\BibitemShut
  {NoStop}%
\bibitem [{\citenamefont {Wang}\ \emph {et~al.}(2021)\citenamefont {Wang},
  \citenamefont {Gao},\ and\ \citenamefont {Xiao}}]{wang2021intrinsic}%
  \BibitemOpen
  \bibfield  {author} {\bibinfo {author} {\bibfnamefont {C.}~\bibnamefont
  {Wang}}, \bibinfo {author} {\bibfnamefont {Y.}~\bibnamefont {Gao}},\ and\
  \bibinfo {author} {\bibfnamefont {D.}~\bibnamefont {Xiao}},\ }\bibfield
  {title} {\bibinfo {title} {Intrinsic nonlinear hall effect in
  antiferromagnetic tetragonal cumnas},\ }\href
  {https://doi.org/10.1103/PhysRevLett.127.277201} {\bibfield  {journal}
  {\bibinfo  {journal} {Physical Review Letters}\ }\textbf {\bibinfo {volume}
  {127}},\ \bibinfo {pages} {277201} (\bibinfo {year} {2021})}\BibitemShut
  {NoStop}%
\bibitem [{\citenamefont {Du}\ \emph {et~al.}(2019)\citenamefont {Du},
  \citenamefont {Wang}, \citenamefont {Li}, \citenamefont {Lu},\ and\
  \citenamefont {Xie}}]{du2019disorder}%
  \BibitemOpen
  \bibfield  {author} {\bibinfo {author} {\bibfnamefont {Z.}~\bibnamefont
  {Du}}, \bibinfo {author} {\bibfnamefont {C.}~\bibnamefont {Wang}}, \bibinfo
  {author} {\bibfnamefont {S.}~\bibnamefont {Li}}, \bibinfo {author}
  {\bibfnamefont {H.-Z.}\ \bibnamefont {Lu}},\ and\ \bibinfo {author}
  {\bibfnamefont {X.}~\bibnamefont {Xie}},\ }\bibfield  {title} {\bibinfo
  {title} {Disorder-induced nonlinear hall effect with time-reversal
  symmetry},\ }\href {https://doi.org/10.1038/s41467-019-10941-3} {\bibfield
  {journal} {\bibinfo  {journal} {Nature Communications}\ }\textbf {\bibinfo
  {volume} {10}},\ \bibinfo {pages} {3047} (\bibinfo {year}
  {2019})}\BibitemShut {NoStop}%
\bibitem [{\citenamefont {Du}\ \emph {et~al.}(2021{\natexlab{b}})\citenamefont
  {Du}, \citenamefont {Wang}, \citenamefont {Sun}, \citenamefont {Lu},\ and\
  \citenamefont {Xie}}]{du2021quantum}%
  \BibitemOpen
  \bibfield  {author} {\bibinfo {author} {\bibfnamefont {Z.}~\bibnamefont
  {Du}}, \bibinfo {author} {\bibfnamefont {C.}~\bibnamefont {Wang}}, \bibinfo
  {author} {\bibfnamefont {H.-P.}\ \bibnamefont {Sun}}, \bibinfo {author}
  {\bibfnamefont {H.-Z.}\ \bibnamefont {Lu}},\ and\ \bibinfo {author}
  {\bibfnamefont {X.}~\bibnamefont {Xie}},\ }\bibfield  {title} {\bibinfo
  {title} {Quantum theory of the nonlinear hall effect},\ }\href
  {https://doi.org/10.1038/s41467-021-25273-4} {\bibfield  {journal} {\bibinfo
  {journal} {Nature communications}\ }\textbf {\bibinfo {volume} {12}},\
  \bibinfo {pages} {5038} (\bibinfo {year} {2021}{\natexlab{b}})}\BibitemShut
  {NoStop}%
\bibitem [{\citenamefont {Tokura}\ and\ \citenamefont
  {Nagaosa}(2018)}]{tokura2018nonreciprocal}%
  \BibitemOpen
  \bibfield  {author} {\bibinfo {author} {\bibfnamefont {Y.}~\bibnamefont
  {Tokura}}\ and\ \bibinfo {author} {\bibfnamefont {N.}~\bibnamefont
  {Nagaosa}},\ }\bibfield  {title} {\bibinfo {title} {Nonreciprocal responses
  from non-centrosymmetric quantum materials},\ }\href
  {https://doi.org/10.1038/s41467-018-05759-4} {\bibfield  {journal} {\bibinfo
  {journal} {Nature communications}\ }\textbf {\bibinfo {volume} {9}},\
  \bibinfo {pages} {3740} (\bibinfo {year} {2018})}\BibitemShut {NoStop}%
\bibitem [{\citenamefont {Provost}\ and\ \citenamefont
  {Vallee}(1980)}]{provost1980riemannian}%
  \BibitemOpen
  \bibfield  {author} {\bibinfo {author} {\bibfnamefont {J.}~\bibnamefont
  {Provost}}\ and\ \bibinfo {author} {\bibfnamefont {G.}~\bibnamefont
  {Vallee}},\ }\bibfield  {title} {\bibinfo {title} {Riemannian structure on
  manifolds of quantum states},\ }\href {https://doi.org/10.1007/BF02193559}
  {\bibfield  {journal} {\bibinfo  {journal} {Communications in Mathematical
  Physics}\ }\textbf {\bibinfo {volume} {76}},\ \bibinfo {pages} {289}
  (\bibinfo {year} {1980})}\BibitemShut {NoStop}%
\bibitem [{\citenamefont {Tsirkin}\ and\ \citenamefont
  {Souza}(2022)}]{tsirkin2022separation}%
  \BibitemOpen
  \bibfield  {author} {\bibinfo {author} {\bibfnamefont {S.}~\bibnamefont
  {Tsirkin}}\ and\ \bibinfo {author} {\bibfnamefont {I.}~\bibnamefont
  {Souza}},\ }\bibfield  {title} {\bibinfo {title} {On the separation of hall
  and ohmic nonlinear responses},\ }\href
  {https://doi.org/10.21468/SciPostPhysCore.5.3.039} {\bibfield  {journal}
  {\bibinfo  {journal} {SciPost Physics Core}\ }\textbf {\bibinfo {volume}
  {5}},\ \bibinfo {pages} {039} (\bibinfo {year} {2022})}\BibitemShut {NoStop}%
\bibitem [{\citenamefont {Holder}\ \emph {et~al.}(2020)\citenamefont {Holder},
  \citenamefont {Kaplan},\ and\ \citenamefont {Yan}}]{holder2020consequences}%
  \BibitemOpen
  \bibfield  {author} {\bibinfo {author} {\bibfnamefont {T.}~\bibnamefont
  {Holder}}, \bibinfo {author} {\bibfnamefont {D.}~\bibnamefont {Kaplan}},\
  and\ \bibinfo {author} {\bibfnamefont {B.}~\bibnamefont {Yan}},\ }\bibfield
  {title} {\bibinfo {title} {Consequences of time-reversal-symmetry breaking in
  the light-matter interaction: Berry curvature, quantum metric, and diabatic
  motion},\ }\href {https://doi.org/10.1103/PhysRevResearch.2.033100}
  {\bibfield  {journal} {\bibinfo  {journal} {Physical Review Research}\
  }\textbf {\bibinfo {volume} {2}},\ \bibinfo {pages} {033100} (\bibinfo {year}
  {2020})}\BibitemShut {NoStop}%
\bibitem [{\citenamefont {Watanabe}\ and\ \citenamefont
  {Yanase}(2020)}]{watanabe2020nonlinear}%
  \BibitemOpen
  \bibfield  {author} {\bibinfo {author} {\bibfnamefont {H.}~\bibnamefont
  {Watanabe}}\ and\ \bibinfo {author} {\bibfnamefont {Y.}~\bibnamefont
  {Yanase}},\ }\bibfield  {title} {\bibinfo {title} {Nonlinear electric
  transport in odd-parity magnetic multipole systems: Application to mn-based
  compounds},\ }\href {https://doi.org/10.1103/PhysRevResearch.2.043081}
  {\bibfield  {journal} {\bibinfo  {journal} {Physical Review Research}\
  }\textbf {\bibinfo {volume} {2}},\ \bibinfo {pages} {043081} (\bibinfo {year}
  {2020})}\BibitemShut {NoStop}%
\bibitem [{\citenamefont {Liu}\ \emph {et~al.}(2021)\citenamefont {Liu},
  \citenamefont {Zhao}, \citenamefont {Huang}, \citenamefont {Wu},
  \citenamefont {Sheng}, \citenamefont {Xiao},\ and\ \citenamefont
  {Yang}}]{liu2021intrinsic}%
  \BibitemOpen
  \bibfield  {author} {\bibinfo {author} {\bibfnamefont {H.}~\bibnamefont
  {Liu}}, \bibinfo {author} {\bibfnamefont {J.}~\bibnamefont {Zhao}}, \bibinfo
  {author} {\bibfnamefont {Y.-X.}\ \bibnamefont {Huang}}, \bibinfo {author}
  {\bibfnamefont {W.}~\bibnamefont {Wu}}, \bibinfo {author} {\bibfnamefont
  {X.-L.}\ \bibnamefont {Sheng}}, \bibinfo {author} {\bibfnamefont
  {C.}~\bibnamefont {Xiao}},\ and\ \bibinfo {author} {\bibfnamefont {S.~A.}\
  \bibnamefont {Yang}},\ }\bibfield  {title} {\bibinfo {title} {Intrinsic
  second-order anomalous hall effect and its application in compensated
  antiferromagnets},\ }\href {https://doi.org/10.1103/PhysRevLett.127.277202}
  {\bibfield  {journal} {\bibinfo  {journal} {Physical Review Letters}\
  }\textbf {\bibinfo {volume} {127}},\ \bibinfo {pages} {277202} (\bibinfo
  {year} {2021})}\BibitemShut {NoStop}%
\bibitem [{\citenamefont {Han}\ \emph {et~al.}(2024)\citenamefont {Han},
  \citenamefont {Uchimura}, \citenamefont {Araki}, \citenamefont {Yoon},
  \citenamefont {Takeuchi}, \citenamefont {Yamane}, \citenamefont {Kanai},
  \citenamefont {Ieda}, \citenamefont {Ohno},\ and\ \citenamefont
  {Fukami}}]{han2024room}%
  \BibitemOpen
  \bibfield  {author} {\bibinfo {author} {\bibfnamefont {J.}~\bibnamefont
  {Han}}, \bibinfo {author} {\bibfnamefont {T.}~\bibnamefont {Uchimura}},
  \bibinfo {author} {\bibfnamefont {Y.}~\bibnamefont {Araki}}, \bibinfo
  {author} {\bibfnamefont {J.-Y.}\ \bibnamefont {Yoon}}, \bibinfo {author}
  {\bibfnamefont {Y.}~\bibnamefont {Takeuchi}}, \bibinfo {author}
  {\bibfnamefont {Y.}~\bibnamefont {Yamane}}, \bibinfo {author} {\bibfnamefont
  {S.}~\bibnamefont {Kanai}}, \bibinfo {author} {\bibfnamefont
  {J.}~\bibnamefont {Ieda}}, \bibinfo {author} {\bibfnamefont {H.}~\bibnamefont
  {Ohno}},\ and\ \bibinfo {author} {\bibfnamefont {S.}~\bibnamefont {Fukami}},\
  }\bibfield  {title} {\bibinfo {title} {Room-temperature flexible manipulation
  of the quantum-metric structure in a topological chiral antiferromagnet},\
  }\href {https://doi.org/10.1038/s41567-024-02476-2} {\bibfield  {journal}
  {\bibinfo  {journal} {Nature Physics}\ ,\ \bibinfo {pages} {1}} (\bibinfo
  {year} {2024})}\BibitemShut {NoStop}%
\bibitem [{\citenamefont {Isobe}\ \emph {et~al.}(2020)\citenamefont {Isobe},
  \citenamefont {Xu},\ and\ \citenamefont {Fu}}]{isobe2020high}%
  \BibitemOpen
  \bibfield  {author} {\bibinfo {author} {\bibfnamefont {H.}~\bibnamefont
  {Isobe}}, \bibinfo {author} {\bibfnamefont {S.-Y.}\ \bibnamefont {Xu}},\ and\
  \bibinfo {author} {\bibfnamefont {L.}~\bibnamefont {Fu}},\ }\bibfield
  {title} {\bibinfo {title} {High-frequency rectification via chiral bloch
  electrons},\ }\href {https://doi.org/10.1126/sciadv.aay2497} {\bibfield
  {journal} {\bibinfo  {journal} {Science advances}\ }\textbf {\bibinfo
  {volume} {6}},\ \bibinfo {pages} {eaay2497} (\bibinfo {year}
  {2020})}\BibitemShut {NoStop}%
\bibitem [{\citenamefont {Berry}(1984)}]{berry1984quantal}%
  \BibitemOpen
  \bibfield  {author} {\bibinfo {author} {\bibfnamefont {M.~V.}\ \bibnamefont
  {Berry}},\ }\bibfield  {title} {\bibinfo {title} {Quantal phase factors
  accompanying adiabatic changes},\ }\href
  {https://doi.org/10.1098/rspa.1984.0023} {\bibfield  {journal} {\bibinfo
  {journal} {Proceedings of the Royal Society of London. A. Mathematical and
  Physical Sciences}\ }\textbf {\bibinfo {volume} {392}},\ \bibinfo {pages}
  {45} (\bibinfo {year} {1984})}\BibitemShut {NoStop}%
\bibitem [{\citenamefont {Xiao}\ \emph {et~al.}(2010)\citenamefont {Xiao},
  \citenamefont {Chang},\ and\ \citenamefont {Niu}}]{xiao2010berry}%
  \BibitemOpen
  \bibfield  {author} {\bibinfo {author} {\bibfnamefont {D.}~\bibnamefont
  {Xiao}}, \bibinfo {author} {\bibfnamefont {M.-C.}\ \bibnamefont {Chang}},\
  and\ \bibinfo {author} {\bibfnamefont {Q.}~\bibnamefont {Niu}},\ }\bibfield
  {title} {\bibinfo {title} {Berry phase effects on electronic properties},\
  }\href {https://doi.org/10.1103/RevModPhys.82.1959} {\bibfield  {journal}
  {\bibinfo  {journal} {Reviews of modern physics}\ }\textbf {\bibinfo {volume}
  {82}},\ \bibinfo {pages} {1959} (\bibinfo {year} {2010})}\BibitemShut
  {NoStop}%
\bibitem [{\citenamefont {Bohm}\ \emph {et~al.}(2013)\citenamefont {Bohm},
  \citenamefont {Mostafazadeh}, \citenamefont {Koizumi}, \citenamefont {Niu},\
  and\ \citenamefont {Zwanziger}}]{bohm2013geometric}%
  \BibitemOpen
  \bibfield  {author} {\bibinfo {author} {\bibfnamefont {A.}~\bibnamefont
  {Bohm}}, \bibinfo {author} {\bibfnamefont {A.}~\bibnamefont {Mostafazadeh}},
  \bibinfo {author} {\bibfnamefont {H.}~\bibnamefont {Koizumi}}, \bibinfo
  {author} {\bibfnamefont {Q.}~\bibnamefont {Niu}},\ and\ \bibinfo {author}
  {\bibfnamefont {J.}~\bibnamefont {Zwanziger}},\ }\href@noop {} {\emph
  {\bibinfo {title} {The Geometric phase in quantum systems: foundations,
  mathematical concepts, and applications in molecular and condensed matter
  physics}}}\ (\bibinfo  {publisher} {Springer Science \& Business Media},\
  \bibinfo {year} {2013})\BibitemShut {NoStop}%
\bibitem [{\citenamefont {Karplus}\ and\ \citenamefont
  {Luttinger}(1954)}]{karplus1954hall}%
  \BibitemOpen
  \bibfield  {author} {\bibinfo {author} {\bibfnamefont {R.}~\bibnamefont
  {Karplus}}\ and\ \bibinfo {author} {\bibfnamefont {J.}~\bibnamefont
  {Luttinger}},\ }\bibfield  {title} {\bibinfo {title} {Hall effect in
  ferromagnetics},\ }\href {https://doi.org/10.1103/PhysRev.95.1154} {\bibfield
   {journal} {\bibinfo  {journal} {Physical Review}\ }\textbf {\bibinfo
  {volume} {95}},\ \bibinfo {pages} {1154} (\bibinfo {year}
  {1954})}\BibitemShut {NoStop}%
\bibitem [{\citenamefont {Low}\ \emph {et~al.}(2015)\citenamefont {Low},
  \citenamefont {Jiang},\ and\ \citenamefont {Guinea}}]{low2015topological}%
  \BibitemOpen
  \bibfield  {author} {\bibinfo {author} {\bibfnamefont {T.}~\bibnamefont
  {Low}}, \bibinfo {author} {\bibfnamefont {Y.}~\bibnamefont {Jiang}},\ and\
  \bibinfo {author} {\bibfnamefont {F.}~\bibnamefont {Guinea}},\ }\bibfield
  {title} {\bibinfo {title} {Topological currents in black phosphorus with
  broken inversion symmetry},\ }\href
  {https://doi.org/10.1103/PhysRevB.92.235447} {\bibfield  {journal} {\bibinfo
  {journal} {Physical Review B}\ }\textbf {\bibinfo {volume} {92}},\ \bibinfo
  {pages} {235447} (\bibinfo {year} {2015})}\BibitemShut {NoStop}%
\bibitem [{\citenamefont {Joseph}\ \emph {et~al.}(2024)\citenamefont {Joseph},
  \citenamefont {Bandyopadhyay},\ and\ \citenamefont
  {Narayan}}]{joseph2024chirality}%
  \BibitemOpen
  \bibfield  {author} {\bibinfo {author} {\bibfnamefont {N.~B.}\ \bibnamefont
  {Joseph}}, \bibinfo {author} {\bibfnamefont {A.}~\bibnamefont
  {Bandyopadhyay}},\ and\ \bibinfo {author} {\bibfnamefont {A.}~\bibnamefont
  {Narayan}},\ }\bibfield  {title} {\bibinfo {title} {Chirality-tunable
  nonlinear hall effect},\ }\href
  {https://doi.org/10.1021/acs.chemmater.4c00427} {\bibfield  {journal}
  {\bibinfo  {journal} {Chemistry of Materials}\ }\textbf {\bibinfo {volume}
  {36}},\ \bibinfo {pages} {8602} (\bibinfo {year} {2024})}\BibitemShut
  {NoStop}%
\bibitem [{\citenamefont {Zhu}\ and\ \citenamefont
  {Yakobson}(2024)}]{zhu2024creating}%
  \BibitemOpen
  \bibfield  {author} {\bibinfo {author} {\bibfnamefont {H.}~\bibnamefont
  {Zhu}}\ and\ \bibinfo {author} {\bibfnamefont {B.~I.}\ \bibnamefont
  {Yakobson}},\ }\bibfield  {title} {\bibinfo {title} {Creating chirality in
  the nearly two dimensions},\ }\href
  {https://doi.org/10.1038/s41563-024-01814-2} {\bibfield  {journal} {\bibinfo
  {journal} {Nature Materials}\ }\textbf {\bibinfo {volume} {23}},\ \bibinfo
  {pages} {316} (\bibinfo {year} {2024})}\BibitemShut {NoStop}%
\bibitem [{\citenamefont {Peshcherenko}\ \emph {et~al.}(2024)\citenamefont
  {Peshcherenko}, \citenamefont {Felser},\ and\ \citenamefont
  {Zhang}}]{peshcherenko2024quantized}%
  \BibitemOpen
  \bibfield  {author} {\bibinfo {author} {\bibfnamefont {N.}~\bibnamefont
  {Peshcherenko}}, \bibinfo {author} {\bibfnamefont {C.}~\bibnamefont
  {Felser}},\ and\ \bibinfo {author} {\bibfnamefont {Y.}~\bibnamefont
  {Zhang}},\ }\bibfield  {title} {\bibinfo {title} {Quantized nonlinear hall
  effect from chiral monopole},\ }\href
  {https://doi.org/10.1103/PhysRevB.110.155143} {\bibfield  {journal} {\bibinfo
   {journal} {Physical Review B}\ }\textbf {\bibinfo {volume} {110}},\ \bibinfo
  {pages} {155143} (\bibinfo {year} {2024})}\BibitemShut {NoStop}%
\bibitem [{\citenamefont {Li}\ \emph {et~al.}(2024{\natexlab{c}})\citenamefont
  {Li}, \citenamefont {Zhang}, \citenamefont {Zhou}, \citenamefont {Ma},
  \citenamefont {Lei}, \citenamefont {Jin}, \citenamefont {He}, \citenamefont
  {Li}, \citenamefont {Law},\ and\ \citenamefont {Wang}}]{li2024quantum}%
  \BibitemOpen
  \bibfield  {author} {\bibinfo {author} {\bibfnamefont {H.}~\bibnamefont
  {Li}}, \bibinfo {author} {\bibfnamefont {C.}~\bibnamefont {Zhang}}, \bibinfo
  {author} {\bibfnamefont {C.}~\bibnamefont {Zhou}}, \bibinfo {author}
  {\bibfnamefont {C.}~\bibnamefont {Ma}}, \bibinfo {author} {\bibfnamefont
  {X.}~\bibnamefont {Lei}}, \bibinfo {author} {\bibfnamefont {Z.}~\bibnamefont
  {Jin}}, \bibinfo {author} {\bibfnamefont {H.}~\bibnamefont {He}}, \bibinfo
  {author} {\bibfnamefont {B.}~\bibnamefont {Li}}, \bibinfo {author}
  {\bibfnamefont {K.~T.}\ \bibnamefont {Law}},\ and\ \bibinfo {author}
  {\bibfnamefont {J.}~\bibnamefont {Wang}},\ }\bibfield  {title} {\bibinfo
  {title} {Quantum geometry quadrupole-induced third-order nonlinear transport
  in antiferromagnetic topological insulator mnbi2te4},\ }\href
  {https://doi.org/10.1038/s41467-024-52206-8} {\bibfield  {journal} {\bibinfo
  {journal} {Nature Communications}\ }\textbf {\bibinfo {volume} {15}},\
  \bibinfo {pages} {7779} (\bibinfo {year} {2024}{\natexlab{c}})}\BibitemShut
  {NoStop}%
\bibitem [{\citenamefont {Sankar}\ \emph {et~al.}(2024)\citenamefont {Sankar},
  \citenamefont {Liu}, \citenamefont {Zhang}, \citenamefont {Li}, \citenamefont
  {Chen}, \citenamefont {Gao}, \citenamefont {Zheng}, \citenamefont {Lin},
  \citenamefont {Qian}, \citenamefont {Yu} \emph
  {et~al.}}]{sankar2024experimental}%
  \BibitemOpen
  \bibfield  {author} {\bibinfo {author} {\bibfnamefont {S.}~\bibnamefont
  {Sankar}}, \bibinfo {author} {\bibfnamefont {R.}~\bibnamefont {Liu}},
  \bibinfo {author} {\bibfnamefont {C.-P.}\ \bibnamefont {Zhang}}, \bibinfo
  {author} {\bibfnamefont {Q.-F.}\ \bibnamefont {Li}}, \bibinfo {author}
  {\bibfnamefont {C.}~\bibnamefont {Chen}}, \bibinfo {author} {\bibfnamefont
  {X.-J.}\ \bibnamefont {Gao}}, \bibinfo {author} {\bibfnamefont
  {J.}~\bibnamefont {Zheng}}, \bibinfo {author} {\bibfnamefont {Y.-H.}\
  \bibnamefont {Lin}}, \bibinfo {author} {\bibfnamefont {K.}~\bibnamefont
  {Qian}}, \bibinfo {author} {\bibfnamefont {R.-P.}\ \bibnamefont {Yu}}, \emph
  {et~al.},\ }\bibfield  {title} {\bibinfo {title} {Experimental evidence for a
  berry curvature quadrupole in an antiferromagnet},\ }\href
  {https://doi.org/10.1103/PhysRevX.14.021046} {\bibfield  {journal} {\bibinfo
  {journal} {Physical Review X}\ }\textbf {\bibinfo {volume} {14}},\ \bibinfo
  {pages} {021046} (\bibinfo {year} {2024})}\BibitemShut {NoStop}%
\bibitem [{\citenamefont {Mak}\ \emph {et~al.}(2014)\citenamefont {Mak},
  \citenamefont {McGill}, \citenamefont {Park},\ and\ \citenamefont
  {McEuen}}]{mak2014valley}%
  \BibitemOpen
  \bibfield  {author} {\bibinfo {author} {\bibfnamefont {K.~F.}\ \bibnamefont
  {Mak}}, \bibinfo {author} {\bibfnamefont {K.~L.}\ \bibnamefont {McGill}},
  \bibinfo {author} {\bibfnamefont {J.}~\bibnamefont {Park}},\ and\ \bibinfo
  {author} {\bibfnamefont {P.~L.}\ \bibnamefont {McEuen}},\ }\bibfield  {title}
  {\bibinfo {title} {The valley hall effect in mos2 transistors},\ }\href
  {https://doi.org/10.1126/science.1250140} {\bibfield  {journal} {\bibinfo
  {journal} {Science}\ }\textbf {\bibinfo {volume} {344}},\ \bibinfo {pages}
  {1489} (\bibinfo {year} {2014})}\BibitemShut {NoStop}%
\bibitem [{\citenamefont {Xu}\ and\ \citenamefont
  {Moore}(2006)}]{xu2006stability}%
  \BibitemOpen
  \bibfield  {author} {\bibinfo {author} {\bibfnamefont {C.}~\bibnamefont
  {Xu}}\ and\ \bibinfo {author} {\bibfnamefont {J.~E.}\ \bibnamefont {Moore}},\
  }\bibfield  {title} {\bibinfo {title} {Stability of the quantum spin hall
  effect: Effects of interactions, disorder, and z 2 topology},\ }\href
  {https://doi.org/10.1103/PhysRevB.73.045322} {\bibfield  {journal} {\bibinfo
  {journal} {Physical Review B—Condensed Matter and Materials Physics}\
  }\textbf {\bibinfo {volume} {73}},\ \bibinfo {pages} {045322} (\bibinfo
  {year} {2006})}\BibitemShut {NoStop}%
\bibitem [{\citenamefont {Berger}(1970)}]{berger1970side}%
  \BibitemOpen
  \bibfield  {author} {\bibinfo {author} {\bibfnamefont {L.}~\bibnamefont
  {Berger}},\ }\bibfield  {title} {\bibinfo {title} {Side-jump mechanism for
  the hall effect of ferromagnets},\ }\href
  {https://doi.org/10.1103/PhysRevB.2.4559} {\bibfield  {journal} {\bibinfo
  {journal} {Physical Review B}\ }\textbf {\bibinfo {volume} {2}},\ \bibinfo
  {pages} {4559} (\bibinfo {year} {1970})}\BibitemShut {NoStop}%
\bibitem [{\citenamefont {Smit}(1955)}]{smit1955spontaneous}%
  \BibitemOpen
  \bibfield  {author} {\bibinfo {author} {\bibfnamefont {J.}~\bibnamefont
  {Smit}},\ }\bibfield  {title} {\bibinfo {title} {The spontaneous hall effect
  in ferromagnetics i},\ }\href {https://doi.org/10.1016/S0031-8914(55)92596-9}
  {\bibfield  {journal} {\bibinfo  {journal} {Physica}\ }\textbf {\bibinfo
  {volume} {21}},\ \bibinfo {pages} {877} (\bibinfo {year} {1955})}\BibitemShut
  {NoStop}%
\bibitem [{\citenamefont {Smit}(1958)}]{smit1958spontaneous}%
  \BibitemOpen
  \bibfield  {author} {\bibinfo {author} {\bibfnamefont {J.}~\bibnamefont
  {Smit}},\ }\bibfield  {title} {\bibinfo {title} {The spontaneous hall effect
  in ferromagnetics ii},\ }\href
  {https://doi.org/10.1016/S0031-8914(58)93541-9} {\bibfield  {journal}
  {\bibinfo  {journal} {Physica}\ }\textbf {\bibinfo {volume} {24}},\ \bibinfo
  {pages} {39} (\bibinfo {year} {1958})}\BibitemShut {NoStop}%
\bibitem [{\citenamefont {Cheng}\ \emph {et~al.}(2024)\citenamefont {Cheng},
  \citenamefont {Gao}, \citenamefont {Zheng}, \citenamefont {Chen},
  \citenamefont {Liu}, \citenamefont {Zhang}, \citenamefont {Zhu},
  \citenamefont {Li}, \citenamefont {Li},\ and\ \citenamefont
  {Zeng}}]{cheng2024giant}%
  \BibitemOpen
  \bibfield  {author} {\bibinfo {author} {\bibfnamefont {B.}~\bibnamefont
  {Cheng}}, \bibinfo {author} {\bibfnamefont {Y.}~\bibnamefont {Gao}}, \bibinfo
  {author} {\bibfnamefont {Z.}~\bibnamefont {Zheng}}, \bibinfo {author}
  {\bibfnamefont {S.}~\bibnamefont {Chen}}, \bibinfo {author} {\bibfnamefont
  {Z.}~\bibnamefont {Liu}}, \bibinfo {author} {\bibfnamefont {L.}~\bibnamefont
  {Zhang}}, \bibinfo {author} {\bibfnamefont {Q.}~\bibnamefont {Zhu}}, \bibinfo
  {author} {\bibfnamefont {H.}~\bibnamefont {Li}}, \bibinfo {author}
  {\bibfnamefont {L.}~\bibnamefont {Li}},\ and\ \bibinfo {author}
  {\bibfnamefont {C.}~\bibnamefont {Zeng}},\ }\bibfield  {title} {\bibinfo
  {title} {Giant nonlinear hall and wireless rectification effects at room
  temperature in the elemental semiconductor tellurium},\ }\href
  {https://doi.org/10.1038/s41467-024-49706-y} {\bibfield  {journal} {\bibinfo
  {journal} {Nature Communications}\ }\textbf {\bibinfo {volume} {15}},\
  \bibinfo {pages} {5513} (\bibinfo {year} {2024})}\BibitemShut {NoStop}%
\bibitem [{\citenamefont {Lu}\ \emph {et~al.}(2024)\citenamefont {Lu},
  \citenamefont {Zhang}, \citenamefont {Wang}, \citenamefont {Zhao},
  \citenamefont {Zhou}, \citenamefont {Gao}, \citenamefont {Chen},
  \citenamefont {Law},\ and\ \citenamefont {Loh}}]{lu2024nonlinear}%
  \BibitemOpen
  \bibfield  {author} {\bibinfo {author} {\bibfnamefont {X.~F.}\ \bibnamefont
  {Lu}}, \bibinfo {author} {\bibfnamefont {C.-P.}\ \bibnamefont {Zhang}},
  \bibinfo {author} {\bibfnamefont {N.}~\bibnamefont {Wang}}, \bibinfo {author}
  {\bibfnamefont {D.}~\bibnamefont {Zhao}}, \bibinfo {author} {\bibfnamefont
  {X.}~\bibnamefont {Zhou}}, \bibinfo {author} {\bibfnamefont {W.}~\bibnamefont
  {Gao}}, \bibinfo {author} {\bibfnamefont {X.~H.}\ \bibnamefont {Chen}},
  \bibinfo {author} {\bibfnamefont {K.}~\bibnamefont {Law}},\ and\ \bibinfo
  {author} {\bibfnamefont {K.~P.}\ \bibnamefont {Loh}},\ }\bibfield  {title}
  {\bibinfo {title} {Nonlinear transport and radio frequency rectification in
  bitebr at room temperature},\ }\href
  {https://doi.org/10.1038/s41467-023-44439-w} {\bibfield  {journal} {\bibinfo
  {journal} {Nature communications}\ }\textbf {\bibinfo {volume} {15}},\
  \bibinfo {pages} {245} (\bibinfo {year} {2024})}\BibitemShut {NoStop}%
\bibitem [{\citenamefont {He}\ \emph {et~al.}(2021)\citenamefont {He},
  \citenamefont {Isobe}, \citenamefont {Zhu}, \citenamefont {Hsu},
  \citenamefont {Fu},\ and\ \citenamefont {Yang}}]{he2021quantum}%
  \BibitemOpen
  \bibfield  {author} {\bibinfo {author} {\bibfnamefont {P.}~\bibnamefont
  {He}}, \bibinfo {author} {\bibfnamefont {H.}~\bibnamefont {Isobe}}, \bibinfo
  {author} {\bibfnamefont {D.}~\bibnamefont {Zhu}}, \bibinfo {author}
  {\bibfnamefont {C.-H.}\ \bibnamefont {Hsu}}, \bibinfo {author} {\bibfnamefont
  {L.}~\bibnamefont {Fu}},\ and\ \bibinfo {author} {\bibfnamefont
  {H.}~\bibnamefont {Yang}},\ }\bibfield  {title} {\bibinfo {title} {Quantum
  frequency doubling in the topological insulator bi2se3},\ }\href
  {https://doi.org/10.1038/s41467-021-20983-1} {\bibfield  {journal} {\bibinfo
  {journal} {Nature Communications}\ }\textbf {\bibinfo {volume} {12}},\
  \bibinfo {pages} {698} (\bibinfo {year} {2021})}\BibitemShut {NoStop}%
\bibitem [{\citenamefont {Novoselov}\ \emph {et~al.}(2004)\citenamefont
  {Novoselov}, \citenamefont {Geim}, \citenamefont {Morozov}, \citenamefont
  {Jiang}, \citenamefont {Zhang}, \citenamefont {Dubonos}, \citenamefont
  {Grigorieva},\ and\ \citenamefont {Firsov}}]{novoselov2004electric}%
  \BibitemOpen
  \bibfield  {author} {\bibinfo {author} {\bibfnamefont {K.~S.}\ \bibnamefont
  {Novoselov}}, \bibinfo {author} {\bibfnamefont {A.~K.}\ \bibnamefont {Geim}},
  \bibinfo {author} {\bibfnamefont {S.~V.}\ \bibnamefont {Morozov}}, \bibinfo
  {author} {\bibfnamefont {D.-e.}\ \bibnamefont {Jiang}}, \bibinfo {author}
  {\bibfnamefont {Y.}~\bibnamefont {Zhang}}, \bibinfo {author} {\bibfnamefont
  {S.~V.}\ \bibnamefont {Dubonos}}, \bibinfo {author} {\bibfnamefont {I.~V.}\
  \bibnamefont {Grigorieva}},\ and\ \bibinfo {author} {\bibfnamefont {A.~A.}\
  \bibnamefont {Firsov}},\ }\bibfield  {title} {\bibinfo {title} {Electric
  field effect in atomically thin carbon films},\ }\href
  {https://doi.org/10.1126/science.1102896} {\bibfield  {journal} {\bibinfo
  {journal} {science}\ }\textbf {\bibinfo {volume} {306}},\ \bibinfo {pages}
  {666} (\bibinfo {year} {2004})}\BibitemShut {NoStop}%
\bibitem [{\citenamefont {Yang}\ \emph {et~al.}(2022)\citenamefont {Yang},
  \citenamefont {Valenzuela}, \citenamefont {Chshiev}, \citenamefont {Couet},
  \citenamefont {Dieny}, \citenamefont {Dlubak}, \citenamefont {Fert},
  \citenamefont {Garello}, \citenamefont {Jamet}, \citenamefont {Jeong} \emph
  {et~al.}}]{yang2022two}%
  \BibitemOpen
  \bibfield  {author} {\bibinfo {author} {\bibfnamefont {H.}~\bibnamefont
  {Yang}}, \bibinfo {author} {\bibfnamefont {S.~O.}\ \bibnamefont
  {Valenzuela}}, \bibinfo {author} {\bibfnamefont {M.}~\bibnamefont {Chshiev}},
  \bibinfo {author} {\bibfnamefont {S.}~\bibnamefont {Couet}}, \bibinfo
  {author} {\bibfnamefont {B.}~\bibnamefont {Dieny}}, \bibinfo {author}
  {\bibfnamefont {B.}~\bibnamefont {Dlubak}}, \bibinfo {author} {\bibfnamefont
  {A.}~\bibnamefont {Fert}}, \bibinfo {author} {\bibfnamefont {K.}~\bibnamefont
  {Garello}}, \bibinfo {author} {\bibfnamefont {M.}~\bibnamefont {Jamet}},
  \bibinfo {author} {\bibfnamefont {D.-E.}\ \bibnamefont {Jeong}}, \emph
  {et~al.},\ }\bibfield  {title} {\bibinfo {title} {Two-dimensional materials
  prospects for non-volatile spintronic memories},\ }\href
  {https://doi.org/10.1038/s41586-022-04768-0} {\bibfield  {journal} {\bibinfo
  {journal} {Nature}\ }\textbf {\bibinfo {volume} {606}},\ \bibinfo {pages}
  {663} (\bibinfo {year} {2022})}\BibitemShut {NoStop}%
\bibitem [{\citenamefont {Li}\ \emph {et~al.}(2017)\citenamefont {Li},
  \citenamefont {Tao}, \citenamefont {Chen}, \citenamefont {Fang},
  \citenamefont {Li}, \citenamefont {Wang}, \citenamefont {Xu},\ and\
  \citenamefont {Zhu}}]{li2017graphene}%
  \BibitemOpen
  \bibfield  {author} {\bibinfo {author} {\bibfnamefont {X.}~\bibnamefont
  {Li}}, \bibinfo {author} {\bibfnamefont {L.}~\bibnamefont {Tao}}, \bibinfo
  {author} {\bibfnamefont {Z.}~\bibnamefont {Chen}}, \bibinfo {author}
  {\bibfnamefont {H.}~\bibnamefont {Fang}}, \bibinfo {author} {\bibfnamefont
  {X.}~\bibnamefont {Li}}, \bibinfo {author} {\bibfnamefont {X.}~\bibnamefont
  {Wang}}, \bibinfo {author} {\bibfnamefont {J.-B.}\ \bibnamefont {Xu}},\ and\
  \bibinfo {author} {\bibfnamefont {H.}~\bibnamefont {Zhu}},\ }\bibfield
  {title} {\bibinfo {title} {Graphene and related two-dimensional materials:
  Structure-property relationships for electronics and optoelectronics},\
  }\bibfield  {journal} {\bibinfo  {journal} {Applied Physics Reviews}\
  }\textbf {\bibinfo {volume} {4}},\ \href {https://doi.org/10.1063/1.4983646}
  {10.1063/1.4983646} (\bibinfo {year} {2017})\BibitemShut {NoStop}%
\bibitem [{\citenamefont {Chia}\ and\ \citenamefont
  {Pumera}(2018)}]{chia2018characteristics}%
  \BibitemOpen
  \bibfield  {author} {\bibinfo {author} {\bibfnamefont {X.}~\bibnamefont
  {Chia}}\ and\ \bibinfo {author} {\bibfnamefont {M.}~\bibnamefont {Pumera}},\
  }\bibfield  {title} {\bibinfo {title} {Characteristics and performance of
  two-dimensional materials for electrocatalysis},\ }\href
  {https://doi.org/10.1038/s41929-018-0181-7} {\bibfield  {journal} {\bibinfo
  {journal} {Nature Catalysis}\ }\textbf {\bibinfo {volume} {1}},\ \bibinfo
  {pages} {909} (\bibinfo {year} {2018})}\BibitemShut {NoStop}%
\bibitem [{\citenamefont {Zhang}\ \emph {et~al.}(2022)\citenamefont {Zhang},
  \citenamefont {Wang}, \citenamefont {Cao}, \citenamefont {Wang},
  \citenamefont {Zhou}, \citenamefont {Watanabe}, \citenamefont {Taniguchi},
  \citenamefont {Yan},\ and\ \citenamefont {Gao}}]{zhang2022controlled}%
  \BibitemOpen
  \bibfield  {author} {\bibinfo {author} {\bibfnamefont {Z.}~\bibnamefont
  {Zhang}}, \bibinfo {author} {\bibfnamefont {N.}~\bibnamefont {Wang}},
  \bibinfo {author} {\bibfnamefont {N.}~\bibnamefont {Cao}}, \bibinfo {author}
  {\bibfnamefont {A.}~\bibnamefont {Wang}}, \bibinfo {author} {\bibfnamefont
  {X.}~\bibnamefont {Zhou}}, \bibinfo {author} {\bibfnamefont {K.}~\bibnamefont
  {Watanabe}}, \bibinfo {author} {\bibfnamefont {T.}~\bibnamefont {Taniguchi}},
  \bibinfo {author} {\bibfnamefont {B.}~\bibnamefont {Yan}},\ and\ \bibinfo
  {author} {\bibfnamefont {W.-b.}\ \bibnamefont {Gao}},\ }\bibfield  {title}
  {\bibinfo {title} {Controlled large non-reciprocal charge transport in an
  intrinsic magnetic topological insulator mnbi2te4},\ }\href
  {https://doi.org/10.1038/s41467-022-33705-y} {\bibfield  {journal} {\bibinfo
  {journal} {Nature communications}\ }\textbf {\bibinfo {volume} {13}},\
  \bibinfo {pages} {6191} (\bibinfo {year} {2022})}\BibitemShut {NoStop}%
\bibitem [{\citenamefont {Yasuda}\ \emph {et~al.}(2020)\citenamefont {Yasuda},
  \citenamefont {Morimoto}, \citenamefont {Yoshimi}, \citenamefont {Mogi},
  \citenamefont {Tsukazaki}, \citenamefont {Kawamura}, \citenamefont
  {Takahashi}, \citenamefont {Kawasaki}, \citenamefont {Nagaosa},\ and\
  \citenamefont {Tokura}}]{yasuda2020large}%
  \BibitemOpen
  \bibfield  {author} {\bibinfo {author} {\bibfnamefont {K.}~\bibnamefont
  {Yasuda}}, \bibinfo {author} {\bibfnamefont {T.}~\bibnamefont {Morimoto}},
  \bibinfo {author} {\bibfnamefont {R.}~\bibnamefont {Yoshimi}}, \bibinfo
  {author} {\bibfnamefont {M.}~\bibnamefont {Mogi}}, \bibinfo {author}
  {\bibfnamefont {A.}~\bibnamefont {Tsukazaki}}, \bibinfo {author}
  {\bibfnamefont {M.}~\bibnamefont {Kawamura}}, \bibinfo {author}
  {\bibfnamefont {K.~S.}\ \bibnamefont {Takahashi}}, \bibinfo {author}
  {\bibfnamefont {M.}~\bibnamefont {Kawasaki}}, \bibinfo {author}
  {\bibfnamefont {N.}~\bibnamefont {Nagaosa}},\ and\ \bibinfo {author}
  {\bibfnamefont {Y.}~\bibnamefont {Tokura}},\ }\bibfield  {title} {\bibinfo
  {title} {Large non-reciprocal charge transport mediated by quantum anomalous
  hall edge states},\ }\href {https://doi.org/10.1038/s41565-020-0733-2}
  {\bibfield  {journal} {\bibinfo  {journal} {Nature Nanotechnology}\ }\textbf
  {\bibinfo {volume} {15}},\ \bibinfo {pages} {831} (\bibinfo {year}
  {2020})}\BibitemShut {NoStop}%
\bibitem [{\citenamefont {Dean}\ \emph {et~al.}(2013)\citenamefont {Dean},
  \citenamefont {Wang}, \citenamefont {Maher}, \citenamefont {Forsythe},
  \citenamefont {Ghahari}, \citenamefont {Gao}, \citenamefont {Katoch},
  \citenamefont {Ishigami}, \citenamefont {Moon}, \citenamefont {Koshino} \emph
  {et~al.}}]{dean2013hofstadter}%
  \BibitemOpen
  \bibfield  {author} {\bibinfo {author} {\bibfnamefont {C.~R.}\ \bibnamefont
  {Dean}}, \bibinfo {author} {\bibfnamefont {L.}~\bibnamefont {Wang}}, \bibinfo
  {author} {\bibfnamefont {P.}~\bibnamefont {Maher}}, \bibinfo {author}
  {\bibfnamefont {C.}~\bibnamefont {Forsythe}}, \bibinfo {author}
  {\bibfnamefont {F.}~\bibnamefont {Ghahari}}, \bibinfo {author} {\bibfnamefont
  {Y.}~\bibnamefont {Gao}}, \bibinfo {author} {\bibfnamefont {J.}~\bibnamefont
  {Katoch}}, \bibinfo {author} {\bibfnamefont {M.}~\bibnamefont {Ishigami}},
  \bibinfo {author} {\bibfnamefont {P.}~\bibnamefont {Moon}}, \bibinfo {author}
  {\bibfnamefont {M.}~\bibnamefont {Koshino}}, \emph {et~al.},\ }\bibfield
  {title} {\bibinfo {title} {Hofstadter’s butterfly and the fractal quantum
  hall effect in moir{\'e} superlattices},\ }\href
  {https://doi.org/10.1038/nature12186} {\bibfield  {journal} {\bibinfo
  {journal} {Nature}\ }\textbf {\bibinfo {volume} {497}},\ \bibinfo {pages}
  {598} (\bibinfo {year} {2013})}\BibitemShut {NoStop}%
\bibitem [{\citenamefont {Novoselov}\ \emph {et~al.}(2016)\citenamefont
  {Novoselov}, \citenamefont {Mishchenko}, \citenamefont {Carvalho},\ and\
  \citenamefont {Castro~Neto}}]{novoselov20162d}%
  \BibitemOpen
  \bibfield  {author} {\bibinfo {author} {\bibfnamefont {K.~S.}\ \bibnamefont
  {Novoselov}}, \bibinfo {author} {\bibfnamefont {A.}~\bibnamefont
  {Mishchenko}}, \bibinfo {author} {\bibfnamefont {A.}~\bibnamefont
  {Carvalho}},\ and\ \bibinfo {author} {\bibfnamefont {A.}~\bibnamefont
  {Castro~Neto}},\ }\bibfield  {title} {\bibinfo {title} {2d materials and van
  der waals heterostructures},\ }\href
  {https://doi.org/10.1126/science.aac9439} {\bibfield  {journal} {\bibinfo
  {journal} {Science}\ }\textbf {\bibinfo {volume} {353}},\ \bibinfo {pages}
  {aac9439} (\bibinfo {year} {2016})}\BibitemShut {NoStop}%
\bibitem [{\citenamefont {Tong}\ \emph {et~al.}(2017)\citenamefont {Tong},
  \citenamefont {Yu}, \citenamefont {Zhu}, \citenamefont {Wang}, \citenamefont
  {Xu},\ and\ \citenamefont {Yao}}]{tong2017topological}%
  \BibitemOpen
  \bibfield  {author} {\bibinfo {author} {\bibfnamefont {Q.}~\bibnamefont
  {Tong}}, \bibinfo {author} {\bibfnamefont {H.}~\bibnamefont {Yu}}, \bibinfo
  {author} {\bibfnamefont {Q.}~\bibnamefont {Zhu}}, \bibinfo {author}
  {\bibfnamefont {Y.}~\bibnamefont {Wang}}, \bibinfo {author} {\bibfnamefont
  {X.}~\bibnamefont {Xu}},\ and\ \bibinfo {author} {\bibfnamefont
  {W.}~\bibnamefont {Yao}},\ }\bibfield  {title} {\bibinfo {title} {Topological
  mosaics in moir{\'e} superlattices of van der waals heterobilayers},\ }\href
  {https://doi.org/10.1038/nphys3968} {\bibfield  {journal} {\bibinfo
  {journal} {Nature Physics}\ }\textbf {\bibinfo {volume} {13}},\ \bibinfo
  {pages} {356} (\bibinfo {year} {2017})}\BibitemShut {NoStop}%
\bibitem [{\citenamefont {Finney}\ \emph {et~al.}(2019)\citenamefont {Finney},
  \citenamefont {Yankowitz}, \citenamefont {Muraleetharan}, \citenamefont
  {Watanabe}, \citenamefont {Taniguchi}, \citenamefont {Dean},\ and\
  \citenamefont {Hone}}]{finney2019tunable}%
  \BibitemOpen
  \bibfield  {author} {\bibinfo {author} {\bibfnamefont {N.~R.}\ \bibnamefont
  {Finney}}, \bibinfo {author} {\bibfnamefont {M.}~\bibnamefont {Yankowitz}},
  \bibinfo {author} {\bibfnamefont {L.}~\bibnamefont {Muraleetharan}}, \bibinfo
  {author} {\bibfnamefont {K.}~\bibnamefont {Watanabe}}, \bibinfo {author}
  {\bibfnamefont {T.}~\bibnamefont {Taniguchi}}, \bibinfo {author}
  {\bibfnamefont {C.~R.}\ \bibnamefont {Dean}},\ and\ \bibinfo {author}
  {\bibfnamefont {J.}~\bibnamefont {Hone}},\ }\bibfield  {title} {\bibinfo
  {title} {Tunable crystal symmetry in graphene--boron nitride heterostructures
  with coexisting moir{\'e} superlattices},\ }\href
  {https://doi.org/10.1038/s41565-019-0547-2} {\bibfield  {journal} {\bibinfo
  {journal} {Nature nanotechnology}\ }\textbf {\bibinfo {volume} {14}},\
  \bibinfo {pages} {1029} (\bibinfo {year} {2019})}\BibitemShut {NoStop}%
\bibitem [{\citenamefont {Meng}\ \emph {et~al.}(2024)\citenamefont {Meng},
  \citenamefont {Li}, \citenamefont {Gao}, \citenamefont {Bi}, \citenamefont
  {Chen}, \citenamefont {Qin}, \citenamefont {Qiu}, \citenamefont {Xu},
  \citenamefont {Huang}, \citenamefont {Wu}, \citenamefont {Luo},\ and\
  \citenamefont {Yuan}}]{Berry-anomalous-Hall-infomat2024}%
  \BibitemOpen
  \bibfield  {author} {\bibinfo {author} {\bibfnamefont {K.}~\bibnamefont
  {Meng}}, \bibinfo {author} {\bibfnamefont {Z.}~\bibnamefont {Li}}, \bibinfo
  {author} {\bibfnamefont {Z.}~\bibnamefont {Gao}}, \bibinfo {author}
  {\bibfnamefont {X.}~\bibnamefont {Bi}}, \bibinfo {author} {\bibfnamefont
  {P.}~\bibnamefont {Chen}}, \bibinfo {author} {\bibfnamefont {F.}~\bibnamefont
  {Qin}}, \bibinfo {author} {\bibfnamefont {C.}~\bibnamefont {Qiu}}, \bibinfo
  {author} {\bibfnamefont {L.}~\bibnamefont {Xu}}, \bibinfo {author}
  {\bibfnamefont {J.}~\bibnamefont {Huang}}, \bibinfo {author} {\bibfnamefont
  {J.}~\bibnamefont {Wu}}, \bibinfo {author} {\bibfnamefont {F.}~\bibnamefont
  {Luo}},\ and\ \bibinfo {author} {\bibfnamefont {H.}~\bibnamefont {Yuan}},\
  }\bibfield  {title} {\bibinfo {title} {Gate-tunable berry curvature in van
  der waals itinerant ferromagnetic crte},\ }\href
  {https://doi.org/https://doi.org/10.1002/inf2.12524} {\bibfield  {journal}
  {\bibinfo  {journal} {InfoMat}\ }\textbf {\bibinfo {volume} {6}},\ \bibinfo
  {pages} {e12524} (\bibinfo {year} {2024})},\ \Eprint
  {https://arxiv.org/abs/https://onlinelibrary.wiley.com/doi/pdf/10.1002/inf2.12524}
  {https://onlinelibrary.wiley.com/doi/pdf/10.1002/inf2.12524} \BibitemShut
  {NoStop}%
\bibitem [{\citenamefont {Tian}\ \emph {et~al.}(2009)\citenamefont {Tian},
  \citenamefont {Ye},\ and\ \citenamefont {Jin}}]{tian2009proper}%
  \BibitemOpen
  \bibfield  {author} {\bibinfo {author} {\bibfnamefont {Y.}~\bibnamefont
  {Tian}}, \bibinfo {author} {\bibfnamefont {L.}~\bibnamefont {Ye}},\ and\
  \bibinfo {author} {\bibfnamefont {X.}~\bibnamefont {Jin}},\ }\bibfield
  {title} {\bibinfo {title} {Proper scaling of the anomalous hall effect},\
  }\href {https://doi.org/10.1103/PhysRevLett.103.087206} {\bibfield  {journal}
  {\bibinfo  {journal} {Phys. Rev. Lett.}\ }\textbf {\bibinfo {volume} {103}},\
  \bibinfo {pages} {087206} (\bibinfo {year} {2009})}\BibitemShut {NoStop}%
\bibitem [{\citenamefont {Liu}\ \emph {et~al.}(2018)\citenamefont {Liu},
  \citenamefont {Sun}, \citenamefont {Kumar}, \citenamefont {Muechler},
  \citenamefont {Sun}, \citenamefont {Jiao}, \citenamefont {Yang},
  \citenamefont {Liu}, \citenamefont {Liang}, \citenamefont {Xu} \emph
  {et~al.}}]{liu2018giant}%
  \BibitemOpen
  \bibfield  {author} {\bibinfo {author} {\bibfnamefont {E.}~\bibnamefont
  {Liu}}, \bibinfo {author} {\bibfnamefont {Y.}~\bibnamefont {Sun}}, \bibinfo
  {author} {\bibfnamefont {N.}~\bibnamefont {Kumar}}, \bibinfo {author}
  {\bibfnamefont {L.}~\bibnamefont {Muechler}}, \bibinfo {author}
  {\bibfnamefont {A.}~\bibnamefont {Sun}}, \bibinfo {author} {\bibfnamefont
  {L.}~\bibnamefont {Jiao}}, \bibinfo {author} {\bibfnamefont {S.-Y.}\
  \bibnamefont {Yang}}, \bibinfo {author} {\bibfnamefont {D.}~\bibnamefont
  {Liu}}, \bibinfo {author} {\bibfnamefont {A.}~\bibnamefont {Liang}}, \bibinfo
  {author} {\bibfnamefont {Q.}~\bibnamefont {Xu}}, \emph {et~al.},\ }\bibfield
  {title} {\bibinfo {title} {Giant anomalous hall effect in a ferromagnetic
  kagome-lattice semimetal},\ }\href
  {https://doi.org/10.1038/s41567-018-0234-5} {\bibfield  {journal} {\bibinfo
  {journal} {Nature physics}\ }\textbf {\bibinfo {volume} {14}},\ \bibinfo
  {pages} {1125} (\bibinfo {year} {2018})}\BibitemShut {NoStop}%
\bibitem [{\citenamefont {Deng}\ \emph {et~al.}(2020)\citenamefont {Deng},
  \citenamefont {Yu}, \citenamefont {Shi}, \citenamefont {Guo}, \citenamefont
  {Xu}, \citenamefont {Wang}, \citenamefont {Chen},\ and\ \citenamefont
  {Zhang}}]{deng2020quantum}%
  \BibitemOpen
  \bibfield  {author} {\bibinfo {author} {\bibfnamefont {Y.}~\bibnamefont
  {Deng}}, \bibinfo {author} {\bibfnamefont {Y.}~\bibnamefont {Yu}}, \bibinfo
  {author} {\bibfnamefont {M.~Z.}\ \bibnamefont {Shi}}, \bibinfo {author}
  {\bibfnamefont {Z.}~\bibnamefont {Guo}}, \bibinfo {author} {\bibfnamefont
  {Z.}~\bibnamefont {Xu}}, \bibinfo {author} {\bibfnamefont {J.}~\bibnamefont
  {Wang}}, \bibinfo {author} {\bibfnamefont {X.~H.}\ \bibnamefont {Chen}},\
  and\ \bibinfo {author} {\bibfnamefont {Y.}~\bibnamefont {Zhang}},\ }\bibfield
   {title} {\bibinfo {title} {Quantum anomalous hall effect in intrinsic
  magnetic topological insulator mnbi2te4},\ }\href
  {https://doi.org/10.1126/science.aax8156} {\bibfield  {journal} {\bibinfo
  {journal} {Science}\ }\textbf {\bibinfo {volume} {367}},\ \bibinfo {pages}
  {895} (\bibinfo {year} {2020})}\BibitemShut {NoStop}%
\bibitem [{\citenamefont {Gao}\ \emph {et~al.}(2024)\citenamefont {Gao},
  \citenamefont {Chen}, \citenamefont {Ghosh}, \citenamefont {Qiu},
  \citenamefont {Liu}, \citenamefont {Onishi}, \citenamefont {Hu},
  \citenamefont {Qian}, \citenamefont {B{\'e}rub{\'e}}, \citenamefont {Dinh}
  \emph {et~al.}}]{gao2024antiferromagnetic}%
  \BibitemOpen
  \bibfield  {author} {\bibinfo {author} {\bibfnamefont {A.}~\bibnamefont
  {Gao}}, \bibinfo {author} {\bibfnamefont {S.-W.}\ \bibnamefont {Chen}},
  \bibinfo {author} {\bibfnamefont {B.}~\bibnamefont {Ghosh}}, \bibinfo
  {author} {\bibfnamefont {J.-X.}\ \bibnamefont {Qiu}}, \bibinfo {author}
  {\bibfnamefont {Y.-F.}\ \bibnamefont {Liu}}, \bibinfo {author} {\bibfnamefont
  {Y.}~\bibnamefont {Onishi}}, \bibinfo {author} {\bibfnamefont
  {C.}~\bibnamefont {Hu}}, \bibinfo {author} {\bibfnamefont {T.}~\bibnamefont
  {Qian}}, \bibinfo {author} {\bibfnamefont {D.}~\bibnamefont
  {B{\'e}rub{\'e}}}, \bibinfo {author} {\bibfnamefont {T.}~\bibnamefont
  {Dinh}}, \emph {et~al.},\ }\bibfield  {title} {\bibinfo {title} {An
  antiferromagnetic diode effect in even-layered mnbi2te4},\ }\href
  {https://doi.org/10.1038/s41928-024-01219-8} {\bibfield  {journal} {\bibinfo
  {journal} {Nature Electronics}\ ,\ \bibinfo {pages} {1}} (\bibinfo {year}
  {2024})}\BibitemShut {NoStop}%
\bibitem [{\citenamefont {He}\ \emph {et~al.}(2022)\citenamefont {He},
  \citenamefont {Koon}, \citenamefont {Isobe}, \citenamefont {Tan},
  \citenamefont {Hu}, \citenamefont {Neto}, \citenamefont {Fu},\ and\
  \citenamefont {Yang}}]{he2022graphene}%
  \BibitemOpen
  \bibfield  {author} {\bibinfo {author} {\bibfnamefont {P.}~\bibnamefont
  {He}}, \bibinfo {author} {\bibfnamefont {G.~K.~W.}\ \bibnamefont {Koon}},
  \bibinfo {author} {\bibfnamefont {H.}~\bibnamefont {Isobe}}, \bibinfo
  {author} {\bibfnamefont {J.~Y.}\ \bibnamefont {Tan}}, \bibinfo {author}
  {\bibfnamefont {J.}~\bibnamefont {Hu}}, \bibinfo {author} {\bibfnamefont
  {A.~H.~C.}\ \bibnamefont {Neto}}, \bibinfo {author} {\bibfnamefont
  {L.}~\bibnamefont {Fu}},\ and\ \bibinfo {author} {\bibfnamefont
  {H.}~\bibnamefont {Yang}},\ }\bibfield  {title} {\bibinfo {title} {Graphene
  moir{\'e} superlattices with giant quantum nonlinearity of chiral bloch
  electrons},\ }\href {https://doi.org/10.1038/s41565-021-01060-6} {\bibfield
  {journal} {\bibinfo  {journal} {Nature nanotechnology}\ }\textbf {\bibinfo
  {volume} {17}},\ \bibinfo {pages} {378} (\bibinfo {year} {2022})}\BibitemShut
  {NoStop}%
\bibitem [{\citenamefont {Zhang}\ \emph
  {et~al.}(2024{\natexlab{b}})\citenamefont {Zhang}, \citenamefont {Ju},
  \citenamefont {Kim}, \citenamefont {Cui}, \citenamefont {Keum}, \citenamefont
  {Park},\ and\ \citenamefont {Lee}}]{zhang2024broken}%
  \BibitemOpen
  \bibfield  {author} {\bibinfo {author} {\bibfnamefont {K.-X.}\ \bibnamefont
  {Zhang}}, \bibinfo {author} {\bibfnamefont {H.}~\bibnamefont {Ju}}, \bibinfo
  {author} {\bibfnamefont {H.}~\bibnamefont {Kim}}, \bibinfo {author}
  {\bibfnamefont {J.}~\bibnamefont {Cui}}, \bibinfo {author} {\bibfnamefont
  {J.}~\bibnamefont {Keum}}, \bibinfo {author} {\bibfnamefont {J.-G.}\
  \bibnamefont {Park}},\ and\ \bibinfo {author} {\bibfnamefont {J.~S.}\
  \bibnamefont {Lee}},\ }\bibfield  {title} {\bibinfo {title} {Broken inversion
  symmetry in van der waals topological ferromagnetic metal iron germanium
  telluride},\ }\href {https://doi.org/10.1002/adma.202312824} {\bibfield
  {journal} {\bibinfo  {journal} {Advanced Materials}\ }\textbf {\bibinfo
  {volume} {36}},\ \bibinfo {pages} {2312824} (\bibinfo {year}
  {2024}{\natexlab{b}})}\BibitemShut {NoStop}%
\bibitem [{\citenamefont {Wang}\ \emph
  {et~al.}(2024{\natexlab{a}})\citenamefont {Wang}, \citenamefont {Li},
  \citenamefont {Zhang}, \citenamefont {Chen}, \citenamefont {Xie},
  \citenamefont {Li}, \citenamefont {Fei}, \citenamefont {Zhang},\ and\
  \citenamefont {Song}}]{wang2024nonlinear}%
  \BibitemOpen
  \bibfield  {author} {\bibinfo {author} {\bibfnamefont {S.}~\bibnamefont
  {Wang}}, \bibinfo {author} {\bibfnamefont {X.}~\bibnamefont {Li}}, \bibinfo
  {author} {\bibfnamefont {H.}~\bibnamefont {Zhang}}, \bibinfo {author}
  {\bibfnamefont {B.}~\bibnamefont {Chen}}, \bibinfo {author} {\bibfnamefont
  {H.}~\bibnamefont {Xie}}, \bibinfo {author} {\bibfnamefont {C.}~\bibnamefont
  {Li}}, \bibinfo {author} {\bibfnamefont {F.}~\bibnamefont {Fei}}, \bibinfo
  {author} {\bibfnamefont {S.}~\bibnamefont {Zhang}},\ and\ \bibinfo {author}
  {\bibfnamefont {F.}~\bibnamefont {Song}},\ }\bibfield  {title} {\bibinfo
  {title} {Nonlinear hall effect and scaling law in sb-doped topological
  insulator mnbi4te7},\ }\bibfield  {journal} {\bibinfo  {journal} {Applied
  Physics Letters}\ }\textbf {\bibinfo {volume} {124}},\ \href
  {https://doi.org/10.1063/5.0202692} {10.1063/5.0202692} (\bibinfo {year}
  {2024}{\natexlab{a}})\BibitemShut {NoStop}%
\bibitem [{\citenamefont {Lesne}\ \emph {et~al.}(2023)\citenamefont {Lesne},
  \citenamefont {Saǧlam}, \citenamefont {Battilomo}, \citenamefont {Mercaldo},
  \citenamefont {van Thiel}, \citenamefont {Filippozzi}, \citenamefont {Noce},
  \citenamefont {Cuoco}, \citenamefont {Steele}, \citenamefont {Ortix} \emph
  {et~al.}}]{lesne2023designing}%
  \BibitemOpen
  \bibfield  {author} {\bibinfo {author} {\bibfnamefont {E.}~\bibnamefont
  {Lesne}}, \bibinfo {author} {\bibfnamefont {Y.~G.}\ \bibnamefont {Saǧlam}},
  \bibinfo {author} {\bibfnamefont {R.}~\bibnamefont {Battilomo}}, \bibinfo
  {author} {\bibfnamefont {M.~T.}\ \bibnamefont {Mercaldo}}, \bibinfo {author}
  {\bibfnamefont {T.~C.}\ \bibnamefont {van Thiel}}, \bibinfo {author}
  {\bibfnamefont {U.}~\bibnamefont {Filippozzi}}, \bibinfo {author}
  {\bibfnamefont {C.}~\bibnamefont {Noce}}, \bibinfo {author} {\bibfnamefont
  {M.}~\bibnamefont {Cuoco}}, \bibinfo {author} {\bibfnamefont {G.~A.}\
  \bibnamefont {Steele}}, \bibinfo {author} {\bibfnamefont {C.}~\bibnamefont
  {Ortix}}, \emph {et~al.},\ }\bibfield  {title} {\bibinfo {title} {Designing
  spin and orbital sources of berry curvature at oxide interfaces},\ }\href
  {https://doi.org/10.1038/s41563-023-01498-0} {\bibfield  {journal} {\bibinfo
  {journal} {Nature Materials}\ }\textbf {\bibinfo {volume} {22}},\ \bibinfo
  {pages} {576} (\bibinfo {year} {2023})}\BibitemShut {NoStop}%
\bibitem [{\citenamefont {Trama}\ \emph {et~al.}(2022)\citenamefont {Trama},
  \citenamefont {Cataudella}, \citenamefont {Perroni}, \citenamefont {Romeo},\
  and\ \citenamefont {Citro}}]{trama2022gate}%
  \BibitemOpen
  \bibfield  {author} {\bibinfo {author} {\bibfnamefont {M.}~\bibnamefont
  {Trama}}, \bibinfo {author} {\bibfnamefont {V.}~\bibnamefont {Cataudella}},
  \bibinfo {author} {\bibfnamefont {C.}~\bibnamefont {Perroni}}, \bibinfo
  {author} {\bibfnamefont {F.}~\bibnamefont {Romeo}},\ and\ \bibinfo {author}
  {\bibfnamefont {R.}~\bibnamefont {Citro}},\ }\bibfield  {title} {\bibinfo
  {title} {Gate tunable anomalous hall effect: Berry curvature probe at oxides
  interfaces},\ }\href {https://doi.org/10.1103/PhysRevB.106.075430} {\bibfield
   {journal} {\bibinfo  {journal} {Physical Review B}\ }\textbf {\bibinfo
  {volume} {106}},\ \bibinfo {pages} {075430} (\bibinfo {year}
  {2022})}\BibitemShut {NoStop}%
\bibitem [{\citenamefont {Groenendijk}\ \emph {et~al.}(2020)\citenamefont
  {Groenendijk}, \citenamefont {Autieri}, \citenamefont {van Thiel},
  \citenamefont {Brzezicki}, \citenamefont {Hortensius}, \citenamefont
  {Afanasiev}, \citenamefont {Gauquelin}, \citenamefont {Barone}, \citenamefont
  {Van~den Bos}, \citenamefont {van Aert} \emph
  {et~al.}}]{groenendijk2020berry}%
  \BibitemOpen
  \bibfield  {author} {\bibinfo {author} {\bibfnamefont {D.~J.}\ \bibnamefont
  {Groenendijk}}, \bibinfo {author} {\bibfnamefont {C.}~\bibnamefont
  {Autieri}}, \bibinfo {author} {\bibfnamefont {T.~C.}\ \bibnamefont {van
  Thiel}}, \bibinfo {author} {\bibfnamefont {W.}~\bibnamefont {Brzezicki}},
  \bibinfo {author} {\bibfnamefont {J.}~\bibnamefont {Hortensius}}, \bibinfo
  {author} {\bibfnamefont {D.}~\bibnamefont {Afanasiev}}, \bibinfo {author}
  {\bibfnamefont {N.}~\bibnamefont {Gauquelin}}, \bibinfo {author}
  {\bibfnamefont {P.}~\bibnamefont {Barone}}, \bibinfo {author} {\bibfnamefont
  {K.}~\bibnamefont {Van~den Bos}}, \bibinfo {author} {\bibfnamefont
  {S.}~\bibnamefont {van Aert}}, \emph {et~al.},\ }\bibfield  {title} {\bibinfo
  {title} {Berry phase engineering at oxide interfaces},\ }\href
  {https://doi.org/10.1103/PhysRevResearch.2.023404} {\bibfield  {journal}
  {\bibinfo  {journal} {Physical Review Research}\ }\textbf {\bibinfo {volume}
  {2}},\ \bibinfo {pages} {023404} (\bibinfo {year} {2020})}\BibitemShut
  {NoStop}%
\bibitem [{\citenamefont {Yankowitz}\ \emph {et~al.}(2019)\citenamefont
  {Yankowitz}, \citenamefont {Ma}, \citenamefont {Jarillo-Herrero},\ and\
  \citenamefont {LeRoy}}]{yankowitz2019van}%
  \BibitemOpen
  \bibfield  {author} {\bibinfo {author} {\bibfnamefont {M.}~\bibnamefont
  {Yankowitz}}, \bibinfo {author} {\bibfnamefont {Q.}~\bibnamefont {Ma}},
  \bibinfo {author} {\bibfnamefont {P.}~\bibnamefont {Jarillo-Herrero}},\ and\
  \bibinfo {author} {\bibfnamefont {B.~J.}\ \bibnamefont {LeRoy}},\ }\bibfield
  {title} {\bibinfo {title} {van der waals heterostructures combining graphene
  and hexagonal boron nitride},\ }\href
  {https://doi.org/10.1038/s42254-018-0016-0} {\bibfield  {journal} {\bibinfo
  {journal} {Nature Reviews Physics}\ }\textbf {\bibinfo {volume} {1}},\
  \bibinfo {pages} {112} (\bibinfo {year} {2019})}\BibitemShut {NoStop}%
\bibitem [{\citenamefont {Li}\ \emph {et~al.}(2023)\citenamefont {Li},
  \citenamefont {Huang}, \citenamefont {Zhou}, \citenamefont {Xu},
  \citenamefont {Qin}, \citenamefont {Chen}, \citenamefont {Sun}, \citenamefont
  {Liu}, \citenamefont {Sui}, \citenamefont {Qiu} \emph
  {et~al.}}]{li2023anisotropic}%
  \BibitemOpen
  \bibfield  {author} {\bibinfo {author} {\bibfnamefont {Z.}~\bibnamefont
  {Li}}, \bibinfo {author} {\bibfnamefont {J.}~\bibnamefont {Huang}}, \bibinfo
  {author} {\bibfnamefont {L.}~\bibnamefont {Zhou}}, \bibinfo {author}
  {\bibfnamefont {Z.}~\bibnamefont {Xu}}, \bibinfo {author} {\bibfnamefont
  {F.}~\bibnamefont {Qin}}, \bibinfo {author} {\bibfnamefont {P.}~\bibnamefont
  {Chen}}, \bibinfo {author} {\bibfnamefont {X.}~\bibnamefont {Sun}}, \bibinfo
  {author} {\bibfnamefont {G.}~\bibnamefont {Liu}}, \bibinfo {author}
  {\bibfnamefont {C.}~\bibnamefont {Sui}}, \bibinfo {author} {\bibfnamefont
  {C.}~\bibnamefont {Qiu}}, \emph {et~al.},\ }\bibfield  {title} {\bibinfo
  {title} {An anisotropic van der waals dielectric for symmetry engineering in
  functionalized heterointerfaces},\ }\href
  {https://doi.org/10.1038/s41467-023-41295-6} {\bibfield  {journal} {\bibinfo
  {journal} {Nature Communications}\ }\textbf {\bibinfo {volume} {14}},\
  \bibinfo {pages} {5568} (\bibinfo {year} {2023})}\BibitemShut {NoStop}%
\bibitem [{\citenamefont {Wang}\ \emph {et~al.}(2013)\citenamefont {Wang},
  \citenamefont {Meric}, \citenamefont {Huang}, \citenamefont {Gao},
  \citenamefont {Gao}, \citenamefont {Tran}, \citenamefont {Taniguchi},
  \citenamefont {Watanabe}, \citenamefont {Campos}, \citenamefont {Muller}
  \emph {et~al.}}]{wang2013one}%
  \BibitemOpen
  \bibfield  {author} {\bibinfo {author} {\bibfnamefont {L.}~\bibnamefont
  {Wang}}, \bibinfo {author} {\bibfnamefont {I.}~\bibnamefont {Meric}},
  \bibinfo {author} {\bibfnamefont {P.}~\bibnamefont {Huang}}, \bibinfo
  {author} {\bibfnamefont {Q.}~\bibnamefont {Gao}}, \bibinfo {author}
  {\bibfnamefont {Y.}~\bibnamefont {Gao}}, \bibinfo {author} {\bibfnamefont
  {H.}~\bibnamefont {Tran}}, \bibinfo {author} {\bibfnamefont {T.}~\bibnamefont
  {Taniguchi}}, \bibinfo {author} {\bibfnamefont {K.}~\bibnamefont {Watanabe}},
  \bibinfo {author} {\bibfnamefont {L.}~\bibnamefont {Campos}}, \bibinfo
  {author} {\bibfnamefont {D.}~\bibnamefont {Muller}}, \emph {et~al.},\
  }\bibfield  {title} {\bibinfo {title} {One-dimensional electrical contact to
  a two-dimensional material},\ }\href
  {https://doi.org/10.1126/science.1244358} {\bibfield  {journal} {\bibinfo
  {journal} {Science}\ }\textbf {\bibinfo {volume} {342}},\ \bibinfo {pages}
  {614} (\bibinfo {year} {2013})}\BibitemShut {NoStop}%
\bibitem [{\citenamefont {Kinoshita}\ \emph {et~al.}(2019)\citenamefont
  {Kinoshita}, \citenamefont {Moriya}, \citenamefont {Onodera}, \citenamefont
  {Wakafuji}, \citenamefont {Masubuchi}, \citenamefont {Watanabe},
  \citenamefont {Taniguchi},\ and\ \citenamefont {Machida}}]{kinoshita2019dry}%
  \BibitemOpen
  \bibfield  {author} {\bibinfo {author} {\bibfnamefont {K.}~\bibnamefont
  {Kinoshita}}, \bibinfo {author} {\bibfnamefont {R.}~\bibnamefont {Moriya}},
  \bibinfo {author} {\bibfnamefont {M.}~\bibnamefont {Onodera}}, \bibinfo
  {author} {\bibfnamefont {Y.}~\bibnamefont {Wakafuji}}, \bibinfo {author}
  {\bibfnamefont {S.}~\bibnamefont {Masubuchi}}, \bibinfo {author}
  {\bibfnamefont {K.}~\bibnamefont {Watanabe}}, \bibinfo {author}
  {\bibfnamefont {T.}~\bibnamefont {Taniguchi}},\ and\ \bibinfo {author}
  {\bibfnamefont {T.}~\bibnamefont {Machida}},\ }\bibfield  {title} {\bibinfo
  {title} {Dry release transfer of graphene and few-layer h-bn by utilizing
  thermoplasticity of polypropylene carbonate},\ }\href
  {https://doi.org/10.1038/s41699-019-0104-8} {\bibfield  {journal} {\bibinfo
  {journal} {npj 2D Materials and Applications}\ }\textbf {\bibinfo {volume}
  {3}},\ \bibinfo {pages} {22} (\bibinfo {year} {2019})}\BibitemShut {NoStop}%
\bibitem [{\citenamefont {Huang}\ \emph
  {et~al.}(2023{\natexlab{a}})\citenamefont {Huang}, \citenamefont {Wu},
  \citenamefont {Zhang}, \citenamefont {Feng}, \citenamefont {Zhou},
  \citenamefont {Wang}, \citenamefont {Chen}, \citenamefont {Cheng},
  \citenamefont {Sun}, \citenamefont {Meng} \emph
  {et~al.}}]{huang2023intrinsic}%
  \BibitemOpen
  \bibfield  {author} {\bibinfo {author} {\bibfnamefont {M.}~\bibnamefont
  {Huang}}, \bibinfo {author} {\bibfnamefont {Z.}~\bibnamefont {Wu}}, \bibinfo
  {author} {\bibfnamefont {X.}~\bibnamefont {Zhang}}, \bibinfo {author}
  {\bibfnamefont {X.}~\bibnamefont {Feng}}, \bibinfo {author} {\bibfnamefont
  {Z.}~\bibnamefont {Zhou}}, \bibinfo {author} {\bibfnamefont {S.}~\bibnamefont
  {Wang}}, \bibinfo {author} {\bibfnamefont {Y.}~\bibnamefont {Chen}}, \bibinfo
  {author} {\bibfnamefont {C.}~\bibnamefont {Cheng}}, \bibinfo {author}
  {\bibfnamefont {K.}~\bibnamefont {Sun}}, \bibinfo {author} {\bibfnamefont
  {Z.~Y.}\ \bibnamefont {Meng}}, \emph {et~al.},\ }\bibfield  {title} {\bibinfo
  {title} {Intrinsic nonlinear hall effect and gate-switchable berry curvature
  sliding in twisted bilayer graphene},\ }\href
  {https://doi.org/10.1103/PhysRevLett.131.066301} {\bibfield  {journal}
  {\bibinfo  {journal} {Physical Review Letters}\ }\textbf {\bibinfo {volume}
  {131}},\ \bibinfo {pages} {066301} (\bibinfo {year}
  {2023}{\natexlab{a}})}\BibitemShut {NoStop}%
\bibitem [{\citenamefont {Kim}\ \emph {et~al.}(2016)\citenamefont {Kim},
  \citenamefont {Yankowitz}, \citenamefont {Fallahazad}, \citenamefont {Kang},
  \citenamefont {Movva}, \citenamefont {Huang}, \citenamefont {Larentis},
  \citenamefont {Corbet}, \citenamefont {Taniguchi}, \citenamefont {Watanabe}
  \emph {et~al.}}]{kim2016van}%
  \BibitemOpen
  \bibfield  {author} {\bibinfo {author} {\bibfnamefont {K.}~\bibnamefont
  {Kim}}, \bibinfo {author} {\bibfnamefont {M.}~\bibnamefont {Yankowitz}},
  \bibinfo {author} {\bibfnamefont {B.}~\bibnamefont {Fallahazad}}, \bibinfo
  {author} {\bibfnamefont {S.}~\bibnamefont {Kang}}, \bibinfo {author}
  {\bibfnamefont {H.~C.}\ \bibnamefont {Movva}}, \bibinfo {author}
  {\bibfnamefont {S.}~\bibnamefont {Huang}}, \bibinfo {author} {\bibfnamefont
  {S.}~\bibnamefont {Larentis}}, \bibinfo {author} {\bibfnamefont {C.~M.}\
  \bibnamefont {Corbet}}, \bibinfo {author} {\bibfnamefont {T.}~\bibnamefont
  {Taniguchi}}, \bibinfo {author} {\bibfnamefont {K.}~\bibnamefont {Watanabe}},
  \emph {et~al.},\ }\bibfield  {title} {\bibinfo {title} {van der waals
  heterostructures with high accuracy rotational alignment},\ }\href
  {https://doi.org/10.1021/acs.nanolett.5b05263} {\bibfield  {journal}
  {\bibinfo  {journal} {Nano letters}\ }\textbf {\bibinfo {volume} {16}},\
  \bibinfo {pages} {1989} (\bibinfo {year} {2016})}\BibitemShut {NoStop}%
\bibitem [{\citenamefont {Saito}\ \emph {et~al.}(2020)\citenamefont {Saito},
  \citenamefont {Ge}, \citenamefont {Watanabe}, \citenamefont {Taniguchi},\
  and\ \citenamefont {Young}}]{saito2020independent}%
  \BibitemOpen
  \bibfield  {author} {\bibinfo {author} {\bibfnamefont {Y.}~\bibnamefont
  {Saito}}, \bibinfo {author} {\bibfnamefont {J.}~\bibnamefont {Ge}}, \bibinfo
  {author} {\bibfnamefont {K.}~\bibnamefont {Watanabe}}, \bibinfo {author}
  {\bibfnamefont {T.}~\bibnamefont {Taniguchi}},\ and\ \bibinfo {author}
  {\bibfnamefont {A.~F.}\ \bibnamefont {Young}},\ }\bibfield  {title} {\bibinfo
  {title} {Independent superconductors and correlated insulators in twisted
  bilayer graphene},\ }\href {https://doi.org/10.1038/s41567-020-0928-3}
  {\bibfield  {journal} {\bibinfo  {journal} {Nature Physics}\ }\textbf
  {\bibinfo {volume} {16}},\ \bibinfo {pages} {926} (\bibinfo {year}
  {2020})}\BibitemShut {NoStop}%
\bibitem [{\citenamefont {Tian}\ \emph {et~al.}(2023)\citenamefont {Tian},
  \citenamefont {Gao}, \citenamefont {Zhang}, \citenamefont {Che},
  \citenamefont {Xu}, \citenamefont {Cheung}, \citenamefont {Watanabe},
  \citenamefont {Taniguchi}, \citenamefont {Randeria}, \citenamefont {Zhang}
  \emph {et~al.}}]{tian2023evidence}%
  \BibitemOpen
  \bibfield  {author} {\bibinfo {author} {\bibfnamefont {H.}~\bibnamefont
  {Tian}}, \bibinfo {author} {\bibfnamefont {X.}~\bibnamefont {Gao}}, \bibinfo
  {author} {\bibfnamefont {Y.}~\bibnamefont {Zhang}}, \bibinfo {author}
  {\bibfnamefont {S.}~\bibnamefont {Che}}, \bibinfo {author} {\bibfnamefont
  {T.}~\bibnamefont {Xu}}, \bibinfo {author} {\bibfnamefont {P.}~\bibnamefont
  {Cheung}}, \bibinfo {author} {\bibfnamefont {K.}~\bibnamefont {Watanabe}},
  \bibinfo {author} {\bibfnamefont {T.}~\bibnamefont {Taniguchi}}, \bibinfo
  {author} {\bibfnamefont {M.}~\bibnamefont {Randeria}}, \bibinfo {author}
  {\bibfnamefont {F.}~\bibnamefont {Zhang}}, \emph {et~al.},\ }\bibfield
  {title} {\bibinfo {title} {Evidence for dirac flat band superconductivity
  enabled by quantum geometry},\ }\href
  {https://doi.org/10.1038/s41586-022-05576-2} {\bibfield  {journal} {\bibinfo
  {journal} {Nature}\ }\textbf {\bibinfo {volume} {614}},\ \bibinfo {pages}
  {440} (\bibinfo {year} {2023})}\BibitemShut {NoStop}%
\bibitem [{\citenamefont {Park}\ \emph {et~al.}(2021)\citenamefont {Park},
  \citenamefont {Cao}, \citenamefont {Watanabe}, \citenamefont {Taniguchi},\
  and\ \citenamefont {Jarillo-Herrero}}]{park2021tunable}%
  \BibitemOpen
  \bibfield  {author} {\bibinfo {author} {\bibfnamefont {J.~M.}\ \bibnamefont
  {Park}}, \bibinfo {author} {\bibfnamefont {Y.}~\bibnamefont {Cao}}, \bibinfo
  {author} {\bibfnamefont {K.}~\bibnamefont {Watanabe}}, \bibinfo {author}
  {\bibfnamefont {T.}~\bibnamefont {Taniguchi}},\ and\ \bibinfo {author}
  {\bibfnamefont {P.}~\bibnamefont {Jarillo-Herrero}},\ }\bibfield  {title}
  {\bibinfo {title} {Tunable strongly coupled superconductivity in magic-angle
  twisted trilayer graphene},\ }\href
  {https://doi.org/10.1038/s41586-021-03192-0} {\bibfield  {journal} {\bibinfo
  {journal} {Nature}\ }\textbf {\bibinfo {volume} {590}},\ \bibinfo {pages}
  {249} (\bibinfo {year} {2021})}\BibitemShut {NoStop}%
\bibitem [{\citenamefont {McGilly}\ \emph {et~al.}(2020)\citenamefont
  {McGilly}, \citenamefont {Kerelsky}, \citenamefont {Finney}, \citenamefont
  {Shapovalov}, \citenamefont {Shih}, \citenamefont {Ghiotto}, \citenamefont
  {Zeng}, \citenamefont {Moore}, \citenamefont {Wu}, \citenamefont {Bai} \emph
  {et~al.}}]{mcgilly2020visualization}%
  \BibitemOpen
  \bibfield  {author} {\bibinfo {author} {\bibfnamefont {L.~J.}\ \bibnamefont
  {McGilly}}, \bibinfo {author} {\bibfnamefont {A.}~\bibnamefont {Kerelsky}},
  \bibinfo {author} {\bibfnamefont {N.~R.}\ \bibnamefont {Finney}}, \bibinfo
  {author} {\bibfnamefont {K.}~\bibnamefont {Shapovalov}}, \bibinfo {author}
  {\bibfnamefont {E.-M.}\ \bibnamefont {Shih}}, \bibinfo {author}
  {\bibfnamefont {A.}~\bibnamefont {Ghiotto}}, \bibinfo {author} {\bibfnamefont
  {Y.}~\bibnamefont {Zeng}}, \bibinfo {author} {\bibfnamefont {S.~L.}\
  \bibnamefont {Moore}}, \bibinfo {author} {\bibfnamefont {W.}~\bibnamefont
  {Wu}}, \bibinfo {author} {\bibfnamefont {Y.}~\bibnamefont {Bai}}, \emph
  {et~al.},\ }\bibfield  {title} {\bibinfo {title} {Visualization of moir{\'e}
  superlattices},\ }\href {https://doi.org/10.1038/s41565-020-0708-3}
  {\bibfield  {journal} {\bibinfo  {journal} {Nature Nanotechnology}\ }\textbf
  {\bibinfo {volume} {15}},\ \bibinfo {pages} {580} (\bibinfo {year}
  {2020})}\BibitemShut {NoStop}%
\bibitem [{\citenamefont {Qiu}\ \emph {et~al.}(2021)\citenamefont {Qiu},
  \citenamefont {Gong}, \citenamefont {Wang}, \citenamefont {Zhang},
  \citenamefont {Yang}, \citenamefont {Wang},\ and\ \citenamefont
  {Xiong}}]{qiu2021recent}%
  \BibitemOpen
  \bibfield  {author} {\bibinfo {author} {\bibfnamefont {D.}~\bibnamefont
  {Qiu}}, \bibinfo {author} {\bibfnamefont {C.}~\bibnamefont {Gong}}, \bibinfo
  {author} {\bibfnamefont {S.}~\bibnamefont {Wang}}, \bibinfo {author}
  {\bibfnamefont {M.}~\bibnamefont {Zhang}}, \bibinfo {author} {\bibfnamefont
  {C.}~\bibnamefont {Yang}}, \bibinfo {author} {\bibfnamefont {X.}~\bibnamefont
  {Wang}},\ and\ \bibinfo {author} {\bibfnamefont {J.}~\bibnamefont {Xiong}},\
  }\bibfield  {title} {\bibinfo {title} {Recent advances in 2d
  superconductors},\ }\href {https://doi.org/10.1002/adma.202006124} {\bibfield
   {journal} {\bibinfo  {journal} {Advanced Materials}\ }\textbf {\bibinfo
  {volume} {33}},\ \bibinfo {pages} {2006124} (\bibinfo {year}
  {2021})}\BibitemShut {NoStop}%
\bibitem [{\citenamefont {Balents}\ \emph {et~al.}(2020)\citenamefont
  {Balents}, \citenamefont {Dean}, \citenamefont {Efetov},\ and\ \citenamefont
  {Young}}]{balents2020superconductivity}%
  \BibitemOpen
  \bibfield  {author} {\bibinfo {author} {\bibfnamefont {L.}~\bibnamefont
  {Balents}}, \bibinfo {author} {\bibfnamefont {C.~R.}\ \bibnamefont {Dean}},
  \bibinfo {author} {\bibfnamefont {D.~K.}\ \bibnamefont {Efetov}},\ and\
  \bibinfo {author} {\bibfnamefont {A.~F.}\ \bibnamefont {Young}},\ }\bibfield
  {title} {\bibinfo {title} {Superconductivity and strong correlations in
  moir{\'e} flat bands},\ }\href {https://doi.org/10.1038/s41567-020-0906-9}
  {\bibfield  {journal} {\bibinfo  {journal} {Nature Physics}\ }\textbf
  {\bibinfo {volume} {16}},\ \bibinfo {pages} {725} (\bibinfo {year}
  {2020})}\BibitemShut {NoStop}%
\bibitem [{\citenamefont {Mak}\ and\ \citenamefont
  {Shan}(2022)}]{mak2022semiconductor}%
  \BibitemOpen
  \bibfield  {author} {\bibinfo {author} {\bibfnamefont {K.~F.}\ \bibnamefont
  {Mak}}\ and\ \bibinfo {author} {\bibfnamefont {J.}~\bibnamefont {Shan}},\
  }\bibfield  {title} {\bibinfo {title} {Semiconductor moir{\'e} materials},\
  }\href {https://doi.org/10.1038/s41565-022-01165-6} {\bibfield  {journal}
  {\bibinfo  {journal} {Nature Nanotechnology}\ }\textbf {\bibinfo {volume}
  {17}},\ \bibinfo {pages} {686} (\bibinfo {year} {2022})}\BibitemShut
  {NoStop}%
\bibitem [{\citenamefont {Xu}\ \emph {et~al.}(2020)\citenamefont {Xu},
  \citenamefont {Liu}, \citenamefont {Rhodes}, \citenamefont {Watanabe},
  \citenamefont {Taniguchi}, \citenamefont {Hone}, \citenamefont {Elser},
  \citenamefont {Mak},\ and\ \citenamefont {Shan}}]{xu2020correlated}%
  \BibitemOpen
  \bibfield  {author} {\bibinfo {author} {\bibfnamefont {Y.}~\bibnamefont
  {Xu}}, \bibinfo {author} {\bibfnamefont {S.}~\bibnamefont {Liu}}, \bibinfo
  {author} {\bibfnamefont {D.~A.}\ \bibnamefont {Rhodes}}, \bibinfo {author}
  {\bibfnamefont {K.}~\bibnamefont {Watanabe}}, \bibinfo {author}
  {\bibfnamefont {T.}~\bibnamefont {Taniguchi}}, \bibinfo {author}
  {\bibfnamefont {J.}~\bibnamefont {Hone}}, \bibinfo {author} {\bibfnamefont
  {V.}~\bibnamefont {Elser}}, \bibinfo {author} {\bibfnamefont {K.~F.}\
  \bibnamefont {Mak}},\ and\ \bibinfo {author} {\bibfnamefont {J.}~\bibnamefont
  {Shan}},\ }\bibfield  {title} {\bibinfo {title} {Correlated insulating states
  at fractional fillings of moir{\'e} superlattices},\ }\href
  {https://doi.org/10.1038/s41586-020-2868-6} {\bibfield  {journal} {\bibinfo
  {journal} {Nature}\ }\textbf {\bibinfo {volume} {587}},\ \bibinfo {pages}
  {214} (\bibinfo {year} {2020})}\BibitemShut {NoStop}%
\bibitem [{\citenamefont {Huang}\ \emph {et~al.}(2021)\citenamefont {Huang},
  \citenamefont {Wang}, \citenamefont {Miao}, \citenamefont {Wang},
  \citenamefont {Li}, \citenamefont {Lian}, \citenamefont {Taniguchi},
  \citenamefont {Watanabe}, \citenamefont {Okamoto}, \citenamefont {Xiao} \emph
  {et~al.}}]{huang2021correlated}%
  \BibitemOpen
  \bibfield  {author} {\bibinfo {author} {\bibfnamefont {X.}~\bibnamefont
  {Huang}}, \bibinfo {author} {\bibfnamefont {T.}~\bibnamefont {Wang}},
  \bibinfo {author} {\bibfnamefont {S.}~\bibnamefont {Miao}}, \bibinfo {author}
  {\bibfnamefont {C.}~\bibnamefont {Wang}}, \bibinfo {author} {\bibfnamefont
  {Z.}~\bibnamefont {Li}}, \bibinfo {author} {\bibfnamefont {Z.}~\bibnamefont
  {Lian}}, \bibinfo {author} {\bibfnamefont {T.}~\bibnamefont {Taniguchi}},
  \bibinfo {author} {\bibfnamefont {K.}~\bibnamefont {Watanabe}}, \bibinfo
  {author} {\bibfnamefont {S.}~\bibnamefont {Okamoto}}, \bibinfo {author}
  {\bibfnamefont {D.}~\bibnamefont {Xiao}}, \emph {et~al.},\ }\bibfield
  {title} {\bibinfo {title} {Correlated insulating states at fractional
  fillings of the ws2/wse2 moir{\'e} lattice},\ }\href
  {https://doi.org/10.1038/s41567-021-01171-w} {\bibfield  {journal} {\bibinfo
  {journal} {Nature Physics}\ }\textbf {\bibinfo {volume} {17}},\ \bibinfo
  {pages} {715} (\bibinfo {year} {2021})}\BibitemShut {NoStop}%
\bibitem [{\citenamefont {Andrei}\ \emph {et~al.}(2021)\citenamefont {Andrei},
  \citenamefont {Efetov}, \citenamefont {Jarillo-Herrero}, \citenamefont
  {MacDonald}, \citenamefont {Mak}, \citenamefont {Senthil}, \citenamefont
  {Tutuc}, \citenamefont {Yazdani},\ and\ \citenamefont
  {Young}}]{andrei2021marvels}%
  \BibitemOpen
  \bibfield  {author} {\bibinfo {author} {\bibfnamefont {E.~Y.}\ \bibnamefont
  {Andrei}}, \bibinfo {author} {\bibfnamefont {D.~K.}\ \bibnamefont {Efetov}},
  \bibinfo {author} {\bibfnamefont {P.}~\bibnamefont {Jarillo-Herrero}},
  \bibinfo {author} {\bibfnamefont {A.~H.}\ \bibnamefont {MacDonald}}, \bibinfo
  {author} {\bibfnamefont {K.~F.}\ \bibnamefont {Mak}}, \bibinfo {author}
  {\bibfnamefont {T.}~\bibnamefont {Senthil}}, \bibinfo {author} {\bibfnamefont
  {E.}~\bibnamefont {Tutuc}}, \bibinfo {author} {\bibfnamefont
  {A.}~\bibnamefont {Yazdani}},\ and\ \bibinfo {author} {\bibfnamefont {A.~F.}\
  \bibnamefont {Young}},\ }\bibfield  {title} {\bibinfo {title} {The marvels of
  moir{\'e} materials},\ }\href {https://doi.org/10.1038/s41578-021-00284-1}
  {\bibfield  {journal} {\bibinfo  {journal} {Nature Reviews Materials}\
  }\textbf {\bibinfo {volume} {6}},\ \bibinfo {pages} {201} (\bibinfo {year}
  {2021})}\BibitemShut {NoStop}%
\bibitem [{\citenamefont {Ma}\ \emph {et~al.}(2022)\citenamefont {Ma},
  \citenamefont {Chen}, \citenamefont {Yananose}, \citenamefont {Zhou},
  \citenamefont {Wang}, \citenamefont {Li}, \citenamefont {Zhu}, \citenamefont
  {Wu}, \citenamefont {Xu}, \citenamefont {Yu} \emph {et~al.}}]{ma2022growth}%
  \BibitemOpen
  \bibfield  {author} {\bibinfo {author} {\bibfnamefont {T.}~\bibnamefont
  {Ma}}, \bibinfo {author} {\bibfnamefont {H.}~\bibnamefont {Chen}}, \bibinfo
  {author} {\bibfnamefont {K.}~\bibnamefont {Yananose}}, \bibinfo {author}
  {\bibfnamefont {X.}~\bibnamefont {Zhou}}, \bibinfo {author} {\bibfnamefont
  {L.}~\bibnamefont {Wang}}, \bibinfo {author} {\bibfnamefont {R.}~\bibnamefont
  {Li}}, \bibinfo {author} {\bibfnamefont {Z.}~\bibnamefont {Zhu}}, \bibinfo
  {author} {\bibfnamefont {Z.}~\bibnamefont {Wu}}, \bibinfo {author}
  {\bibfnamefont {Q.-H.}\ \bibnamefont {Xu}}, \bibinfo {author} {\bibfnamefont
  {J.}~\bibnamefont {Yu}}, \emph {et~al.},\ }\bibfield  {title} {\bibinfo
  {title} {Growth of bilayer mote2 single crystals with strong non-linear hall
  effect},\ }\href {https://doi.org/10.1038/s41467-022-33201-3} {\bibfield
  {journal} {\bibinfo  {journal} {Nature communications}\ }\textbf {\bibinfo
  {volume} {13}},\ \bibinfo {pages} {5465} (\bibinfo {year}
  {2022})}\BibitemShut {NoStop}%
\bibitem [{\citenamefont {Su{\'a}rez-Rodr{\'\i}guez}\ \emph
  {et~al.}(2024{\natexlab{a}})\citenamefont {Su{\'a}rez-Rodr{\'\i}guez},
  \citenamefont {Mart{\'\i}n-Garc{\'\i}a}, \citenamefont {Skowro{\'n}ski},
  \citenamefont {Calavalle}, \citenamefont {Tsirkin}, \citenamefont {Souza},
  \citenamefont {De~Juan}, \citenamefont {Chuvilin}, \citenamefont {Fert},
  \citenamefont {Gobbi} \emph {et~al.}}]{suarez2024odd}%
  \BibitemOpen
  \bibfield  {author} {\bibinfo {author} {\bibfnamefont {M.}~\bibnamefont
  {Su{\'a}rez-Rodr{\'\i}guez}}, \bibinfo {author} {\bibfnamefont
  {B.}~\bibnamefont {Mart{\'\i}n-Garc{\'\i}a}}, \bibinfo {author}
  {\bibfnamefont {W.}~\bibnamefont {Skowro{\'n}ski}}, \bibinfo {author}
  {\bibfnamefont {F.}~\bibnamefont {Calavalle}}, \bibinfo {author}
  {\bibfnamefont {S.~S.}\ \bibnamefont {Tsirkin}}, \bibinfo {author}
  {\bibfnamefont {I.}~\bibnamefont {Souza}}, \bibinfo {author} {\bibfnamefont
  {F.}~\bibnamefont {De~Juan}}, \bibinfo {author} {\bibfnamefont
  {A.}~\bibnamefont {Chuvilin}}, \bibinfo {author} {\bibfnamefont
  {A.}~\bibnamefont {Fert}}, \bibinfo {author} {\bibfnamefont {M.}~\bibnamefont
  {Gobbi}}, \emph {et~al.},\ }\bibfield  {title} {\bibinfo {title} {Odd
  nonlinear conductivity under spatial inversion in chiral tellurium},\ }\href
  {https://doi.org/10.1103/PhysRevLett.132.046303} {\bibfield  {journal}
  {\bibinfo  {journal} {Physical Review Letters}\ }\textbf {\bibinfo {volume}
  {132}},\ \bibinfo {pages} {046303} (\bibinfo {year}
  {2024}{\natexlab{a}})}\BibitemShut {NoStop}%
\bibitem [{\citenamefont {Makushko}\ \emph {et~al.}(2024)\citenamefont
  {Makushko}, \citenamefont {Kovalev}, \citenamefont {Zabila}, \citenamefont
  {Ilyakov}, \citenamefont {Ponomaryov}, \citenamefont {Arshad}, \citenamefont
  {Prajapati}, \citenamefont {de~Oliveira}, \citenamefont {Deinert},
  \citenamefont {Chekhonin} \emph {et~al.}}]{makushko2024tunable}%
  \BibitemOpen
  \bibfield  {author} {\bibinfo {author} {\bibfnamefont {P.}~\bibnamefont
  {Makushko}}, \bibinfo {author} {\bibfnamefont {S.}~\bibnamefont {Kovalev}},
  \bibinfo {author} {\bibfnamefont {Y.}~\bibnamefont {Zabila}}, \bibinfo
  {author} {\bibfnamefont {I.}~\bibnamefont {Ilyakov}}, \bibinfo {author}
  {\bibfnamefont {A.}~\bibnamefont {Ponomaryov}}, \bibinfo {author}
  {\bibfnamefont {A.}~\bibnamefont {Arshad}}, \bibinfo {author} {\bibfnamefont
  {G.~L.}\ \bibnamefont {Prajapati}}, \bibinfo {author} {\bibfnamefont {T.~V.}\
  \bibnamefont {de~Oliveira}}, \bibinfo {author} {\bibfnamefont {J.-C.}\
  \bibnamefont {Deinert}}, \bibinfo {author} {\bibfnamefont {P.}~\bibnamefont
  {Chekhonin}}, \emph {et~al.},\ }\bibfield  {title} {\bibinfo {title} {A
  tunable room-temperature nonlinear hall effect in elemental bismuth thin
  films},\ }\href {https://doi.org/10.1038/s41928-024-01118-y} {\bibfield
  {journal} {\bibinfo  {journal} {Nature Electronics}\ }\textbf {\bibinfo
  {volume} {7}},\ \bibinfo {pages} {207} (\bibinfo {year} {2024})}\BibitemShut
  {NoStop}%
\bibitem [{\citenamefont {Min}\ \emph {et~al.}(2023)\citenamefont {Min},
  \citenamefont {Tan}, \citenamefont {Xie}, \citenamefont {Miao}, \citenamefont
  {Zhang}, \citenamefont {Lee}, \citenamefont {Gopalan}, \citenamefont {Liu},
  \citenamefont {Alem}, \citenamefont {Yan} \emph {et~al.}}]{min2023strong}%
  \BibitemOpen
  \bibfield  {author} {\bibinfo {author} {\bibfnamefont {L.}~\bibnamefont
  {Min}}, \bibinfo {author} {\bibfnamefont {H.}~\bibnamefont {Tan}}, \bibinfo
  {author} {\bibfnamefont {Z.}~\bibnamefont {Xie}}, \bibinfo {author}
  {\bibfnamefont {L.}~\bibnamefont {Miao}}, \bibinfo {author} {\bibfnamefont
  {R.}~\bibnamefont {Zhang}}, \bibinfo {author} {\bibfnamefont {S.~H.}\
  \bibnamefont {Lee}}, \bibinfo {author} {\bibfnamefont {V.}~\bibnamefont
  {Gopalan}}, \bibinfo {author} {\bibfnamefont {C.-X.}\ \bibnamefont {Liu}},
  \bibinfo {author} {\bibfnamefont {N.}~\bibnamefont {Alem}}, \bibinfo {author}
  {\bibfnamefont {B.}~\bibnamefont {Yan}}, \emph {et~al.},\ }\bibfield  {title}
  {\bibinfo {title} {Strong room-temperature bulk nonlinear hall effect in a
  spin-valley locked dirac material},\ }\href
  {https://doi.org/10.1038/s41467-023-35989-0} {\bibfield  {journal} {\bibinfo
  {journal} {Nature communications}\ }\textbf {\bibinfo {volume} {14}},\
  \bibinfo {pages} {364} (\bibinfo {year} {2023})}\BibitemShut {NoStop}%
\bibitem [{\citenamefont {Huang}\ \emph
  {et~al.}(2023{\natexlab{b}})\citenamefont {Huang}, \citenamefont {Wu},
  \citenamefont {Hu}, \citenamefont {Cai}, \citenamefont {Li}, \citenamefont
  {An}, \citenamefont {Feng}, \citenamefont {Ye}, \citenamefont {Lin},
  \citenamefont {Law} \emph {et~al.}}]{huang2023giant}%
  \BibitemOpen
  \bibfield  {author} {\bibinfo {author} {\bibfnamefont {M.}~\bibnamefont
  {Huang}}, \bibinfo {author} {\bibfnamefont {Z.}~\bibnamefont {Wu}}, \bibinfo
  {author} {\bibfnamefont {J.}~\bibnamefont {Hu}}, \bibinfo {author}
  {\bibfnamefont {X.}~\bibnamefont {Cai}}, \bibinfo {author} {\bibfnamefont
  {E.}~\bibnamefont {Li}}, \bibinfo {author} {\bibfnamefont {L.}~\bibnamefont
  {An}}, \bibinfo {author} {\bibfnamefont {X.}~\bibnamefont {Feng}}, \bibinfo
  {author} {\bibfnamefont {Z.}~\bibnamefont {Ye}}, \bibinfo {author}
  {\bibfnamefont {N.}~\bibnamefont {Lin}}, \bibinfo {author} {\bibfnamefont
  {K.~T.}\ \bibnamefont {Law}}, \emph {et~al.},\ }\bibfield  {title} {\bibinfo
  {title} {Giant nonlinear hall effect in twisted bilayer wse2},\ }\href
  {https://doi.org/10.1093/nsr/nwac232} {\bibfield  {journal} {\bibinfo
  {journal} {National Science Review}\ }\textbf {\bibinfo {volume} {10}},\
  \bibinfo {pages} {nwac232} (\bibinfo {year}
  {2023}{\natexlab{b}})}\BibitemShut {NoStop}%
\bibitem [{\citenamefont {Duan}\ \emph {et~al.}(2022)\citenamefont {Duan},
  \citenamefont {Jian}, \citenamefont {Gao}, \citenamefont {Peng},
  \citenamefont {Zhong}, \citenamefont {Feng}, \citenamefont {Mao},\ and\
  \citenamefont {Yao}}]{duan2022giant}%
  \BibitemOpen
  \bibfield  {author} {\bibinfo {author} {\bibfnamefont {J.}~\bibnamefont
  {Duan}}, \bibinfo {author} {\bibfnamefont {Y.}~\bibnamefont {Jian}}, \bibinfo
  {author} {\bibfnamefont {Y.}~\bibnamefont {Gao}}, \bibinfo {author}
  {\bibfnamefont {H.}~\bibnamefont {Peng}}, \bibinfo {author} {\bibfnamefont
  {J.}~\bibnamefont {Zhong}}, \bibinfo {author} {\bibfnamefont
  {Q.}~\bibnamefont {Feng}}, \bibinfo {author} {\bibfnamefont {J.}~\bibnamefont
  {Mao}},\ and\ \bibinfo {author} {\bibfnamefont {Y.}~\bibnamefont {Yao}},\
  }\bibfield  {title} {\bibinfo {title} {Giant second-order nonlinear hall
  effect in twisted bilayer graphene},\ }\href
  {https://doi.org/10.1103/PhysRevLett.129.186801} {\bibfield  {journal}
  {\bibinfo  {journal} {Physical Review Letters}\ }\textbf {\bibinfo {volume}
  {129}},\ \bibinfo {pages} {186801} (\bibinfo {year} {2022})}\BibitemShut
  {NoStop}%
\bibitem [{\citenamefont {Sinha}\ \emph {et~al.}(2022)\citenamefont {Sinha},
  \citenamefont {Adak}, \citenamefont {Chakraborty}, \citenamefont {Das},
  \citenamefont {Debnath}, \citenamefont {Sangani}, \citenamefont {Watanabe},
  \citenamefont {Taniguchi}, \citenamefont {Waghmare}, \citenamefont {Agarwal}
  \emph {et~al.}}]{sinha2022berry}%
  \BibitemOpen
  \bibfield  {author} {\bibinfo {author} {\bibfnamefont {S.}~\bibnamefont
  {Sinha}}, \bibinfo {author} {\bibfnamefont {P.~C.}\ \bibnamefont {Adak}},
  \bibinfo {author} {\bibfnamefont {A.}~\bibnamefont {Chakraborty}}, \bibinfo
  {author} {\bibfnamefont {K.}~\bibnamefont {Das}}, \bibinfo {author}
  {\bibfnamefont {K.}~\bibnamefont {Debnath}}, \bibinfo {author} {\bibfnamefont
  {L.~V.}\ \bibnamefont {Sangani}}, \bibinfo {author} {\bibfnamefont
  {K.}~\bibnamefont {Watanabe}}, \bibinfo {author} {\bibfnamefont
  {T.}~\bibnamefont {Taniguchi}}, \bibinfo {author} {\bibfnamefont {U.~V.}\
  \bibnamefont {Waghmare}}, \bibinfo {author} {\bibfnamefont {A.}~\bibnamefont
  {Agarwal}}, \emph {et~al.},\ }\bibfield  {title} {\bibinfo {title} {Berry
  curvature dipole senses topological transition in a moir{\'e} superlattice},\
  }\href {https://doi.org/10.1038/s41567-022-01606-y} {\bibfield  {journal}
  {\bibinfo  {journal} {Nature Physics}\ }\textbf {\bibinfo {volume} {18}},\
  \bibinfo {pages} {765} (\bibinfo {year} {2022})}\BibitemShut {NoStop}%
\bibitem [{\citenamefont {Wang}\ \emph
  {et~al.}(2024{\natexlab{b}})\citenamefont {Wang}, \citenamefont {You},
  \citenamefont {Wang}, \citenamefont {Zhou}, \citenamefont {Zhang},
  \citenamefont {Lai}, \citenamefont {Feng}, \citenamefont {Lin}, \citenamefont
  {Chang},\ and\ \citenamefont {Gao}}]{wang2024non}%
  \BibitemOpen
  \bibfield  {author} {\bibinfo {author} {\bibfnamefont {N.}~\bibnamefont
  {Wang}}, \bibinfo {author} {\bibfnamefont {J.-Y.}\ \bibnamefont {You}},
  \bibinfo {author} {\bibfnamefont {A.}~\bibnamefont {Wang}}, \bibinfo {author}
  {\bibfnamefont {X.}~\bibnamefont {Zhou}}, \bibinfo {author} {\bibfnamefont
  {Z.}~\bibnamefont {Zhang}}, \bibinfo {author} {\bibfnamefont
  {S.}~\bibnamefont {Lai}}, \bibinfo {author} {\bibfnamefont {Y.-P.}\
  \bibnamefont {Feng}}, \bibinfo {author} {\bibfnamefont {H.}~\bibnamefont
  {Lin}}, \bibinfo {author} {\bibfnamefont {G.}~\bibnamefont {Chang}},\ and\
  \bibinfo {author} {\bibfnamefont {W.-b.}\ \bibnamefont {Gao}},\ }\bibfield
  {title} {\bibinfo {title} {Non-centrosymmetric topological phase probed by
  non-linear hall effect},\ }\href {https://doi.org/10.1093/nsr/nwad103}
  {\bibfield  {journal} {\bibinfo  {journal} {National Science Review}\
  }\textbf {\bibinfo {volume} {11}},\ \bibinfo {pages} {nwad103} (\bibinfo
  {year} {2024}{\natexlab{b}})}\BibitemShut {NoStop}%
\bibitem [{\citenamefont {Qin}\ \emph {et~al.}(2021)\citenamefont {Qin},
  \citenamefont {Zhu}, \citenamefont {Ye}, \citenamefont {Xu}, \citenamefont
  {Song}, \citenamefont {Liang}, \citenamefont {Liu},\ and\ \citenamefont
  {Liao}}]{qin2021strain}%
  \BibitemOpen
  \bibfield  {author} {\bibinfo {author} {\bibfnamefont {M.-S.}\ \bibnamefont
  {Qin}}, \bibinfo {author} {\bibfnamefont {P.-F.}\ \bibnamefont {Zhu}},
  \bibinfo {author} {\bibfnamefont {X.-G.}\ \bibnamefont {Ye}}, \bibinfo
  {author} {\bibfnamefont {W.-Z.}\ \bibnamefont {Xu}}, \bibinfo {author}
  {\bibfnamefont {Z.-H.}\ \bibnamefont {Song}}, \bibinfo {author}
  {\bibfnamefont {J.}~\bibnamefont {Liang}}, \bibinfo {author} {\bibfnamefont
  {K.}~\bibnamefont {Liu}},\ and\ \bibinfo {author} {\bibfnamefont {Z.-M.}\
  \bibnamefont {Liao}},\ }\bibfield  {title} {\bibinfo {title} {Strain tunable
  berry curvature dipole, orbital magnetization and nonlinear hall effect in
  wse2 monolayer},\ }\href {https://doi.org/10.1088/0256-307X/38/1/017301}
  {\bibfield  {journal} {\bibinfo  {journal} {Chinese Physics Letters}\
  }\textbf {\bibinfo {volume} {38}},\ \bibinfo {pages} {017301} (\bibinfo
  {year} {2021})}\BibitemShut {NoStop}%
\bibitem [{\citenamefont {Lai}\ \emph {et~al.}(2021)\citenamefont {Lai},
  \citenamefont {Liu}, \citenamefont {Zhang}, \citenamefont {Zhao},
  \citenamefont {Feng}, \citenamefont {Wang}, \citenamefont {Tang},
  \citenamefont {Liu}, \citenamefont {Novoselov}, \citenamefont {Yang} \emph
  {et~al.}}]{lai2021third}%
  \BibitemOpen
  \bibfield  {author} {\bibinfo {author} {\bibfnamefont {S.}~\bibnamefont
  {Lai}}, \bibinfo {author} {\bibfnamefont {H.}~\bibnamefont {Liu}}, \bibinfo
  {author} {\bibfnamefont {Z.}~\bibnamefont {Zhang}}, \bibinfo {author}
  {\bibfnamefont {J.}~\bibnamefont {Zhao}}, \bibinfo {author} {\bibfnamefont
  {X.}~\bibnamefont {Feng}}, \bibinfo {author} {\bibfnamefont {N.}~\bibnamefont
  {Wang}}, \bibinfo {author} {\bibfnamefont {C.}~\bibnamefont {Tang}}, \bibinfo
  {author} {\bibfnamefont {Y.}~\bibnamefont {Liu}}, \bibinfo {author}
  {\bibfnamefont {K.}~\bibnamefont {Novoselov}}, \bibinfo {author}
  {\bibfnamefont {S.~A.}\ \bibnamefont {Yang}}, \emph {et~al.},\ }\bibfield
  {title} {\bibinfo {title} {Third-order nonlinear hall effect induced by the
  berry-connection polarizability tensor},\ }\href
  {https://doi.org/10.1038/s41565-021-00917-0} {\bibfield  {journal} {\bibinfo
  {journal} {Nature Nanotechnology}\ }\textbf {\bibinfo {volume} {16}},\
  \bibinfo {pages} {869} (\bibinfo {year} {2021})}\BibitemShut {NoStop}%
\bibitem [{\citenamefont {Wang}\ \emph {et~al.}(2022)\citenamefont {Wang},
  \citenamefont {Xiao}, \citenamefont {Liu}, \citenamefont {Zhang},
  \citenamefont {Lai}, \citenamefont {Zhu}, \citenamefont {Cai}, \citenamefont
  {Wang}, \citenamefont {Chen}, \citenamefont {Deng} \emph
  {et~al.}}]{wang2022room}%
  \BibitemOpen
  \bibfield  {author} {\bibinfo {author} {\bibfnamefont {C.}~\bibnamefont
  {Wang}}, \bibinfo {author} {\bibfnamefont {R.-C.}\ \bibnamefont {Xiao}},
  \bibinfo {author} {\bibfnamefont {H.}~\bibnamefont {Liu}}, \bibinfo {author}
  {\bibfnamefont {Z.}~\bibnamefont {Zhang}}, \bibinfo {author} {\bibfnamefont
  {S.}~\bibnamefont {Lai}}, \bibinfo {author} {\bibfnamefont {C.}~\bibnamefont
  {Zhu}}, \bibinfo {author} {\bibfnamefont {H.}~\bibnamefont {Cai}}, \bibinfo
  {author} {\bibfnamefont {N.}~\bibnamefont {Wang}}, \bibinfo {author}
  {\bibfnamefont {S.}~\bibnamefont {Chen}}, \bibinfo {author} {\bibfnamefont
  {Y.}~\bibnamefont {Deng}}, \emph {et~al.},\ }\bibfield  {title} {\bibinfo
  {title} {Room-temperature third-order nonlinear hall effect in weyl semimetal
  tairte4},\ }\href {https://doi.org/10.1093/nsr/nwac020} {\bibfield  {journal}
  {\bibinfo  {journal} {National Science Review}\ }\textbf {\bibinfo {volume}
  {9}},\ \bibinfo {pages} {nwac020} (\bibinfo {year} {2022})}\BibitemShut
  {NoStop}%
\bibitem [{\citenamefont {He}\ \emph {et~al.}(2024)\citenamefont {He},
  \citenamefont {Isobe}, \citenamefont {Koon}, \citenamefont {Tan},
  \citenamefont {Hu}, \citenamefont {Li}, \citenamefont {Nagaosa},\ and\
  \citenamefont {Shen}}]{he2024third}%
  \BibitemOpen
  \bibfield  {author} {\bibinfo {author} {\bibfnamefont {P.}~\bibnamefont
  {He}}, \bibinfo {author} {\bibfnamefont {H.}~\bibnamefont {Isobe}}, \bibinfo
  {author} {\bibfnamefont {G.~K.~W.}\ \bibnamefont {Koon}}, \bibinfo {author}
  {\bibfnamefont {J.~Y.}\ \bibnamefont {Tan}}, \bibinfo {author} {\bibfnamefont
  {J.}~\bibnamefont {Hu}}, \bibinfo {author} {\bibfnamefont {J.}~\bibnamefont
  {Li}}, \bibinfo {author} {\bibfnamefont {N.}~\bibnamefont {Nagaosa}},\ and\
  \bibinfo {author} {\bibfnamefont {J.}~\bibnamefont {Shen}},\ }\bibfield
  {title} {\bibinfo {title} {Third-order nonlinear hall effect in a quantum
  hall system},\ }\href {https://doi.org/10.1038/s41565-024-01730-1} {\bibfield
   {journal} {\bibinfo  {journal} {Nature Nanotechnology}\ ,\ \bibinfo {pages}
  {1}} (\bibinfo {year} {2024})}\BibitemShut {NoStop}%
\bibitem [{\citenamefont {Ye}\ \emph {et~al.}(2022)\citenamefont {Ye},
  \citenamefont {Zhu}, \citenamefont {Xu}, \citenamefont {Zang}, \citenamefont
  {Ye},\ and\ \citenamefont {Liao}}]{ye2022orbital}%
  \BibitemOpen
  \bibfield  {author} {\bibinfo {author} {\bibfnamefont {X.-G.}\ \bibnamefont
  {Ye}}, \bibinfo {author} {\bibfnamefont {P.-F.}\ \bibnamefont {Zhu}},
  \bibinfo {author} {\bibfnamefont {W.-Z.}\ \bibnamefont {Xu}}, \bibinfo
  {author} {\bibfnamefont {Z.}~\bibnamefont {Zang}}, \bibinfo {author}
  {\bibfnamefont {Y.}~\bibnamefont {Ye}},\ and\ \bibinfo {author}
  {\bibfnamefont {Z.-M.}\ \bibnamefont {Liao}},\ }\bibfield  {title} {\bibinfo
  {title} {Orbital polarization and third-order anomalous hall effect in w te
  2},\ }\href {https://doi.org/10.1103/PhysRevB.106.045414} {\bibfield
  {journal} {\bibinfo  {journal} {Physical Review B}\ }\textbf {\bibinfo
  {volume} {106}},\ \bibinfo {pages} {045414} (\bibinfo {year}
  {2022})}\BibitemShut {NoStop}%
\bibitem [{\citenamefont {Chen}\ \emph
  {et~al.}(2024{\natexlab{b}})\citenamefont {Chen}, \citenamefont {Liao},
  \citenamefont {Dong}, \citenamefont {Liu}, \citenamefont {Zhao},
  \citenamefont {Li}, \citenamefont {Wang},\ and\ \citenamefont
  {Liao}}]{chen2024charge}%
  \BibitemOpen
  \bibfield  {author} {\bibinfo {author} {\bibfnamefont {Z.-H.}\ \bibnamefont
  {Chen}}, \bibinfo {author} {\bibfnamefont {X.}~\bibnamefont {Liao}}, \bibinfo
  {author} {\bibfnamefont {J.-W.}\ \bibnamefont {Dong}}, \bibinfo {author}
  {\bibfnamefont {X.-Y.}\ \bibnamefont {Liu}}, \bibinfo {author} {\bibfnamefont
  {T.-Y.}\ \bibnamefont {Zhao}}, \bibinfo {author} {\bibfnamefont
  {D.}~\bibnamefont {Li}}, \bibinfo {author} {\bibfnamefont {A.-Q.}\
  \bibnamefont {Wang}},\ and\ \bibinfo {author} {\bibfnamefont {Z.-M.}\
  \bibnamefont {Liao}},\ }\bibfield  {title} {\bibinfo {title} {Charge density
  wave modulated third-order nonlinear hall effect in 1 t-vse 2 nanosheets},\
  }\href {https://doi.org/10.1103/PhysRevB.110.235135} {\bibfield  {journal}
  {\bibinfo  {journal} {Physical Review B}\ }\textbf {\bibinfo {volume}
  {110}},\ \bibinfo {pages} {235135} (\bibinfo {year}
  {2024}{\natexlab{b}})}\BibitemShut {NoStop}%
\bibitem [{\citenamefont {Li}\ \emph {et~al.}(2024{\natexlab{d}})\citenamefont
  {Li}, \citenamefont {Wang}, \citenamefont {Yang}, \citenamefont {Zhang},
  \citenamefont {Teo}, \citenamefont {Lin}, \citenamefont {He}, \citenamefont
  {Wang}, \citenamefont {Song}, \citenamefont {Tian} \emph
  {et~al.}}]{li2024giant}%
  \BibitemOpen
  \bibfield  {author} {\bibinfo {author} {\bibfnamefont {S.}~\bibnamefont
  {Li}}, \bibinfo {author} {\bibfnamefont {X.}~\bibnamefont {Wang}}, \bibinfo
  {author} {\bibfnamefont {Z.}~\bibnamefont {Yang}}, \bibinfo {author}
  {\bibfnamefont {L.}~\bibnamefont {Zhang}}, \bibinfo {author} {\bibfnamefont
  {S.~L.}\ \bibnamefont {Teo}}, \bibinfo {author} {\bibfnamefont
  {M.}~\bibnamefont {Lin}}, \bibinfo {author} {\bibfnamefont {R.}~\bibnamefont
  {He}}, \bibinfo {author} {\bibfnamefont {N.}~\bibnamefont {Wang}}, \bibinfo
  {author} {\bibfnamefont {P.}~\bibnamefont {Song}}, \bibinfo {author}
  {\bibfnamefont {W.}~\bibnamefont {Tian}}, \emph {et~al.},\ }\bibfield
  {title} {\bibinfo {title} {Giant third-order nonlinear hall effect in misfit
  layer compound (sns) 1.17 (nbs2) 3},\ }\href
  {https://doi.org/10.1021/acsami.3c18319} {\bibfield  {journal} {\bibinfo
  {journal} {ACS Applied Materials \& Interfaces}\ }\textbf {\bibinfo {volume}
  {16}},\ \bibinfo {pages} {11043} (\bibinfo {year}
  {2024}{\natexlab{d}})}\BibitemShut {NoStop}%
\bibitem [{\citenamefont {Xiao}\ \emph
  {et~al.}(2020{\natexlab{a}})\citenamefont {Xiao}, \citenamefont {Shao},
  \citenamefont {Zhang},\ and\ \citenamefont {Jiang}}]{xiao2020two}%
  \BibitemOpen
  \bibfield  {author} {\bibinfo {author} {\bibfnamefont {R.-C.}\ \bibnamefont
  {Xiao}}, \bibinfo {author} {\bibfnamefont {D.-F.}\ \bibnamefont {Shao}},
  \bibinfo {author} {\bibfnamefont {Z.-Q.}\ \bibnamefont {Zhang}},\ and\
  \bibinfo {author} {\bibfnamefont {H.}~\bibnamefont {Jiang}},\ }\bibfield
  {title} {\bibinfo {title} {Two-dimensional metals for piezoelectriclike
  devices based on berry-curvature dipole},\ }\href
  {https://doi.org/10.1103/PhysRevApplied.13.044014} {\bibfield  {journal}
  {\bibinfo  {journal} {Physical Review Applied}\ }\textbf {\bibinfo {volume}
  {13}},\ \bibinfo {pages} {044014} (\bibinfo {year}
  {2020}{\natexlab{a}})}\BibitemShut {NoStop}%
\bibitem [{\citenamefont {Hamamoto}\ \emph {et~al.}(2017)\citenamefont
  {Hamamoto}, \citenamefont {Ezawa}, \citenamefont {Kim}, \citenamefont
  {Morimoto},\ and\ \citenamefont {Nagaosa}}]{hamamoto2017nonlinear}%
  \BibitemOpen
  \bibfield  {author} {\bibinfo {author} {\bibfnamefont {K.}~\bibnamefont
  {Hamamoto}}, \bibinfo {author} {\bibfnamefont {M.}~\bibnamefont {Ezawa}},
  \bibinfo {author} {\bibfnamefont {K.~W.}\ \bibnamefont {Kim}}, \bibinfo
  {author} {\bibfnamefont {T.}~\bibnamefont {Morimoto}},\ and\ \bibinfo
  {author} {\bibfnamefont {N.}~\bibnamefont {Nagaosa}},\ }\bibfield  {title}
  {\bibinfo {title} {Nonlinear spin current generation in noncentrosymmetric
  spin-orbit coupled systems},\ }\href
  {https://doi.org/10.1103/PhysRevB.95.224430} {\bibfield  {journal} {\bibinfo
  {journal} {Physical Review B}\ }\textbf {\bibinfo {volume} {95}},\ \bibinfo
  {pages} {224430} (\bibinfo {year} {2017})}\BibitemShut {NoStop}%
\bibitem [{\citenamefont {Araki}(2018)}]{araki2018strain}%
  \BibitemOpen
  \bibfield  {author} {\bibinfo {author} {\bibfnamefont {Y.}~\bibnamefont
  {Araki}},\ }\bibfield  {title} {\bibinfo {title} {Strain-induced nonlinear
  spin hall effect in topological dirac semimetal},\ }\href
  {https://doi.org/10.1038/s41598-018-33655-w} {\bibfield  {journal} {\bibinfo
  {journal} {Scientific reports}\ }\textbf {\bibinfo {volume} {8}},\ \bibinfo
  {pages} {15236} (\bibinfo {year} {2018})}\BibitemShut {NoStop}%
\bibitem [{\citenamefont {Zeng}\ \emph {et~al.}(2019)\citenamefont {Zeng},
  \citenamefont {Nandy}, \citenamefont {Taraphder},\ and\ \citenamefont
  {Tewari}}]{zeng2019nonlinear}%
  \BibitemOpen
  \bibfield  {author} {\bibinfo {author} {\bibfnamefont {C.}~\bibnamefont
  {Zeng}}, \bibinfo {author} {\bibfnamefont {S.}~\bibnamefont {Nandy}},
  \bibinfo {author} {\bibfnamefont {A.}~\bibnamefont {Taraphder}},\ and\
  \bibinfo {author} {\bibfnamefont {S.}~\bibnamefont {Tewari}},\ }\bibfield
  {title} {\bibinfo {title} {Nonlinear nernst effect in bilayer wte 2},\ }\href
  {https://doi.org/10.1103/PhysRevB.100.245102} {\bibfield  {journal} {\bibinfo
   {journal} {Physical Review B}\ }\textbf {\bibinfo {volume} {100}},\ \bibinfo
  {pages} {245102} (\bibinfo {year} {2019})}\BibitemShut {NoStop}%
\bibitem [{\citenamefont {Zeng}\ \emph {et~al.}(2020)\citenamefont {Zeng},
  \citenamefont {Nandy},\ and\ \citenamefont {Tewari}}]{zeng2020fundamental}%
  \BibitemOpen
  \bibfield  {author} {\bibinfo {author} {\bibfnamefont {C.}~\bibnamefont
  {Zeng}}, \bibinfo {author} {\bibfnamefont {S.}~\bibnamefont {Nandy}},\ and\
  \bibinfo {author} {\bibfnamefont {S.}~\bibnamefont {Tewari}},\ }\bibfield
  {title} {\bibinfo {title} {Fundamental relations for anomalous thermoelectric
  transport coefficients in the nonlinear regime},\ }\href
  {https://doi.org/10.1103/PhysRevResearch.2.032066} {\bibfield  {journal}
  {\bibinfo  {journal} {Physical Review Research}\ }\textbf {\bibinfo {volume}
  {2}},\ \bibinfo {pages} {032066} (\bibinfo {year} {2020})}\BibitemShut
  {NoStop}%
\bibitem [{\citenamefont {Nakai}\ and\ \citenamefont
  {Nagaosa}(2019)}]{nakai2019nonreciprocal}%
  \BibitemOpen
  \bibfield  {author} {\bibinfo {author} {\bibfnamefont {R.}~\bibnamefont
  {Nakai}}\ and\ \bibinfo {author} {\bibfnamefont {N.}~\bibnamefont
  {Nagaosa}},\ }\bibfield  {title} {\bibinfo {title} {Nonreciprocal thermal and
  thermoelectric transport of electrons in noncentrosymmetric crystals},\
  }\href {https://doi.org/10.1103/PhysRevB.99.115201} {\bibfield  {journal}
  {\bibinfo  {journal} {Physical Review B}\ }\textbf {\bibinfo {volume} {99}},\
  \bibinfo {pages} {115201} (\bibinfo {year} {2019})}\BibitemShut {NoStop}%
\bibitem [{\citenamefont {Yu}\ \emph {et~al.}(2019)\citenamefont {Yu},
  \citenamefont {Zhu}, \citenamefont {You}, \citenamefont {Low},\ and\
  \citenamefont {Su}}]{yu2019topological}%
  \BibitemOpen
  \bibfield  {author} {\bibinfo {author} {\bibfnamefont {X.-Q.}\ \bibnamefont
  {Yu}}, \bibinfo {author} {\bibfnamefont {Z.-G.}\ \bibnamefont {Zhu}},
  \bibinfo {author} {\bibfnamefont {J.-S.}\ \bibnamefont {You}}, \bibinfo
  {author} {\bibfnamefont {T.}~\bibnamefont {Low}},\ and\ \bibinfo {author}
  {\bibfnamefont {G.}~\bibnamefont {Su}},\ }\bibfield  {title} {\bibinfo
  {title} {Topological nonlinear anomalous nernst effect in strained transition
  metal dichalcogenides},\ }\href {https://doi.org/10.1103/PhysRevB.99.201410}
  {\bibfield  {journal} {\bibinfo  {journal} {Physical Review B}\ }\textbf
  {\bibinfo {volume} {99}},\ \bibinfo {pages} {201410} (\bibinfo {year}
  {2019})}\BibitemShut {NoStop}%
\bibitem [{\citenamefont {Kumar}\ \emph {et~al.}(2018)\citenamefont {Kumar},
  \citenamefont {Guin}, \citenamefont {Felser},\ and\ \citenamefont
  {Shekhar}}]{kumar2018planar}%
  \BibitemOpen
  \bibfield  {author} {\bibinfo {author} {\bibfnamefont {N.}~\bibnamefont
  {Kumar}}, \bibinfo {author} {\bibfnamefont {S.~N.}\ \bibnamefont {Guin}},
  \bibinfo {author} {\bibfnamefont {C.}~\bibnamefont {Felser}},\ and\ \bibinfo
  {author} {\bibfnamefont {C.}~\bibnamefont {Shekhar}},\ }\bibfield  {title}
  {\bibinfo {title} {Planar hall effect in the weyl semimetal gdptbi},\ }\href
  {https://doi.org/10.1103/PhysRevB.98.041103} {\bibfield  {journal} {\bibinfo
  {journal} {Physical Review B}\ }\textbf {\bibinfo {volume} {98}},\ \bibinfo
  {pages} {041103} (\bibinfo {year} {2018})}\BibitemShut {NoStop}%
\bibitem [{\citenamefont {Burkov}(2017)}]{burkov2017giant}%
  \BibitemOpen
  \bibfield  {author} {\bibinfo {author} {\bibfnamefont {A.}~\bibnamefont
  {Burkov}},\ }\bibfield  {title} {\bibinfo {title} {Giant planar hall effect
  in topological metals},\ }\href {https://doi.org/10.1103/PhysRevB.96.041110}
  {\bibfield  {journal} {\bibinfo  {journal} {Physical Review B}\ }\textbf
  {\bibinfo {volume} {96}},\ \bibinfo {pages} {041110} (\bibinfo {year}
  {2017})}\BibitemShut {NoStop}%
\bibitem [{\citenamefont {Tang}\ \emph {et~al.}(2003)\citenamefont {Tang},
  \citenamefont {Kawakami}, \citenamefont {Awschalom},\ and\ \citenamefont
  {Roukes}}]{tang2003giant}%
  \BibitemOpen
  \bibfield  {author} {\bibinfo {author} {\bibfnamefont {H.}~\bibnamefont
  {Tang}}, \bibinfo {author} {\bibfnamefont {R.}~\bibnamefont {Kawakami}},
  \bibinfo {author} {\bibfnamefont {D.}~\bibnamefont {Awschalom}},\ and\
  \bibinfo {author} {\bibfnamefont {M.}~\bibnamefont {Roukes}},\ }\bibfield
  {title} {\bibinfo {title} {Giant planar hall effect in epitaxial (ga, mn) as
  devices},\ }\href {https://doi.org/10.1103/PhysRevLett.90.107201} {\bibfield
  {journal} {\bibinfo  {journal} {Physical review letters}\ }\textbf {\bibinfo
  {volume} {90}},\ \bibinfo {pages} {107201} (\bibinfo {year}
  {2003})}\BibitemShut {NoStop}%
\bibitem [{\citenamefont {He}\ \emph {et~al.}(2019)\citenamefont {He},
  \citenamefont {Zhang}, \citenamefont {Zhu}, \citenamefont {Shi},
  \citenamefont {Heinonen}, \citenamefont {Vignale},\ and\ \citenamefont
  {Yang}}]{he2019nonlinearl}%
  \BibitemOpen
  \bibfield  {author} {\bibinfo {author} {\bibfnamefont {P.}~\bibnamefont
  {He}}, \bibinfo {author} {\bibfnamefont {S.~S.-L.}\ \bibnamefont {Zhang}},
  \bibinfo {author} {\bibfnamefont {D.}~\bibnamefont {Zhu}}, \bibinfo {author}
  {\bibfnamefont {S.}~\bibnamefont {Shi}}, \bibinfo {author} {\bibfnamefont
  {O.~G.}\ \bibnamefont {Heinonen}}, \bibinfo {author} {\bibfnamefont
  {G.}~\bibnamefont {Vignale}},\ and\ \bibinfo {author} {\bibfnamefont
  {H.}~\bibnamefont {Yang}},\ }\bibfield  {title} {\bibinfo {title} {Nonlinear
  planar hall effect},\ }\href {https://doi.org/10.1103/PhysRevLett.123.016801}
  {\bibfield  {journal} {\bibinfo  {journal} {Physical review letters}\
  }\textbf {\bibinfo {volume} {123}},\ \bibinfo {pages} {016801} (\bibinfo
  {year} {2019})}\BibitemShut {NoStop}%
\bibitem [{\citenamefont {Rao}\ \emph {et~al.}(2021)\citenamefont {Rao},
  \citenamefont {Zhou}, \citenamefont {Wu}, \citenamefont {Duan}, \citenamefont
  {Deng},\ and\ \citenamefont {Wang}}]{rao2021theory}%
  \BibitemOpen
  \bibfield  {author} {\bibinfo {author} {\bibfnamefont {W.}~\bibnamefont
  {Rao}}, \bibinfo {author} {\bibfnamefont {Y.-L.}\ \bibnamefont {Zhou}},
  \bibinfo {author} {\bibfnamefont {Y.-j.}\ \bibnamefont {Wu}}, \bibinfo
  {author} {\bibfnamefont {H.-J.}\ \bibnamefont {Duan}}, \bibinfo {author}
  {\bibfnamefont {M.-X.}\ \bibnamefont {Deng}},\ and\ \bibinfo {author}
  {\bibfnamefont {R.-Q.}\ \bibnamefont {Wang}},\ }\bibfield  {title} {\bibinfo
  {title} {Theory for linear and nonlinear planar hall effect in topological
  insulator thin films},\ }\href {https://doi.org/10.1103/PhysRevB.103.155415}
  {\bibfield  {journal} {\bibinfo  {journal} {Physical Review B}\ }\textbf
  {\bibinfo {volume} {103}},\ \bibinfo {pages} {155415} (\bibinfo {year}
  {2021})}\BibitemShut {NoStop}%
\bibitem [{\citenamefont {Xiao}\ \emph
  {et~al.}(2020{\natexlab{b}})\citenamefont {Xiao}, \citenamefont {Wang},
  \citenamefont {Wang}, \citenamefont {Pemmaraju}, \citenamefont {Wang},
  \citenamefont {Muscher}, \citenamefont {Sie}, \citenamefont {Nyby},
  \citenamefont {Devereaux}, \citenamefont {Qian} \emph
  {et~al.}}]{xiao2020berry}%
  \BibitemOpen
  \bibfield  {author} {\bibinfo {author} {\bibfnamefont {J.}~\bibnamefont
  {Xiao}}, \bibinfo {author} {\bibfnamefont {Y.}~\bibnamefont {Wang}}, \bibinfo
  {author} {\bibfnamefont {H.}~\bibnamefont {Wang}}, \bibinfo {author}
  {\bibfnamefont {C.}~\bibnamefont {Pemmaraju}}, \bibinfo {author}
  {\bibfnamefont {S.}~\bibnamefont {Wang}}, \bibinfo {author} {\bibfnamefont
  {P.}~\bibnamefont {Muscher}}, \bibinfo {author} {\bibfnamefont {E.~J.}\
  \bibnamefont {Sie}}, \bibinfo {author} {\bibfnamefont {C.~M.}\ \bibnamefont
  {Nyby}}, \bibinfo {author} {\bibfnamefont {T.~P.}\ \bibnamefont {Devereaux}},
  \bibinfo {author} {\bibfnamefont {X.}~\bibnamefont {Qian}}, \emph {et~al.},\
  }\bibfield  {title} {\bibinfo {title} {Berry curvature memory through
  electrically driven stacking transitions},\ }\href
  {https://doi.org/10.1038/s41567-020-0947-0} {\bibfield  {journal} {\bibinfo
  {journal} {Nature Physics}\ }\textbf {\bibinfo {volume} {16}},\ \bibinfo
  {pages} {1028} (\bibinfo {year} {2020}{\natexlab{b}})}\BibitemShut {NoStop}%
\bibitem [{\citenamefont {Shao}\ \emph {et~al.}(2020)\citenamefont {Shao},
  \citenamefont {Zhang}, \citenamefont {Gurung}, \citenamefont {Yang},\ and\
  \citenamefont {Tsymbal}}]{shao2020nonlinear}%
  \BibitemOpen
  \bibfield  {author} {\bibinfo {author} {\bibfnamefont {D.-F.}\ \bibnamefont
  {Shao}}, \bibinfo {author} {\bibfnamefont {S.-H.}\ \bibnamefont {Zhang}},
  \bibinfo {author} {\bibfnamefont {G.}~\bibnamefont {Gurung}}, \bibinfo
  {author} {\bibfnamefont {W.}~\bibnamefont {Yang}},\ and\ \bibinfo {author}
  {\bibfnamefont {E.~Y.}\ \bibnamefont {Tsymbal}},\ }\bibfield  {title}
  {\bibinfo {title} {Nonlinear anomalous hall effect for n{\'e}el vector
  detection},\ }\href {https://doi.org/10.1103/PhysRevLett.124.067203}
  {\bibfield  {journal} {\bibinfo  {journal} {Physical Review Letters}\
  }\textbf {\bibinfo {volume} {124}},\ \bibinfo {pages} {067203} (\bibinfo
  {year} {2020})}\BibitemShut {NoStop}%
\bibitem [{\citenamefont {Zhang}\ and\ \citenamefont
  {Fu}(2021)}]{zhang2021terahertz}%
  \BibitemOpen
  \bibfield  {author} {\bibinfo {author} {\bibfnamefont {Y.}~\bibnamefont
  {Zhang}}\ and\ \bibinfo {author} {\bibfnamefont {L.}~\bibnamefont {Fu}},\
  }\bibfield  {title} {\bibinfo {title} {Terahertz detection based on nonlinear
  hall effect without magnetic field},\ }\href
  {https://doi.org/10.1073/pnas.2100736118} {\bibfield  {journal} {\bibinfo
  {journal} {Proceedings of the National Academy of Sciences}\ }\textbf
  {\bibinfo {volume} {118}},\ \bibinfo {pages} {e2100736118} (\bibinfo {year}
  {2021})}\BibitemShut {NoStop}%
\bibitem [{\citenamefont {Su{\'a}rez-Rodr{\'\i}guez}\ \emph
  {et~al.}(2024{\natexlab{b}})\citenamefont {Su{\'a}rez-Rodr{\'\i}guez},
  \citenamefont {Mart{\'\i}n-Garc{\'\i}a}, \citenamefont {Skowro{\'n}ski},
  \citenamefont {Staszek}, \citenamefont {Calavalle}, \citenamefont {Fert},
  \citenamefont {Gobbi}, \citenamefont {Casanova},\ and\ \citenamefont
  {Hueso}}]{suarez2024microscale}%
  \BibitemOpen
  \bibfield  {author} {\bibinfo {author} {\bibfnamefont {M.}~\bibnamefont
  {Su{\'a}rez-Rodr{\'\i}guez}}, \bibinfo {author} {\bibfnamefont
  {B.}~\bibnamefont {Mart{\'\i}n-Garc{\'\i}a}}, \bibinfo {author}
  {\bibfnamefont {W.}~\bibnamefont {Skowro{\'n}ski}}, \bibinfo {author}
  {\bibfnamefont {K.}~\bibnamefont {Staszek}}, \bibinfo {author} {\bibfnamefont
  {F.}~\bibnamefont {Calavalle}}, \bibinfo {author} {\bibfnamefont
  {A.}~\bibnamefont {Fert}}, \bibinfo {author} {\bibfnamefont {M.}~\bibnamefont
  {Gobbi}}, \bibinfo {author} {\bibfnamefont {F.}~\bibnamefont {Casanova}},\
  and\ \bibinfo {author} {\bibfnamefont {L.~E.}\ \bibnamefont {Hueso}},\
  }\bibfield  {title} {\bibinfo {title} {Microscale chiral rectennas for energy
  harvesting},\ }\href {https://doi.org/10.1002/adma.202400729} {\bibfield
  {journal} {\bibinfo  {journal} {Advanced Materials}\ ,\ \bibinfo {pages}
  {2400729}} (\bibinfo {year} {2024}{\natexlab{b}})}\BibitemShut {NoStop}%
\bibitem [{\citenamefont {Muhammad}\ \emph {et~al.}(2022)\citenamefont
  {Muhammad}, \citenamefont {Tiang}, \citenamefont {Wong}, \citenamefont
  {Rambe}, \citenamefont {Adam}, \citenamefont {Smida}, \citenamefont {Waly},
  \citenamefont {Iqbal}, \citenamefont {Abubakar},\ and\ \citenamefont
  {Mohd~Yasin}}]{muhammad2022harvesting}%
  \BibitemOpen
  \bibfield  {author} {\bibinfo {author} {\bibfnamefont {S.}~\bibnamefont
  {Muhammad}}, \bibinfo {author} {\bibfnamefont {J.~J.}\ \bibnamefont {Tiang}},
  \bibinfo {author} {\bibfnamefont {S.~K.}\ \bibnamefont {Wong}}, \bibinfo
  {author} {\bibfnamefont {A.~H.}\ \bibnamefont {Rambe}}, \bibinfo {author}
  {\bibfnamefont {I.}~\bibnamefont {Adam}}, \bibinfo {author} {\bibfnamefont
  {A.}~\bibnamefont {Smida}}, \bibinfo {author} {\bibfnamefont {M.~I.}\
  \bibnamefont {Waly}}, \bibinfo {author} {\bibfnamefont {A.}~\bibnamefont
  {Iqbal}}, \bibinfo {author} {\bibfnamefont {A.~S.}\ \bibnamefont
  {Abubakar}},\ and\ \bibinfo {author} {\bibfnamefont {M.~N.}\ \bibnamefont
  {Mohd~Yasin}},\ }\bibfield  {title} {\bibinfo {title} {Harvesting systems for
  rf energy: trends, challenges, techniques, and tradeoffs},\ }\href
  {https://doi.org/10.3390/electronics11060959} {\bibfield  {journal} {\bibinfo
   {journal} {Electronics}\ }\textbf {\bibinfo {volume} {11}},\ \bibinfo
  {pages} {959} (\bibinfo {year} {2022})}\BibitemShut {NoStop}%
\bibitem [{\citenamefont {Suárez-Rodríguez}\ \emph
  {et~al.}(2024)\citenamefont {Suárez-Rodríguez}, \citenamefont {Juan},
  \citenamefont {Souza}, \citenamefont {Gobbi}, \citenamefont {Casanova},\ and\
  \citenamefont {Hueso}}]{Hueso-review-arxiv2024}%
  \BibitemOpen
  \bibfield  {author} {\bibinfo {author} {\bibfnamefont {M.}~\bibnamefont
  {Suárez-Rodríguez}}, \bibinfo {author} {\bibfnamefont {F.~D.}\ \bibnamefont
  {Juan}}, \bibinfo {author} {\bibfnamefont {I.}~\bibnamefont {Souza}},
  \bibinfo {author} {\bibfnamefont {M.}~\bibnamefont {Gobbi}}, \bibinfo
  {author} {\bibfnamefont {F.}~\bibnamefont {Casanova}},\ and\ \bibinfo
  {author} {\bibfnamefont {L.~E.}\ \bibnamefont {Hueso}},\ }\href
  {https://arxiv.org/abs/2412.05253} {\bibinfo {title} {Non-linear transport in
  non-centrosymmetric systems: From fundamentals to applications}} (\bibinfo
  {year} {2024}),\ \Eprint {https://arxiv.org/abs/2412.05253} {arXiv:2412.05253
  [cond-mat.mes-hall]} \BibitemShut {NoStop}%
\bibitem [{\citenamefont {Qin}\ \emph {et~al.}(2024{\natexlab{a}})\citenamefont
  {Qin}, \citenamefont {Chen},\ and\ \citenamefont {Lee}}]{qin2024light}%
  \BibitemOpen
  \bibfield  {author} {\bibinfo {author} {\bibfnamefont {F.}~\bibnamefont
  {Qin}}, \bibinfo {author} {\bibfnamefont {R.}~\bibnamefont {Chen}},\ and\
  \bibinfo {author} {\bibfnamefont {C.~H.}\ \bibnamefont {Lee}},\ }\bibfield
  {title} {\bibinfo {title} {Light-enhanced nonlinear hall effect},\ }\href
  {https://doi.org/10.1103/10.1038/s42005-024-01820-5} {\bibfield  {journal}
  {\bibinfo  {journal} {Communications Physics}\ }\textbf {\bibinfo {volume}
  {7}},\ \bibinfo {pages} {368} (\bibinfo {year}
  {2024}{\natexlab{a}})}\BibitemShut {NoStop}%
\bibitem [{\citenamefont {Qin}\ \emph {et~al.}(2024{\natexlab{b}})\citenamefont
  {Qin}, \citenamefont {Shen},\ and\ \citenamefont {Lee}}]{qin2024nonlinear}%
  \BibitemOpen
  \bibfield  {author} {\bibinfo {author} {\bibfnamefont {F.}~\bibnamefont
  {Qin}}, \bibinfo {author} {\bibfnamefont {R.}~\bibnamefont {Shen}},\ and\
  \bibinfo {author} {\bibfnamefont {C.~H.}\ \bibnamefont {Lee}},\ }\bibfield
  {title} {\bibinfo {title} {Nonlinear hall effects with an exceptional ring},\
  }\bibfield  {journal} {\bibinfo  {journal} {arXiv preprint arXiv:2411.06509}\
  }\href {https://doi.org/10.48550/arXiv.2411.06509}
  {10.48550/arXiv.2411.06509} (\bibinfo {year}
  {2024}{\natexlab{b}})\BibitemShut {NoStop}%
\end{thebibliography}
%

\end{document}